\pdfoutput=1
\documentclass[iop, apj, twocolappendix, numberedappendix]{emulateapj}

\newif\iflocal
\localfalse

\usepackage{xspace}
\usepackage{amsmath}
\usepackage{framed} 
\usepackage{txfonts}
\usepackage{epstopdf}
\usepackage{color}
\usepackage{rotating}
\usepackage{natbib}
\usepackage{ulem}
\usepackage{xspace}
\usepackage[colorlinks=true,urlcolor=black,linkcolor=blue,citecolor=blue]{hyperref}
\usepackage{sistyle}
\SIthousandsep{,}

\iflocal
\def\includedir{/Users/benedito/University/docs/latex}
\def\figdir{figs}
\else
\def\includedir{.}
\def\figdir{.}
\fi

\usepackage{etoolbox}
\makeatletter
\patchcmd{\NAT@citex}
  {\@citea\NAT@hyper@{\NAT@nmfmt{\NAT@nm}\NAT@date}}
  {\@citea\NAT@nmfmt{\NAT@nm}\NAT@hyper@{\NAT@date}}
  {}
  {}
\patchcmd{\NAT@citex}
  {\@citea\NAT@hyper@{%
     \NAT@nmfmt{\NAT@nm}%
     \hyper@natlinkbreak{\NAT@aysep\NAT@spacechar}{\@citeb\@extra@b@citeb}%
     \NAT@date}}
  {\@citea\NAT@nmfmt{\NAT@nm}%
   \NAT@aysep\NAT@spacechar%
   \NAT@hyper@{\NAT@date}}
  {}
  {}
\patchcmd{\NAT@citex}
  {\@citea\NAT@hyper@{%
     \NAT@nmfmt{\NAT@nm}%
     \hyper@natlinkbreak{\NAT@spacechar\NAT@@open\if*#1*\else#1\NAT@spacechar\fi}%
       {\@citeb\@extra@b@citeb}%
     \NAT@date}}
  {\@citea\NAT@nmfmt{\NAT@nm}%
   \NAT@spacechar\NAT@@open\if*#1*\else#1\NAT@spacechar\fi%
   \NAT@hyper@{\NAT@date}}
  {}
  {}
\makeatother

\newcommand{\kpch}{\>{h^{-1}{\rm kpc}}}
\newcommand{\mpch}{\>h^{-1}{\rm {Mpc}}}

\newcommand{\msunh}{\>h^{-1} M_\odot}

\def\gcm3{\mathrm{g} / \mathrm{cm}^3}

\def\LCDM{$\Lambda$CDM\xspace}
\def\rhoc{\rho_{\rm c}}
\def\rhom{\rho_{\rm m}}

\def\gtsima{$\; \buildrel > \over \sim \;$}
\def\ltsima{$\; \buildrel < \over \sim \;$}
\def\prosima{$\; \buildrel \propto \over \sim \;$}
\def\gsim{\lower.7ex\hbox{\gtsima}}
\def\lsim{\lower.7ex\hbox{\ltsima}}
\def\simgt{\lower.7ex\hbox{\gtsima}}
\def\simlt{\lower.7ex\hbox{\ltsima}}
\def\simpr{\lower.7ex\hbox{\prosima}}

\def\sparta{\textsc{Sparta}\xspace}

\def\gotetra{\textsc{gotetra}\xspace}
\def\colossus{\textsc{Colossus}\xspace}
\def\rockstar{\textsc{Rockstar}\xspace}
\def\consistenttrees{\textsc{Consistent-Trees}\xspace}

\def\planck{Planck\xspace}
\def\wmap{WMAP7\xspace}

\def\erebos{Erebos\xspace}

\def\deltac{\delta_{\rm c}}

\def\mvir{M_{\rm vir}}

\def\mtom{M_{\rm 200m}}
\def\rtom{R_{\rm 200m}}

\def\mtoc{M_{\rm 200c}}

\def\mfoc{M_{\rm 500c}}
\def\rfoc{R_{\rm 500c}}

\def\rsp{R_{\rm sp}}

\def\msp{M_{\rm sp}}

\def\deltasp{\Delta_{\rm sp}}

\shorttitle{Splashback Mass Function}
\shortauthors{Splashback Mass Function}

\journalinfo{The Astrophysical Journal, {\rm 903:87 (15pp), 2020 November 10}}
\submitted{Received 2020 July 18; revised 2020 October 4; accepted 2020 October 6; published 2020 November 6}

\begin{document}


\iflocal
\def\figdir{figs}
\else
\def\figdir{.}
\fi


\defcitealias{diemer_13_scalingrel}{DKM13}
\defcitealias{diemer_14}{DK14}
\defcitealias{diemer_15}{DK15}
\defcitealias{diemer_17_sparta}{Paper I}
\defcitealias{diemer_17_rsp}{Paper II}
\defcitealias{diemer_20_catalogs}{Paper III}


\title{Universal at last? The splashback mass function of dark matter halos}
\author{Benedikt Diemer$^{1,2}$}
\altaffiliation{$^2$NHFP Einstein Fellow}
\affil{
$^1$Department of Astronomy, University of Maryland, College Park, MD 20742, USA; \href{mailto:diemer@umd.edu}{diemer@umd.edu}
}


\begin{abstract}
The mass function of dark matter halos is one of the most fundamental statistics in structure formation. Many theoretical models (such as Press-Schechter theory) are based on the notion that it could be universal, meaning independent of redshift and cosmology, when expressed in the appropriate variables. However, simulations exhibit persistent non-universalities in the mass functions of the virial mass and other commonly used spherical overdensity definitions. We systematically study the universality of mass functions over a wide range of mass definitions, for the first time including the recently proposed splashback mass, $\msp$. We confirm that, in \LCDM cosmologies, all mass definitions exhibit varying levels of non-universality that increase with peak height and reach between 20\% and 500\% at the highest masses we can test. $\mvir$, $\mtom$, and $\msp$ exhibit similar levels of non-universality. There are, however, two regimes where the splashback mass functions are significantly more universal. First, they are universal to 10\% at $z \leq 2$, whereas spherical overdensity definitions experience an evolution due to dark energy. Second, when additionally considering self-similar cosmologies with extreme power spectra, splashback mass functions are remarkably universal (to between 40\% and 60\%) whereas their spherical overdensity counterparts reach non-universalities between 180\% and 450\%. These results strongly support the notion that the splashback radius is a physically motivated definition of the halo boundary. We present a simple and universal fitting formula for splashback mass functions that accurately reproduces our simulation data.
\end{abstract}



\section{Introduction}
\label{sec:intro}

In a universe where galaxies form at the center of dark matter halos, the number of halos as a function of their mass is one of the most basic statistics. This {\it mass function} serves as a crucial connecting point between cosmological models and observations, for instance when counts of galaxy clusters are translated into constraints on cosmological parameters \citep[e.g.,][]{allen_11, kravtsov_12}. The seminal work of \citet{press_74} presented a simple model where peaks in a Gaussian random field collapse if their overdensity passes a certain threshold \citep[see also][]{bond_91}. This model assumes that the collapse threshold, and thus the mass function, are {\it universal}, meaning independent of redshift and cosmology. Here, the abundance of halos is expressed as the multiplicity function, $f(\sigma)$, the fraction of matter that has collapsed into halos within a logarithmic interval in the variance of the underlying Gaussian density field. If true, this universality would allow us to find powerful functional forms for $f(\sigma)$ that would describe the abundance of halos in all cosmologies, without the need to simulate each one separately. 

While the specific form of $f(\sigma)$ suggested by \citet{press_74} was shown to deviate from simulation results \citep[e.g.,][]{gross_98, tormen_98}, the quest for a universal mass function continued. When halo masses are measured using a friends-of-friends (FOF) algorithm \citep{huchra_82, davis_85}, $f(\sigma)$ has been found to be roughly independent of redshift \citep{jenkins_01, white_02, reed_03, warren_06, lukic_07, heitmann_19}. On closer inspection, however, the FOF mass function does exhibit deviations from universality at high redshifts and between cosmologies \citep{reed_07, crocce_10, bhattacharya_11}, with some dependence on the linking length parameter \citep{more_11_fof, courtin_11, juan_14}. Moreover, it is difficult to intuitively understand the physical meaning of FOF masses and to predict them from simple models of halo formation such as top-hat collapse \citep{gunn_72}.

Spherical overdensity (SO) masses are preferable for this purpose, but the relation between FOF and SO masses depends on redshift and cosmology and exhibits large, non-Gaussian scatter \citep{white_01_mass, tinker_08, lukic_09, more_11_fof, courtin_11}. For SO masses, the question of universality depends on the overdensity threshold, a fixed, or varying multiple of the critical or mean density of the universe that results in definitions such as $\mtoc$, $\mvir$, and $\mtom$ \citep{lacey_94, white_01_mass}. The SO mass function has been shown to vary with redshift and cosmology \citep{tinker_08, watson_13_mf, bocquet_16, bocquet_20}, although some claims of universality with redshift have been made for particular definitions \citep[e.g.,][]{despali_16}. At this juncture, one possible way forward is to predict the mass function with increasingly complex models that may eventually capture some non-universality, for instance based on excursion sets \citep[e.g.,][]{maggiore_10, musso_12} or on peak statistics \citep[e.g.][]{bond_96_peakpatch, monaco_02}

In this work, we ask a different question: could the non-universality be an artifact of the SO mass definition? Should we expect universality when using such definitions, and are there alternatives? To this end, we undertake a first exploration of the splashback mass function of halos. The splashback radius, $\rsp$, and its enclosed mass, $\msp$, have been proposed as a more physically motivated halo definition \citep{diemer_14, adhikari_14, more_15}. The splashback radius is located at the first apocenter radius of the most recently infalling dark matter, which is marked by a sharp drop in the density profile. $\rsp$ depends on the mass accretion rate of a halo and should, by analogy with the spherical collapse model, separate infalling from orbiting material \citep{fillmore_84, bertschinger_85, shi_16_rsp}. Given this physical picture, we might hope that the splashback mass corresponds more tightly to the universal collapse threshold assumed in  many mass function models. On the other hand, measuring the splashback radii of individual simulated halos is more complicated than for SO definitions, which could introduce artificial non-universalities. To test the universality of the mass function, we need measurements of SO and splashback masses over a wide range of halo mass, redshift, and cosmology, including universes with radically different power spectra. We have recently presented such a dataset, namely publicly available halo catalogs and merger trees with splashback data computed using the \sparta code \citep[][hereafter \citetalias{diemer_17_sparta}, \citetalias{diemer_17_rsp}, and \citetalias{diemer_20_catalogs}, respectively]{diemer_17_sparta, diemer_17_rsp, diemer_20_catalogs}.

This paper is organized as follows. We discuss the theoretical underpinnings of mass functions in Section~\ref{sec:theory} and describe our simulation data and algorithms in Section~\ref{sec:methods}. We analyze the universality of the mass function in Section~\ref{sec:results}. We further discuss the notion of universality in Section~\ref{sec:discussion} before summarizing our results in Section~\ref{sec:conclusion}.


\section{Theory}
\label{sec:theory}

In this section, we review the universal form of the mass function and the necessary numerical calculations, which can be reproduced with the publicly available code \colossus \citep{diemer_18_colossus}.

\subsection{Universal Variables}
\label{sec:theory:basics}

The mass function, by definition, represents the comoving number density of halos as a function of mass, $n(M)$, in a given cosmology, at a given redshift $z$, and (in practice) in a simulation with a finite box size $L$, particle number $N^3$, and other numerical limitations (Table~\ref{table:sims}). Here, $M$ is understood to represent the mass according to some definition. This basic count $n(M)$, however, is not directly comparable between cosmic times or cosmologies because halos grow over time and because their number depends on cosmological parameters. Instead, we convert to the multiplicity function,
\begin{equation}
\label{eq:fsigma}
\frac{dn}{d \ln(M)} = f(\sigma) \frac{\rho_{\rm m,0}}{M} \frac{d \ln(\sigma^{-1})}{d \ln(M)} \,,
\end{equation}
which is natural in the \citet{press_74} picture \citep[see e.g.][for a derivation]{kravtsov_12}. Here, $\rho_{\rm m,0}$ is the mean matter density of the universe at $z = 0$ and $\sigma = \sigma(R_{\rm L}, z)$ is the variance of the linear density field on the Lagrangian scale of a halo. While mass has commonly been expressed as $\sigma^{-1}$, we choose to convert it to peak height, which compares $\sigma$ to the overdensity at which we expect halos to collapse,
\begin{equation}
\label{eq:nu}
\nu \equiv \frac{\delta_{\rm c}(z)}{\sigma(R_{\rm L}, z)} = \frac{\delta_{\rm c}(z)}{\sigma(R_{\rm L}, z = 0) \times D_+(z)} \,,
\end{equation}
where $\deltac(z)$ is the critical overdensity for collapse and $D_+(z)$ is the linear growth factor. We also convert $f(\sigma)$ to peak height,
\begin{equation}
\label{eq:fnu}
f(\nu) =  \frac{M}{\rho_{\rm m,0}} \frac{dn}{d \ln(\nu)} \,,
\end{equation}
where we have used that $d \ln(\sigma^{-1}) = d \ln(\nu)$ at fixed redshift. The multiplicity function can now be interpreted as the fraction of mass that has collapsed into halos in a logarithmic unit interval in peak height. To investigate the universality of $f(\nu)$, we need to carefully consider the terms in Equation~\ref{eq:nu}. The first ingredient is the variance of the linear power spectrum, $P(k)$,
\begin{equation}
\label{eq:sigma}
\sigma^2(R,z) = \frac{1}{2 \pi^2} \int_{k_{\rm min}}^{\infty} k^2 P(k,z) \left| \widetilde{W}(kR) \right|^2 dk \,.
\end{equation}
We discuss the integration limits in Section~\ref{sec:theory:corrections}. We use a Boltzmann code to compute the power spectra, which are also used to generate the initial conditions of our simulations (Section~\ref{sec:method:data}). \colossus interpolates the tabulated spectra with a cubic spline \citep{diemer_18_colossus}. One important ingredient in Equation~\ref{eq:sigma} is $\widetilde{W}$, the Fourier transform of the top-hat filter in real space,
\begin{equation}
\label{eq:tophat}
\widetilde{W}_{\rm tophat} = \frac{3}{(kR)^3} \left[ \sin(kR) - kR \times \cos(kR) \right] \,.
\end{equation}
We discuss alternative estimates of $P(k)$ and other filter choices in Section~\ref{sec:discussion:nu}. The Lagrangian scale of a halo with mass $M$ is the comoving radius of a sphere that encompasses the halo's mass at the mean density of the universe,
\begin{equation}
\label{eq:mtor}
M_{\rm L} \equiv M = (4 \pi/3) \rho_{\rm m,0} R_{\rm L}^3 \,.
\end{equation}
The final ingredient is the critical overdensity for collapse, which is often set to a constant, $\delta_{\rm c,EdS} \simeq 1.68647$, derived from the spherical top-hat collapse model in an Einstein--de Sitter universe \citep{gunn_72}. If $\delta_{\rm c}(z) \equiv \delta_{\rm c,EdS}$, peak height, and $1 / \sigma$ would be related by a constant factor. However, due to dark energy, the collapse density evolves slightly with redshift, 
\begin{equation}
\label{eq:deltac}
\delta_{\rm c}(z) \simeq \delta_{\rm c,EdS} \Omega_{\rm m}(z)^{0.0055} \,.
\end{equation}
This expression \citep[from][]{mo_10_book} is almost identical in effect to that of \citet{kitayama_96_2}. For the cosmologies considered in this paper, $\deltac$ evolves by less than 1\%, but even small changes in $\nu$ can lead to large changes in the universality of $f(\nu)$. We discuss the impact of this evolution in Section~\ref{sec:discussion:nu}.

\begin{deluxetable*}{lcccccccccccc}
\tablecaption{$N$-body Simulations
\label{table:sims}}
\tablewidth{\textwidth}
\tablehead{
\colhead{Name} &
\colhead{$L\, (\mpch)$} &
\colhead{$N^3$} &
\colhead{$m_{\rm p}\, (\msunh)$} &
\colhead{$\epsilon\, (\kpch)$} &
\colhead{$\epsilon / (L / N)$} &
\colhead{$z_{\rm initial}$} &
\colhead{$z_{\rm final}$} &
\colhead{$N_{\rm snaps}$} &
\colhead{$z_{\rm f-snap}$} &
\colhead{$z_{\rm f-cat}$} &
\colhead{Cosmology} &
\colhead{Reference}
}
\startdata
L2000-WMAP7 & $2000$ & $1024^3$ & $5.6 \times 10^{11}$  & $65$  & $1/30$ & $49$ & $0$ & $100$ & $20$ & $4.2$ & \wmap & \citetalias{diemer_15} \\
L1000-WMAP7 & $1000$ & $1024^3$ & $7.0 \times 10^{10}$ & $33$ & $1/30$ & $49$ & $0$ &  $100$ & $20$ & $6.2$ &  \wmap  & \citetalias{diemer_13_scalingrel} \\
L0500-WMAP7 & $500$  & $1024^3$ & $8.7 \times 10^{9}$  & $14$ & $1/35$  & $49$ & $0$ &  $100$ & $20$ & $8.8$ &  \wmap  & \citetalias{diemer_14} \\
L0250-WMAP7 & $250$  & $1024^3$ & $1.1 \times 10^{9}$  & $5.8$  & $1/42$  & $49$ & $0$ &  $100$ & $20$ & $11.5$ & \wmap & \citetalias{diemer_14} \\
L0125-WMAP7 & $125$  & $1024^3$ & $1.4 \times 10^{8}$  & $2.4$  & $1/51$  & $49$ & $0$ &  $100$ & $20$ & $14.5$ & \wmap & \citetalias{diemer_14} \\
L0063-WMAP7 & $62.5$ & $1024^3$ & $1.7 \times 10^{7}$  & $1.0$  & $1/60$ & $49$ & $0$ &  $100$ & $20$ & $17.6$ & \wmap & \citetalias{diemer_14} \\
L0031-WMAP7 & $31.25$ & $1024^3$ & $2.1 \times 10^{6}$  & $0.25$  & $1/122$ & $49$ & $2$ &  $64$ & $20$ & $20$ & \wmap & \citetalias{diemer_15} \\
L0500-Planck & $500$  & $1024^3$ & $1.0 \times 10^{10}$  & $14$ & $1/35$  & $49$ & $0$ &  $100$ & $20$ & $9.1$ & \planck & \citetalias{diemer_15} \\
L0250-Planck & $250$  & $1024^3$ & $1.3 \times 10^{9}$  & $5.8$  & $1/42$  & $49$ & $0$ &  $100$ & $20$ & $12.3$ & \planck & \citetalias{diemer_15} \\
L0125-Planck & $125$  & $1024^3$ & $1.6 \times 10^{8}$  & $2.4$  & $1/51$  & $49$ & $0$ &  $100$ & $20$ & $15.5$ & \planck & \citetalias{diemer_15} \\
L0100-PL-1.0 & $100$  & $1024^3$ & $2.6 \times 10^{8}$  & $0.5$  & $1/195$  & $119$ & $2$ & $64$ & $20$ & $20$ & PL, $n=-1.0$ & \citetalias{diemer_15} \\
L0100-PL-1.5 & $100$  & $1024^3$ & $2.6 \times 10^{8}$  & $0.5$  & $1/195$  & $99$ & $1$ & $78$ & $20$ & $20$ & PL, $n=-1.5$ & \citetalias{diemer_15} \\
L0100-PL-2.0 & $100$  & $1024^3$ & $2.6 \times 10^{8}$  & $1.0$  & $1/98$  & $49$ & $0.5$ & $100$ & $20$ & $15.5$ & PL, $n=-2.0$ & \citetalias{diemer_15} \\
L0100-PL-2.5 & $100$  & $1024^3$ & $2.6 \times 10^{8}$  & $1.0$  & $1/98$  & $49$ & $0$ & $100$ & $20$ & $5.4$ & PL, $n=-2.5$ & \citetalias{diemer_15}
\enddata
\tablecomments{The $N$-body simulations used in this paper. $L$ denotes the box size in comoving units, $N^3$ the number of particles, $m_{\rm p}$ the particle mass, and $\epsilon$ the force-softening length in physical units. The redshift range of each simulation is determined by the first and last redshifts $z_{\rm initial}$ and $z_{\rm final}$, but snapshots were output only between $z_{\rm f-snap}$ and $z_{\rm final}$. The structure in the earlier snapshots of some simulations is not developed enough to yield any halos yet, the first catalog with halos is output at $z_{\rm f-cat}$. The references correspond to: \citet[][\citetalias{diemer_13_scalingrel}]{diemer_13_scalingrel}, \citet[][\citetalias{diemer_14}]{diemer_14}, and \citet[][\citetalias{diemer_15}]{diemer_15}. Our system for choosing force resolutions is discussed in \citetalias{diemer_14}.}
\end{deluxetable*}

\def\figsize{0.27}
\begin{figure*}
\centering
\includegraphics[trim =  0mm 0mm 0mm 0mm, clip, scale=\figsize]{\figdir/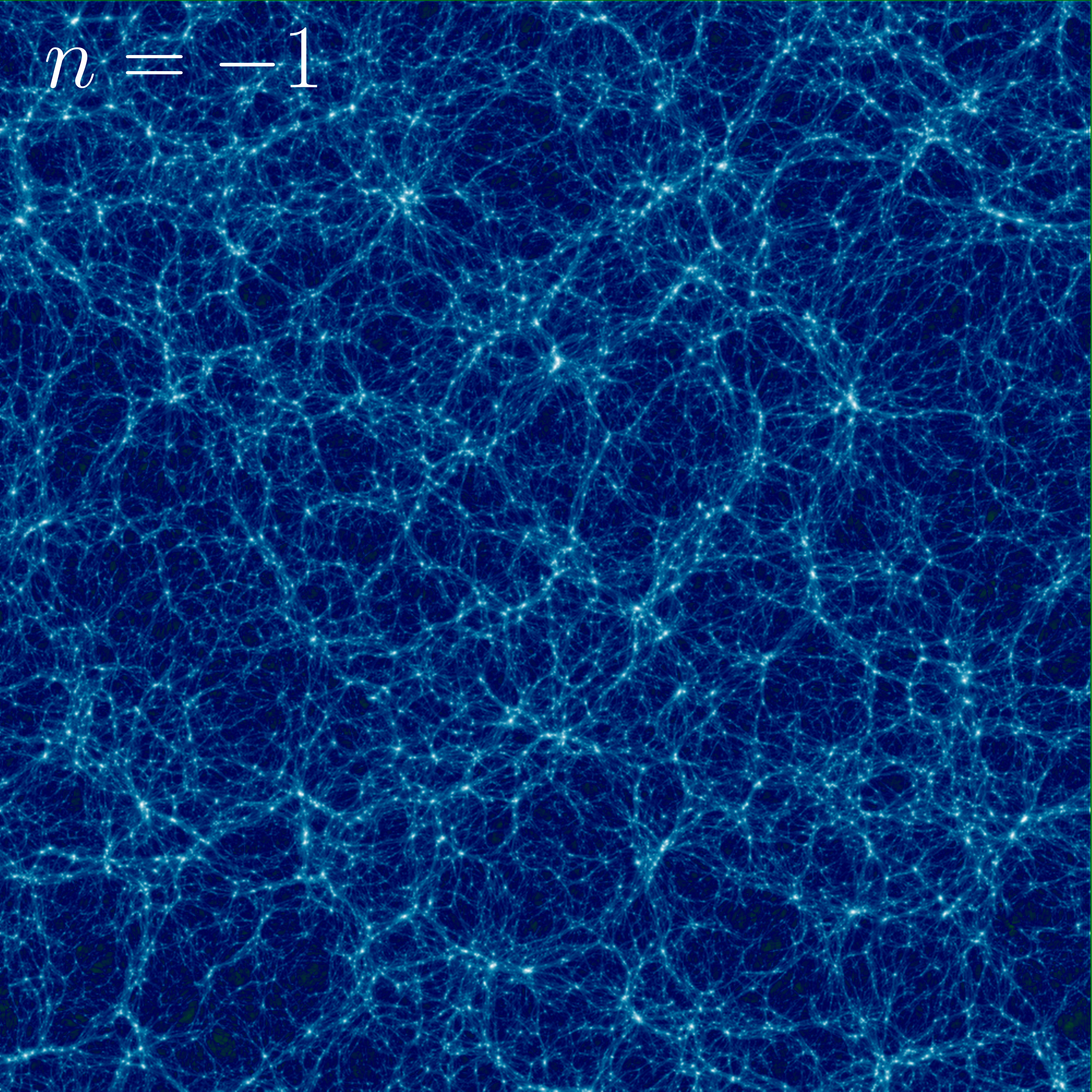}
\hspace{-0.3mm}
\includegraphics[trim =  0mm 0mm 0mm 0mm, clip, scale=\figsize]{\figdir/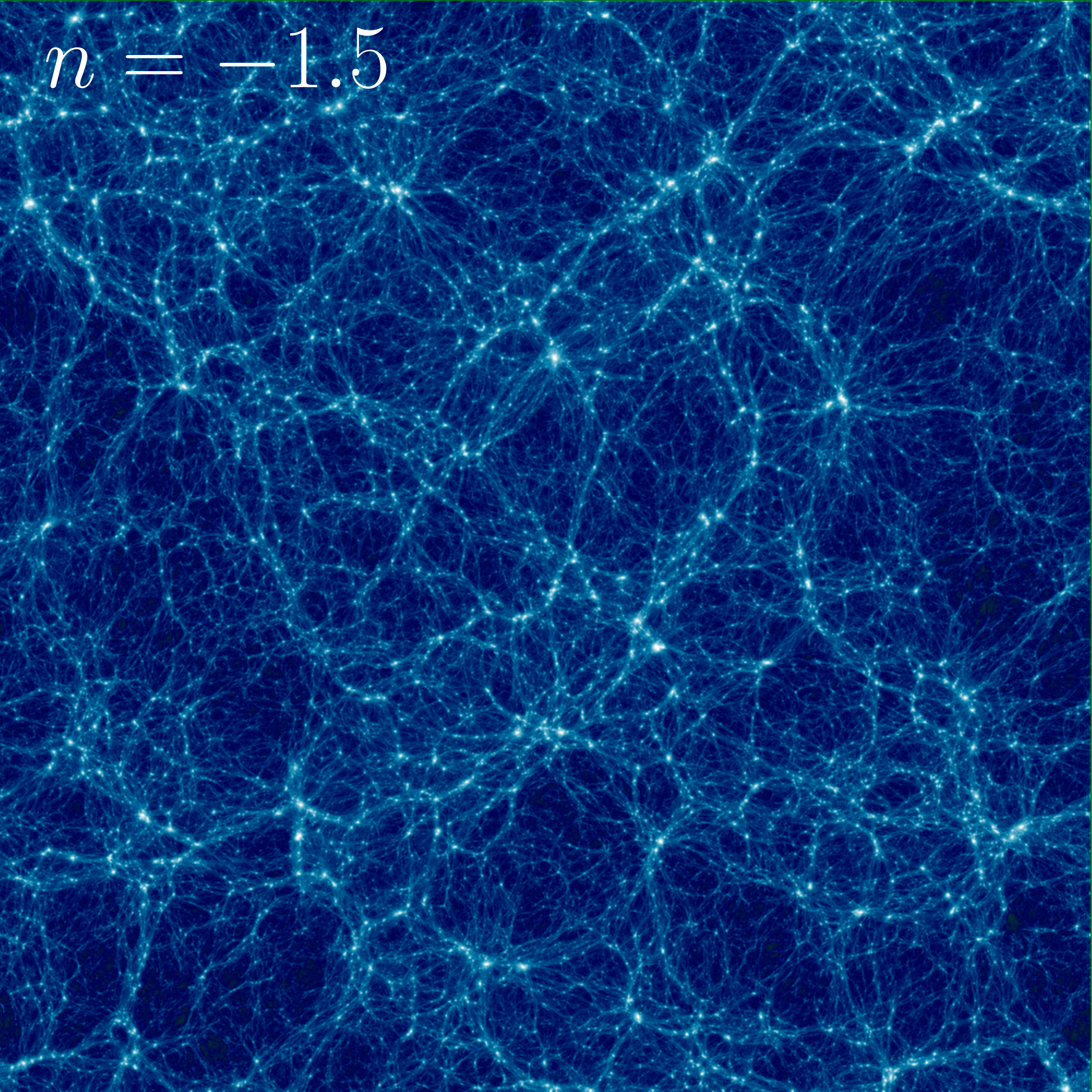} \\
\vspace{0.8mm}
\includegraphics[trim =  0mm 0mm 0mm 0mm, clip, scale=\figsize]{\figdir/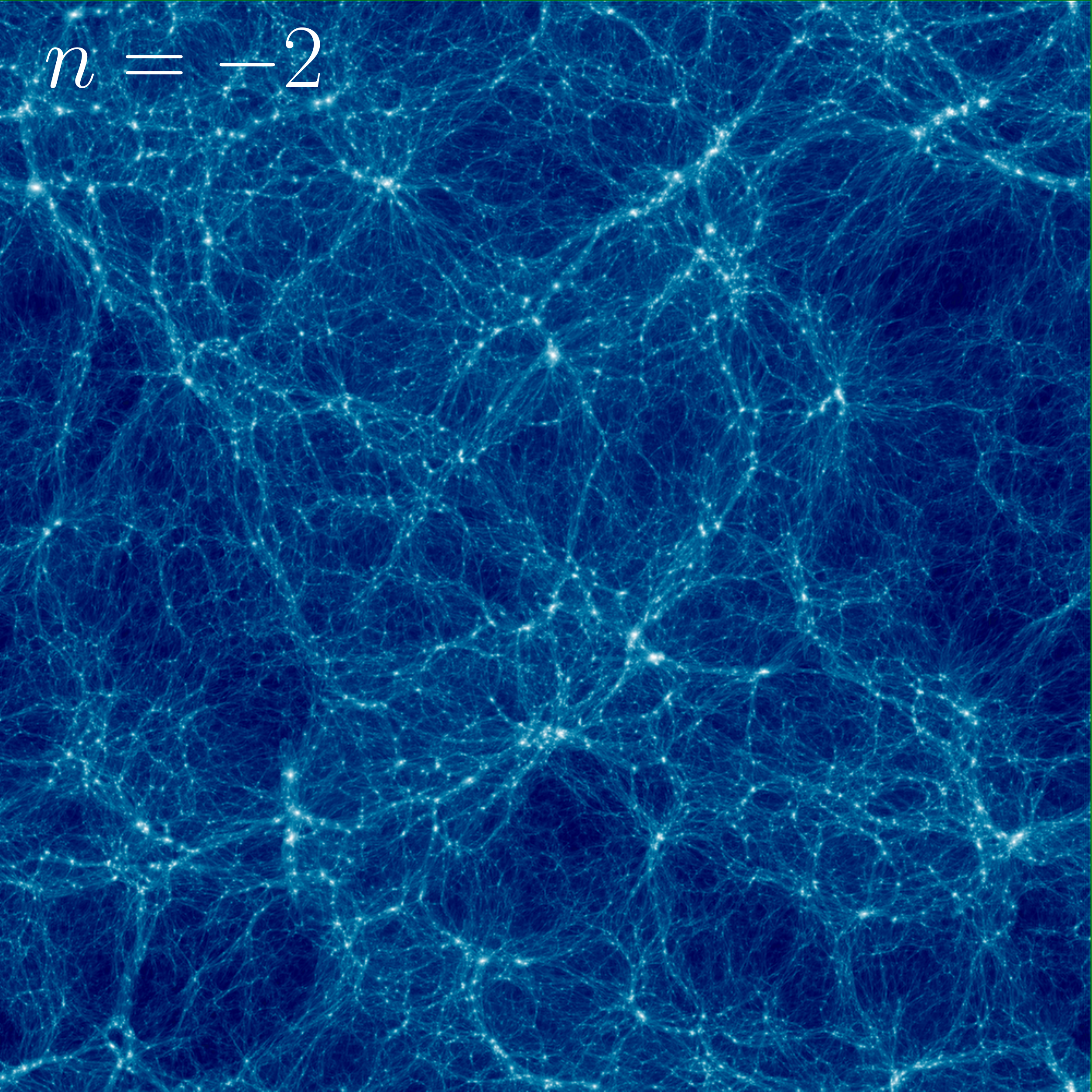}
\hspace{-0.3mm}
\includegraphics[trim =  0mm 0mm 0mm 0mm, clip, scale=\figsize]{\figdir/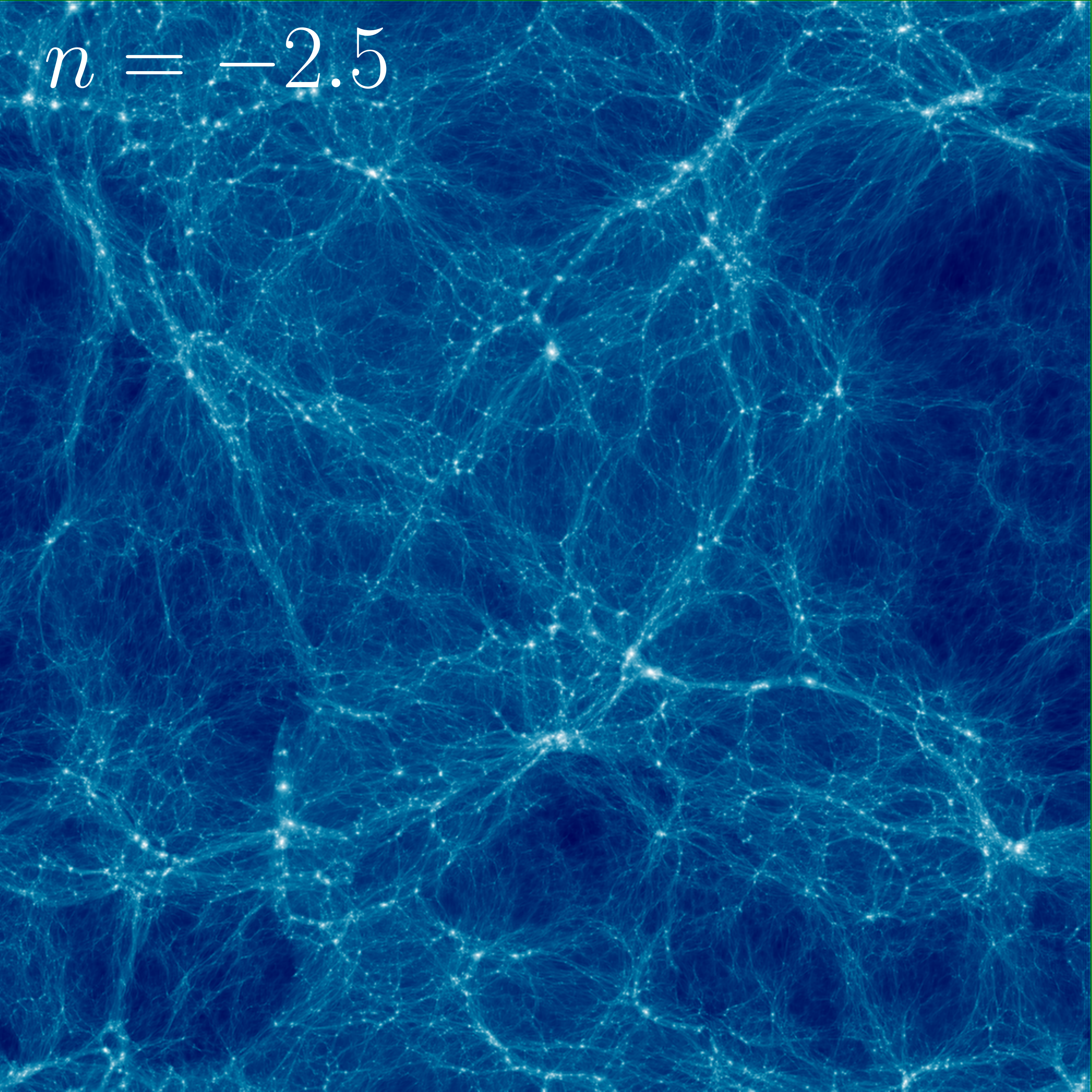}
\caption{Density slices through self-similar universes with power spectrum slopes of $n = -1$, $-1.5$, $-2$, and $-2.5$. The simulations are shown at their final redshifts of $2$, $1$, $0.5$, and $0$, respectively. The slices are $5 \mpch$ thick, the color scale corresponds to the projected density on a logarithmic scale between $3\times 10^{-3}$ and $100$ with respect to the cosmic mean. The simulations were normalized to the same $\sigma_8$ at $z = 0$ but exhibit very different distributions of structure: while shallow power spectra lead to much small-scale and little large-scale structure, the structure in the $n = -2.5$ simulation collapses at almost the same time for large and small scales (a slope of $-3$ would correspond to all scales collapsing at the same time and can thus not be simulated with a finite box). The largest structures look similar in shape because the same random phases were used in all simulations. The visualizations were created using the \gotetra code by P. Mansfield (\href{https://github.com/phil-mansfield/gotetra}{https://github.com/phil-mansfield/gotetra}), which is based on a tetrahedron-based estimate of the density field \citep{kaehler_12, abel_12, hahn_13}. Higher-resolution images of the self-similar and \LCDM simulations are available at  \href{http://www.benediktdiemer.com/visualization}{benediktdiemer.com/visualization}.}
\label{fig:viz_ss}
\end{figure*}

\subsection{Corrections for Finite Volume and Resolution}
\label{sec:theory:corrections}

The $\sigma$ integral in Equation~\ref{eq:sigma} should ideally be taken from zero to infinity because all $k$ scales contribute to the top-hat filter. In simulations, the mass function is affected by their finite box size, which cuts off the power spectrum at a scale $k_{\rm min} = 2 \pi / L$ \citep{tormen_96, yoshida_03_running, bagla_05, power_06, reed_07, lukic_07, gnedin_11_dc}. We note that $k_{\rm min}$ is not taken into account when normalizing the power spectrum to a given $\sigma_8$ because that correct value was used when setting up the initial conditions. The reduction in variance is due to power spectrum modes that were cut off in the initial conditions and is not recovered through renormalization. With $k_{\rm min} > 0$, the $f(\nu)$ form of the multiplicity function has the advantage that it avoids converting $n(M)$ to $f(\sigma)$ via the $d \ln \sigma / d \ln M$ term in Equation~\ref{eq:fsigma}. This term is susceptible to numerical inaccuracies due to the combination of a sharp cut off at $k_{\rm min}$ and the oscillatory shape of the top-hat filter (Equation~\ref{eq:tophat}). Nevertheless, our finite-volume correction does not perfectly account for the real effects of a limited box size. For example, \citet{power_06} show that the mass function can be enhanced at intermediate masses due to halos whose mass is reduced by the missing spectral power. We quantitatively analyze the effects of our finite-volume correction in Section~\ref{sec:discussion:nu}.

The reader may wonder why we did not impose an upper limit on the $\sigma$ integral due to the limited resolution of the simulations. The highest wave mode represented is the Nyquist frequency, $k_{\rm max} = \pi N / L$. This frequency corresponds roughly to a smallest Lagrangian scale that we can resolve, $R_{\rm ny} = 2 L / N$. The scales we wish to resolve will be limited by a minimum number of particles in a halo, $N_{\rm min}$ (Sections~\ref{sec:method:so} and \ref{sec:method:sp}), resulting in a minimum Lagrangian mass $M_{\rm min} = N_{\rm min} \rho_{\rm m,0} L^3 / N^3$ and a minimum Lagrangian radius
\begin{equation}
\label{eq:rmin}
R_{\rm L,min} = \left( \frac{3 N_{\rm min}}{4 \pi} \right)^{1/3} \frac{L}{N} \approx 2.5\, \left( \frac{N_{\rm min}}{500} \right)^{1/3}\, R_{\rm ny} \,.
\end{equation}
We have inserted 500 particles as a typical minimum mass because that is the threshold chosen in Section~\ref{sec:method:so}. Regardless of cosmology, redshift, or box size, the unresolved scale is a factor of a few smaller than the smallest halos we wish to resolve. We have confirmed that the upper cutoff does not change $\sigma$ appreciably (for $N_{\rm min} = 500$), even in cosmologies with shallow power spectra. However, we caution that, due to the large $k$ extent of the top-hat filter, small changes become noticeable for resolution limits of $N_{\rm min} \lsim 100$.

\subsection{Functional Form}
\label{sec:theory:fit}

The field of fitting mass functions began with the Press--Schechter prediction that $f(\nu) \propto \nu \times \exp (-\nu^2 / 2)$. This function provides an impressive fit for a theoretical Ansatz without free parameters but has to be amended to fit simulations in detail. One particularly useful four-parameter extension is
\begin{equation}
\label{eq:fit_tinker}
f(\sigma) = A \left[ \left( \frac{\sigma}{b'} \right)^{-a} + 1 \right] e^{-c' / \sigma^2} \,,
\end{equation}
which was used by \citet{warren_06} and \citet{tinker_08} but is ultimately based on expressions in \citet{sheth_99, sheth_02}. As previously discussed, $\sigma$ and $\nu$ are not related by a constant factor in this work, making it more natural to express the fitting function in terms of peak height,
\begin{equation}
\label{eq:fit_global}
f(\nu) = A \left[ \left( \frac{\nu}{b} \right)^{a} + 1 \right] e^{-c \nu^2}
\end{equation}
with the simple parameter transformations $b = \deltac / b'$ and $c = c' / \deltac^2$. 

 
\section{Simulations and Algorithms}
\label{sec:methods}

In this section, we briefly review our $N$-body simulations. We describe our procedure for computing mass functions and show how to combine the results from multiple simulations. This work is based on the halo catalogs presented in \citetalias{diemer_20_catalogs}, and we refer the reader to that paper for details.

\subsection{Simulation Data}
\label{sec:method:data}

We use the \erebos suite of  $N$-body simulations, whose most important properties we review in Table~\ref{table:sims}. This collection of simulations allows us to explore a \wmap \LCDM cosmology based on that of the Bolshoi simulation \citep[][$\Omega_{\rm m} = 0.27$, $\Omega_{\rm b} = 0.0469$, $h = 0.7$, $\sigma_8 = 0.82$, and $n_{\rm s} = 0.95$]{komatsu_11, klypin_11} as well as a \planck-like cosmology \citep[][$\Omega_{\rm m} = 0.32$, $\Omega_{\rm b} = 0.0491$, $h = 0.67$, $\sigma_8 = 0.834$, and $n_{\rm s} = 0.9624$]{planck_14}. These cosmologies span the likely ranges of those parameters that are most important for structure formation, such as $\Omega_{\rm m}$. We also consider self-similar Einstein--de Sitter universes with power-law initial spectra of slopes $-1$, $-1.5$, $-2$, and $-2.5$. The power spectra for the Lambda Cold Dark Matter (\LCDM) simulations were computed using \textsc{Camb} \citep{lewis_00}, and the initial conditions for all simulations were generated by \textsc{2LPTic} \citep{crocce_06}. The simulations were run with \textsc{Gadget2} \citep{springel_05_gadget2}. 

Given that we are concerned with the question of universality, the self-similar simulations play an important role in our investigation. As these types of universes are rarely visualized, Figure~\ref{fig:viz_ss} shows density slices through our four simulations at their final redshifts (see \citealt{efstathiou_88} and \citealt{elahi_09_thesis} for older examples). The scales at which matter clusters visibly evolve from the shallowest to the steepest power spectrum slopes, spanning a greater range of conditions than in \LCDM where the slopes that are relevant for halo formation range from about $-1.8$ to $-2.8$ \citep[e.g.,][]{diemer_15}. The different slopes translate into a entirely different abundances, formation histories, and structures of halos \citep[e.g.,][]{knollmann_08, schneider_15, ludlow_17}, providing a stringent test for any universal description of the abundance of halos. Another useful feature of self-similar universes is that both power spectrum and time are independent of physical scales, meaning that properties such as the mass function must be universal with redshift except for resolution effects \citep{efstathiou_88, lacey_94, lee_99_mfunc, elahi_09_plsubs, leroy_20, joyce_20}. 

While our simulations span a wide range of cold dark matter (CDM) cosmologies, we do not consider alternative dark matter models, neutrinos, or modified gravity in this paper. Thus, statements about universality with cosmology are understood to refer to CDM universes.

\subsection{Halo Catalogs}
\label{sec:method:catalogs}

\def\figsize{0.57}
\begin{figure*}
\centering
\includegraphics[trim =  3mm 24mm 2mm 2mm, clip, scale=\figsize]{\figdir/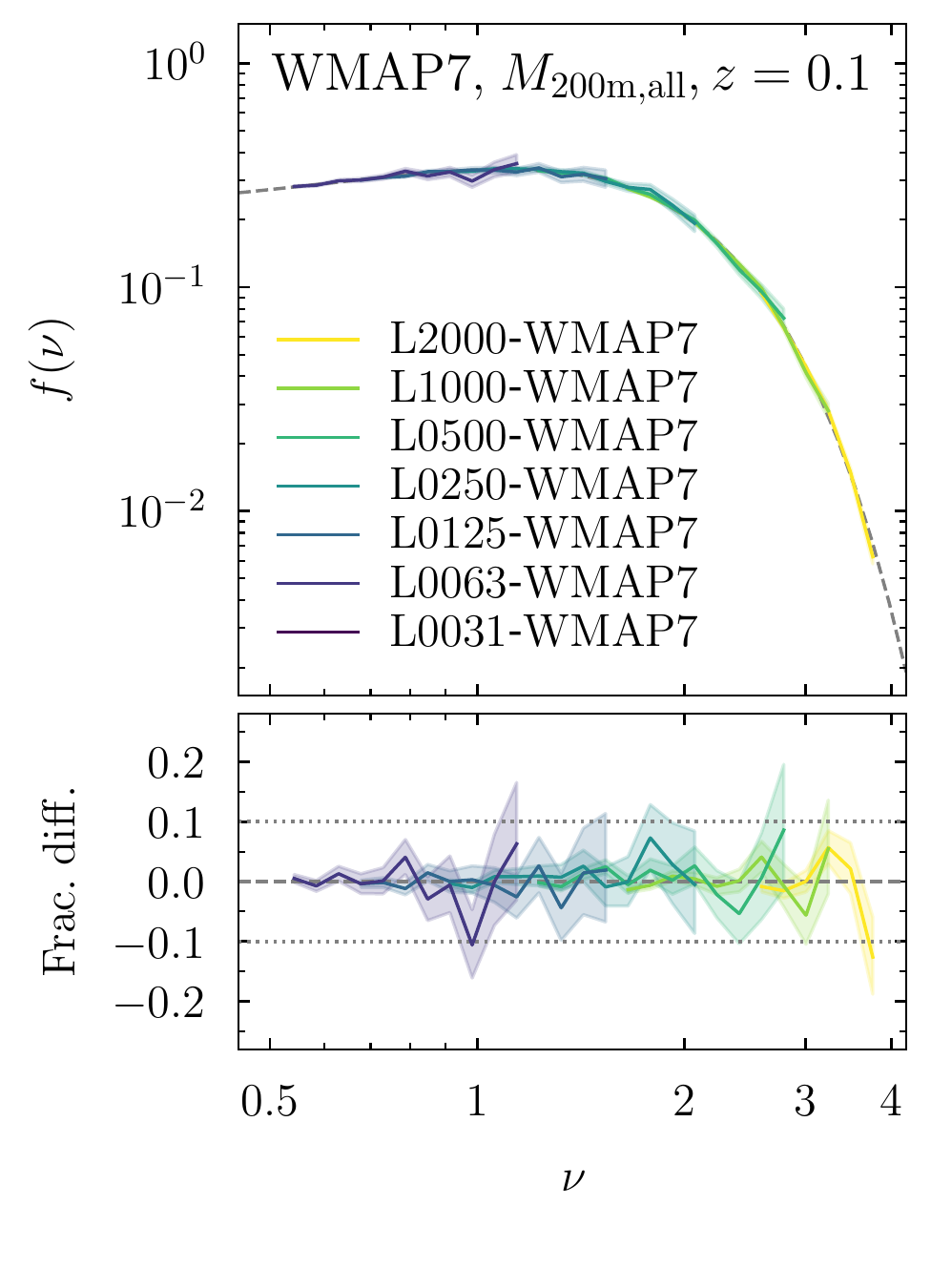}
\includegraphics[trim =  25mm 24mm 2mm 2mm, clip, scale=\figsize]{\figdir/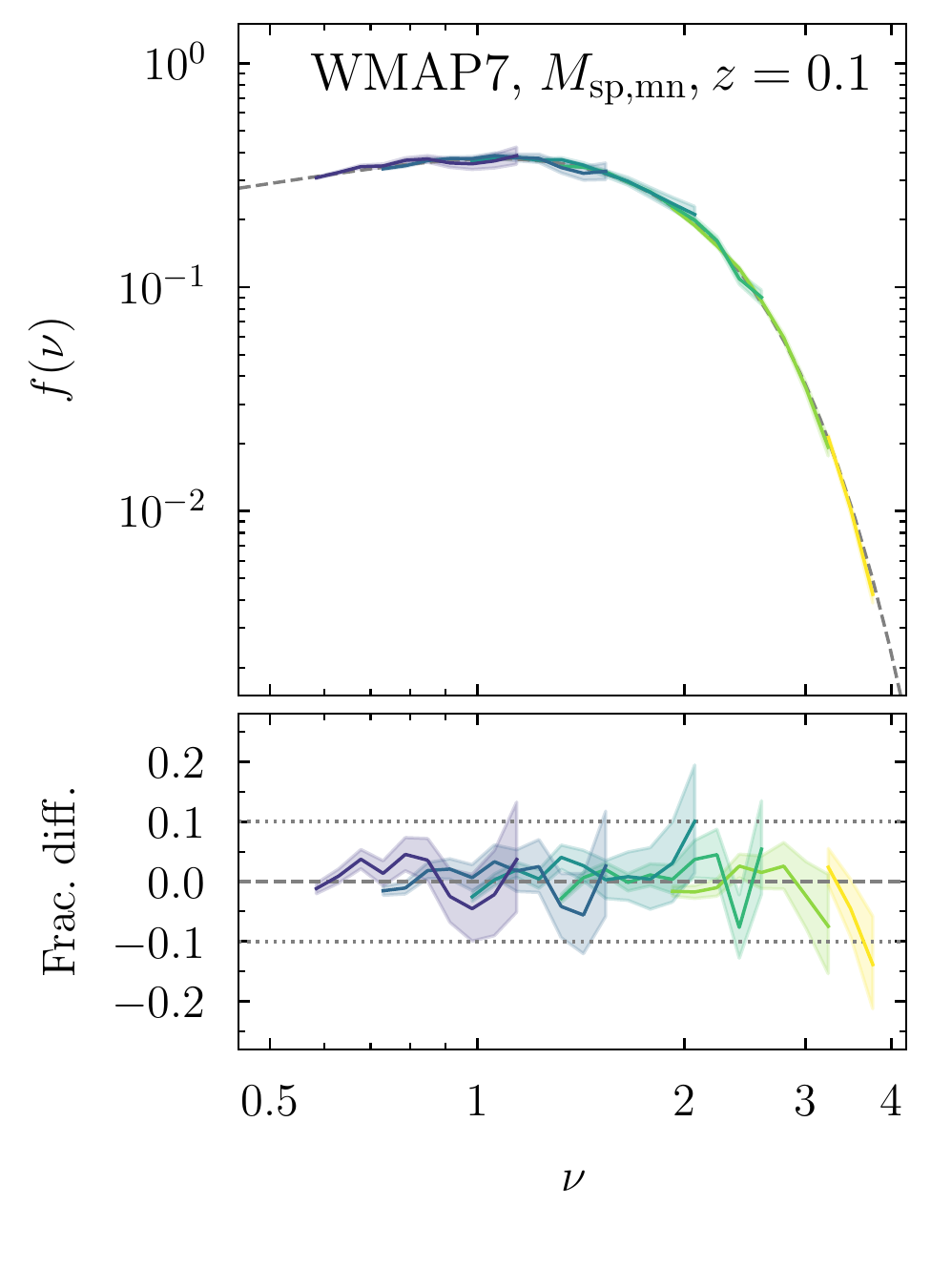}
\includegraphics[trim =  25mm 24mm 2mm 2mm, clip, scale=\figsize]{\figdir/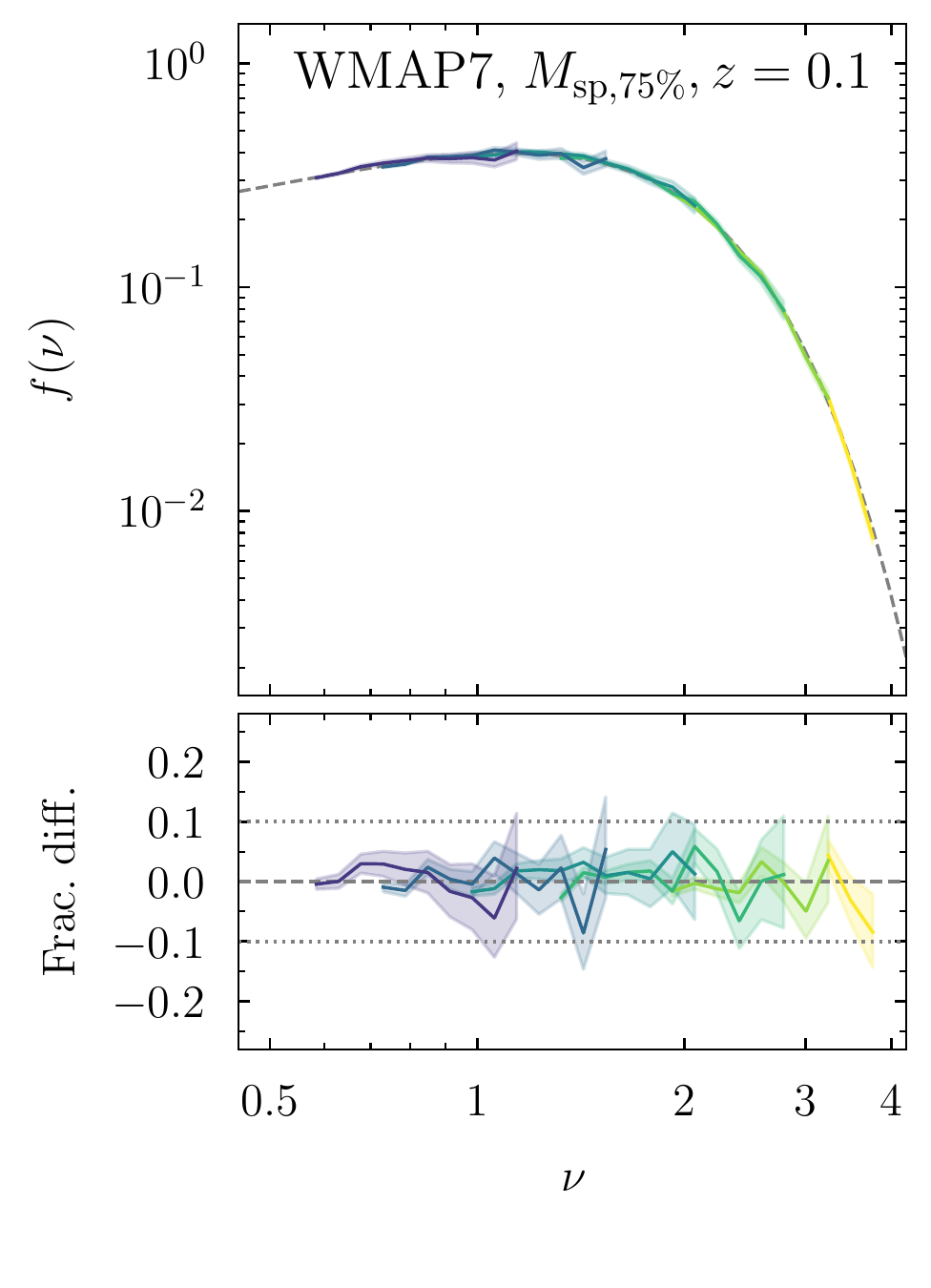}
\includegraphics[trim =  25mm 24mm 2mm 2mm, clip, scale=\figsize]{\figdir/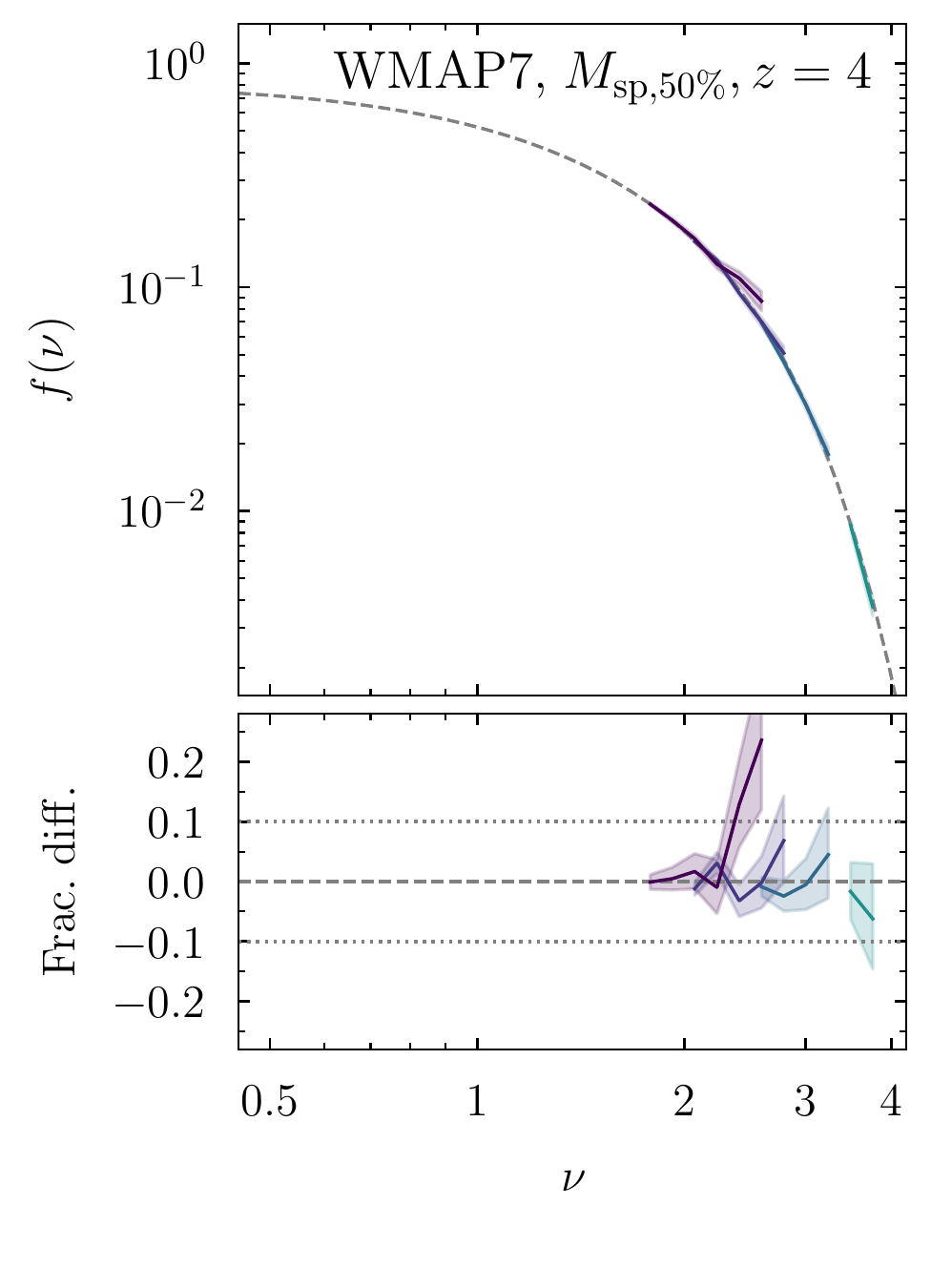}
\includegraphics[trim =  3mm 7mm 2mm 2mm, clip, scale=\figsize]{\figdir/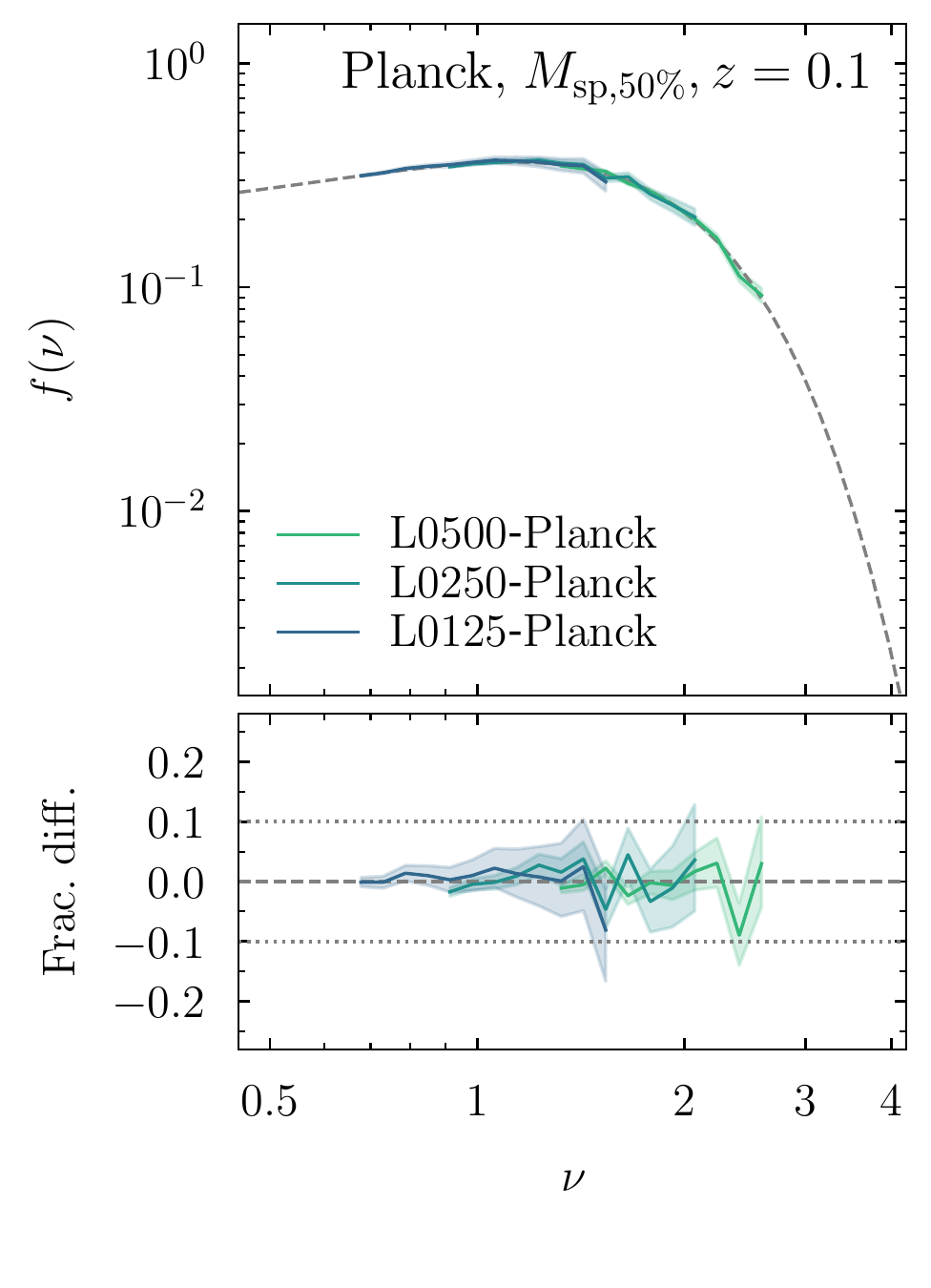}
\includegraphics[trim =  25mm 7mm 2mm 2mm, clip, scale=\figsize]{\figdir/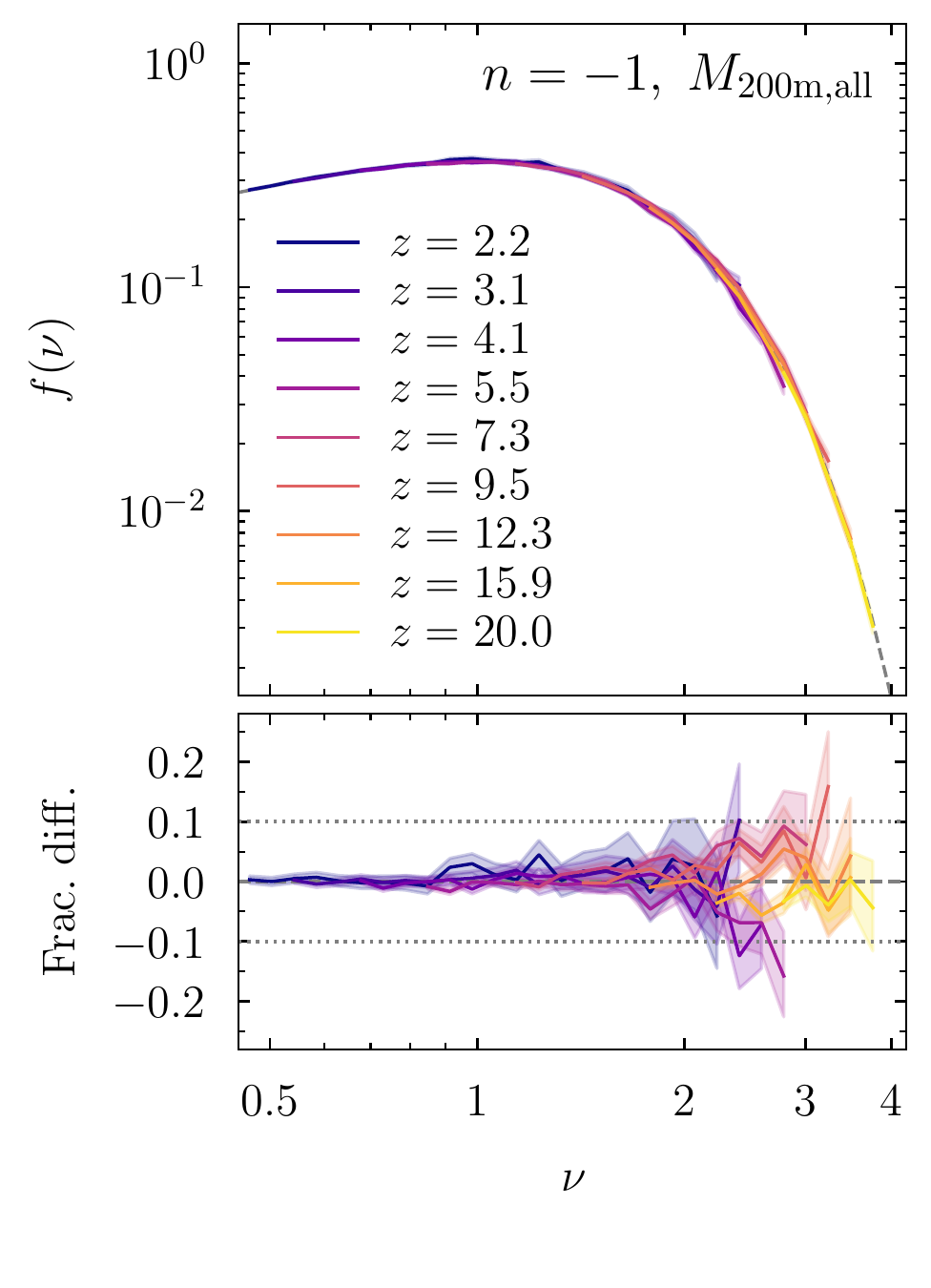}
\includegraphics[trim =  25mm 7mm 2mm 2mm, clip, scale=\figsize]{\figdir/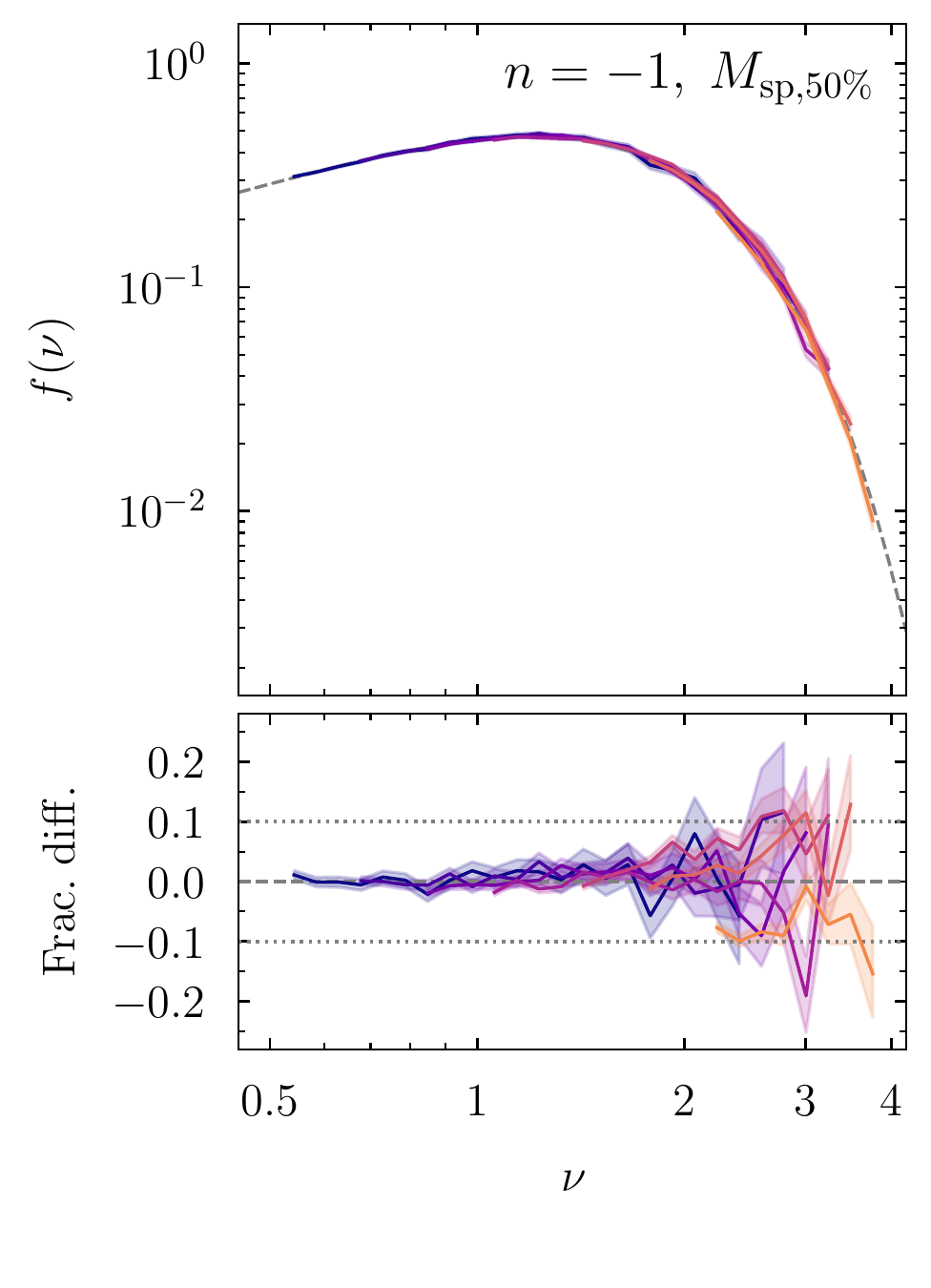}
\includegraphics[trim =  25mm 7mm 2mm 2mm, clip, scale=\figsize]{\figdir/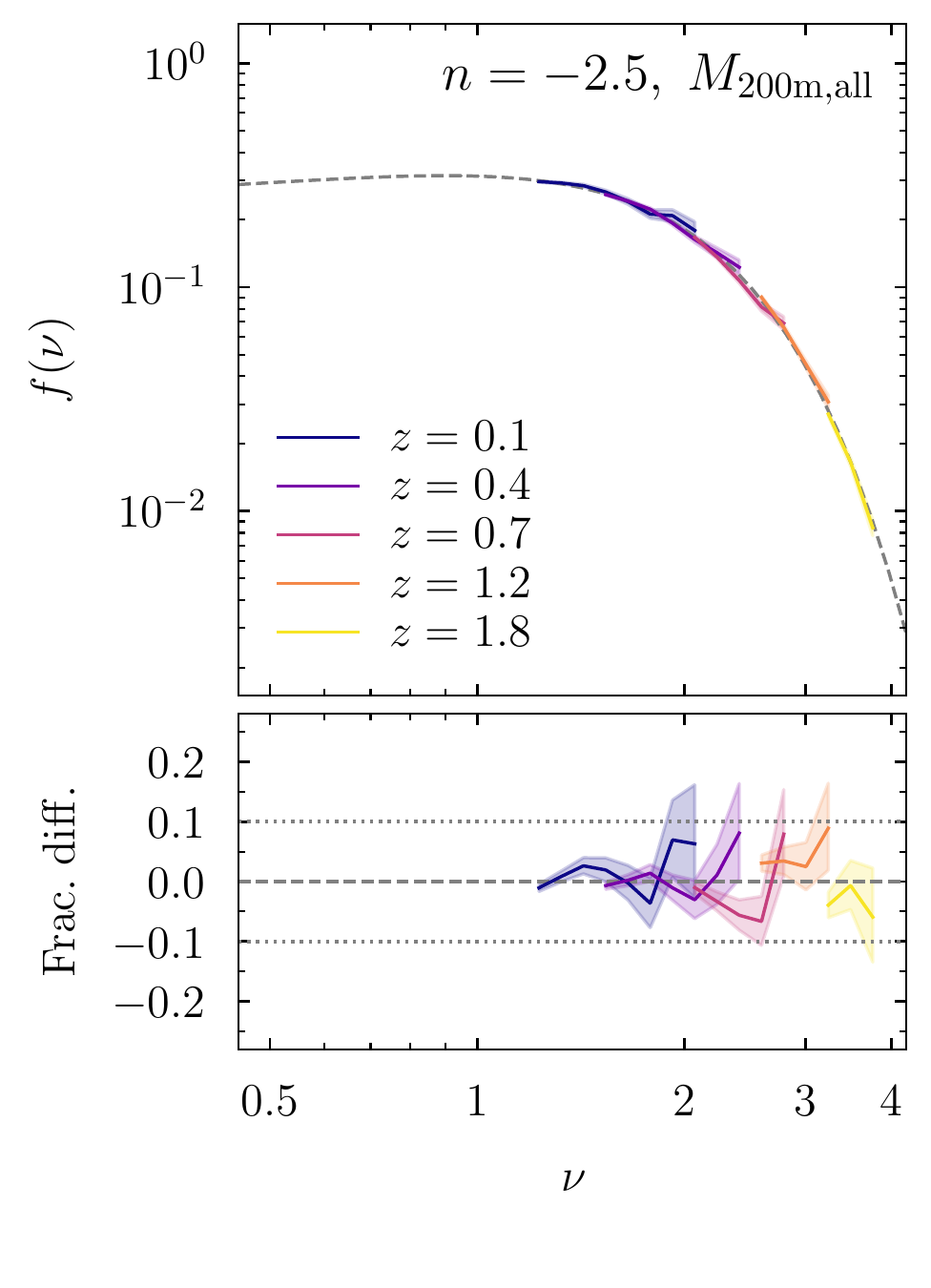}
\caption{Convergence of mass functions for different cosmologies, mass definitions, and redshifts, as labeled in the upper-right corners of each panel. Each panel shows a set of mass functions (lines) and their statistical uncertainties (shaded areas). If converged, the mass functions are expected to overlap, either between different box sizes in \LCDM (top four panels for \wmap, bottom-left panel for \planck) or between different redshifts in a single self-similar universe (three bottom-right panels). The smaller bottom panels show the fractional difference to a fit (using Equation~\ref{eq:fit_global}, gray dashed lines). We show only bins where the mass function is constrained by at least $100$ halos to avoid crowding the plot with noisy measurements. We have picked representative examples of mass definitions and redshifts, the results are similar for other definitions. Generally, the mass functions are converged to within the statistically expected uncertainty, partly because we have imposed the appropriate mass cuts at the low-mass end. We do notice slight convergence issues at high redshift (top right panel) and for the $n = -1$ self-similar simulation (bottom center panels), although it is not clear that the deviations are statistically significant or systematic. In the following figures, we combine the data in each panel into a single sample per cosmology, mass definition, and redshift (the latter only for \LCDM).}
\label{fig:convergence}
\end{figure*}

We use the halo catalogs presented in \citetalias{diemer_20_catalogs} for this work, which we briefly review here. We start with catalogs created by the \rockstar and \consistenttrees codes \citep{behroozi_13_rockstar, behroozi_13_trees}. \rockstar finds the particles in FOF groups in six-dimensional phase space using the default linking length of $0.28$. This linking length matters only insofar as the FOF groups need to include all particles within the SO radii we are interested in, which is the case except for very rare groups that are slightly incomplete at $\rtom$ \citepalias{diemer_20_catalogs}. From the particles in FOF groups, \rockstar computes numerous halo properties, including bound-only SO masses (where gravitationally unbound particles are excluded).

We have run the \sparta code \citepalias{diemer_17_sparta} on all simulations to compute the splashback and all-particle SO masses that will be the focus of this paper. We then combined the \sparta data with the original catalogs, creating enhanced catalogs with a large number of mass definitions. When calculating mass functions, we wish to consider only isolated (host) halos. Thus, the radius (or mass) definition impacts the mass function in two ways: it shifts halos in mass and changes the number of host halos. We have computed separate host--subhalo relations for each mass definition \citepalias{diemer_20_catalogs}.

\subsection{SO Masses}
\label{sec:method:so}

We consider four  SO definitions that span the range of commonly used density thresholds, namely $\mfoc$, $\mtoc$, $\mvir$, and $\mtom$, referring to thresholds of $200$ or $500$ times the critical or mean matter density of the universe. $\mvir$ corresponds to an evolving threshold as approximated by \citet{bryan_98}. Before computing their mass functions, we need to consider the resolution limits of our simulations, the completeness of our catalogs, and the accuracy to which masses are measured. The latter is particularly important when measuring steep mass functions, where scatter can cause Eddington bias \citep{eddington_13}. 

We impose a lower limit of $500$ particles regardless of the overdensity threshold. While this choice leads to slightly different cutoffs compared to a fixed mass definition, it limits the statistical variance due to particle noise \citep{trenti_10, benson_17}. We have derived this limit by overlaying the mass functions from simulations with different resolutions and find that they are converged to within the statistical uncertainties for almost all SO definitions, redshifts, and cosmologies (Figure~\ref{fig:convergence}), with the possible exception of $n = -1$ and $n = -2.5$ which exhibit noticeable, although not highly significant, deviations. \citet{mansfield_20_resolution} concluded that mass functions in the \LCDM\ \erebos simulations are generally converged at about $200$ particles, but we find some residual non-convergence at the low-mass end when using this limit (especially in the self-similar boxes). Moreover, there is no need to push to the smallest halos in each simulation because our staggered box sizes ensure that the mass range is covered relatively evenly (Table~\ref{table:sims}).

When counting halos to compute mass functions (as opposed to taking averages of halo properties), we need to worry about any incompleteness. In rare cases, the computation of SO masses fails because the density never reaches the threshold at the halo center (most commonly for $\mfoc$) or because the density never falls below the threshold at large radii (most common in low-density definitions such as $\mtom$). The latter effect predominantly occurs for all-particle masses, the former also affects bound-only masses. However, both effects are much more important for subhalos and thus do not pose a significant problem for this investigation. We find that less than $10^{-4}$ of our halos are typically affected, with the fraction rising to $10^{-3}$ for $\rfoc$ and in the self-similar simulations. Given the much larger systematic uncertainties, we make no attempt to correct for this issue.

Not all SO measurements are physically meaningful. For instance, the all-particle masses of a small fraction of halos include a large amount of material from another, nearby halo. In \citetalias{diemer_20_catalogs}, we found that this issue is most common in small boxes and at high redshift, which could introduce slight redshift-dependent shifts into our mass functions. We identify the affected halos by comparing the all-particle and bound-only masses of the same definition and use the bound-only mass if the ratio is greater than two. This limit is arbitrary, but there is no clear boundary between ``normal'' mass ratios and pathological cases. Our choice echoes \rockstar's algorithm, which discards halos as unphysical if more than half of their FOF group is unbound. Our procedure shifts the mass functions by up to a few percent, but these shifts are mostly uniform with redshift or improve the universality of SO definitions. Excessively large all-particle radii also include some spurious subhalos that would otherwise be hosts, a small effect that we cannot correct for here.

Despite these issues, all-particle SO masses are preferred over bound-only masses because they are uniquely defined. In contrast, bound-only masses depend intimately on the halo finder and unbinding algorithm \citepalias{diemer_20_catalogs}, making it harder to establish the connection to theoretical models and observations. We have checked that there are no systematic differences between the universalities of bound-only and all-particle mass functions. The figures in the paper show all-particle SO mass functions; labels such as $\mtoc$ refer to the all-particle version unless otherwise specified.

\subsection{Splashback Masses}
\label{sec:method:sp}

\sparta measures splashback masses by following the orbits of all particles in a halo and finding their first apocenter. From the collection of the apocenter times and radii, the splashback radius and mass of a halo are derived by slightly smoothing the distribution in time, taking into account both past and future splashbacks. The splashback radius can then be defined as the mean of the particle distribution, $R_{\rm sp,mn}$, the median, $R_{\rm sp,50\%}$, or higher percentiles such as $R_{\rm sp,75\%}$ and $R_{\rm sp,90\%}$. The latter is the largest percentile computed by \sparta and closest to the non-spherical splashback shells of \citet{mansfield_17}. The resulting enclosed overdensity, $\deltasp$, varies between about $50$ and $400 \rhom$ and depends strongly on mass accretion rate, redshift, and halo mass \citepalias{diemer_17_rsp}. At the final snapshots of a simulation, the \sparta estimates become less accurate because there are no future particle orbits that can be taken into account. \sparta corrects for this bias, but the algorithm does incur increased scatter \citepalias{diemer_17_sparta}. Thus, we restrict ourselves to $z \geq 0.13$ for our fits and convergence tests. 

We impose a lower limit of $1000$ particles on our splashback masses, as suggested in \citetalias{diemer_17_sparta}. Again, we have tested this limit by comparing simulations of different resolutions and find that they are converged, with the previously mentioned caveat regarding high redshifts and the most extreme self-similar universes (Figure~\ref{fig:convergence}). The splashback calculation can fail for a number of reasons, most commonly insufficient particle splashback events. As described in \citetalias{diemer_20_catalogs}, we increase the completeness by interpolating and extrapolating splashback masses across time if possible. If this interpolation fails, we guess a value based on the fitting function of \citetalias{diemer_20_catalogs}. Out of the halos that pass the particle number cut, between 0.1\% and 1\% have masses not directly computed by \sparta, with the fraction increasing toward small box sizes (low halo masses). However, the vast majority of those masses were computed from extrapolation of nearby past or future values, which gives fairly accurate estimates \citepalias{diemer_20_catalogs}. Less than 0.1\% of halos had their masses estimated by the fitting function, whose predictions suffer from large scatter; these numbers do not strongly depend on redshift. In principle, even a small number of poor estimates could up-scatter masses into higher bins with intrinsically low counts. We have confirmed that removing all halos without \sparta-computed splashback masses has no appreciable effect on our results, even for the $n = -1$ self-similar simulation that suffers from the worst incompleteness effects \citepalias{diemer_20_catalogs}. We conclude that the incompleteness of our catalogs does not constitute a serious source of error.

\subsection{Computing Mass Functions}
\label{sec:method:mfunc}

Figure~\ref{fig:convergence} demonstrates that our mass functions for individual simulations are converged at fixed peak height. For the remainder of the paper, we will combine the \LCDM mass functions from multiple simulations into one sample per redshift and cosmology. These combined mass functions cannot be converted between mass and peak height any longer due to the simulation-specific corrections of Section~\ref{sec:theory:corrections}. Thus, we combine different simulations in $\nu$-$f(\nu)$ space. For each comparison, we define a logarithmically spaced set of peak height bins, $\nu_i$. For a given simulation and mass definition, we mask out bins that are not fully above the lower particle number limit. We convert all halo masses to peak height using Equation~\ref{eq:nu} and compute
\begin{equation}
\label{eq:dn}
\left. \frac{dn}{d \ln(\nu)}  \right|_i =  \frac{\sum_{j}^{N_{\rm sim}} n_{i,j} V_j}{\left( \sum_{j}^{N_{\rm sim}} V_j \right) \times \left( \ln \nu_{i+1} - \ln \nu_i \right)} \,,
\end{equation}
where $n_{i,j}$ is the number of halos in simulation $j$ and bin $i$, $V_j$ is the volume of simulation $j$, and $\nu_{i+1}$ and $\nu_{i}$ are the upper and lower edges of the bin. The index $j$ is understood to run only over simulations that contribute to the mass bin (or over only one simulation in the case of Figure~\ref{fig:convergence}). We estimate the fractional error on $f(\nu)$ from the uncertainty on the total number count in each bin,
\begin{equation}
\label{eq:error}
\left. \frac{\sigma_f}{f} \right|_i = \pm \frac{\sqrt{n_i + \frac{1}{4}} \pm \frac{1}{2}}{n_i} \,.
\end{equation}
This estimate of the uncertainty approaches $1/\sqrt{n}$ at large $n$ but does not reach zero for $n = 0$ and reproduces the asymmetry of the Poisson distribution \citep{lukic_07}. It does not, however, include other systematic errors or sample variance \citep{hu_03}.

\def\figsize{0.57}
\begin{figure*}
\centering
\includegraphics[trim =  1mm 24mm 2mm 2mm, clip, scale=\figsize]{\figdir/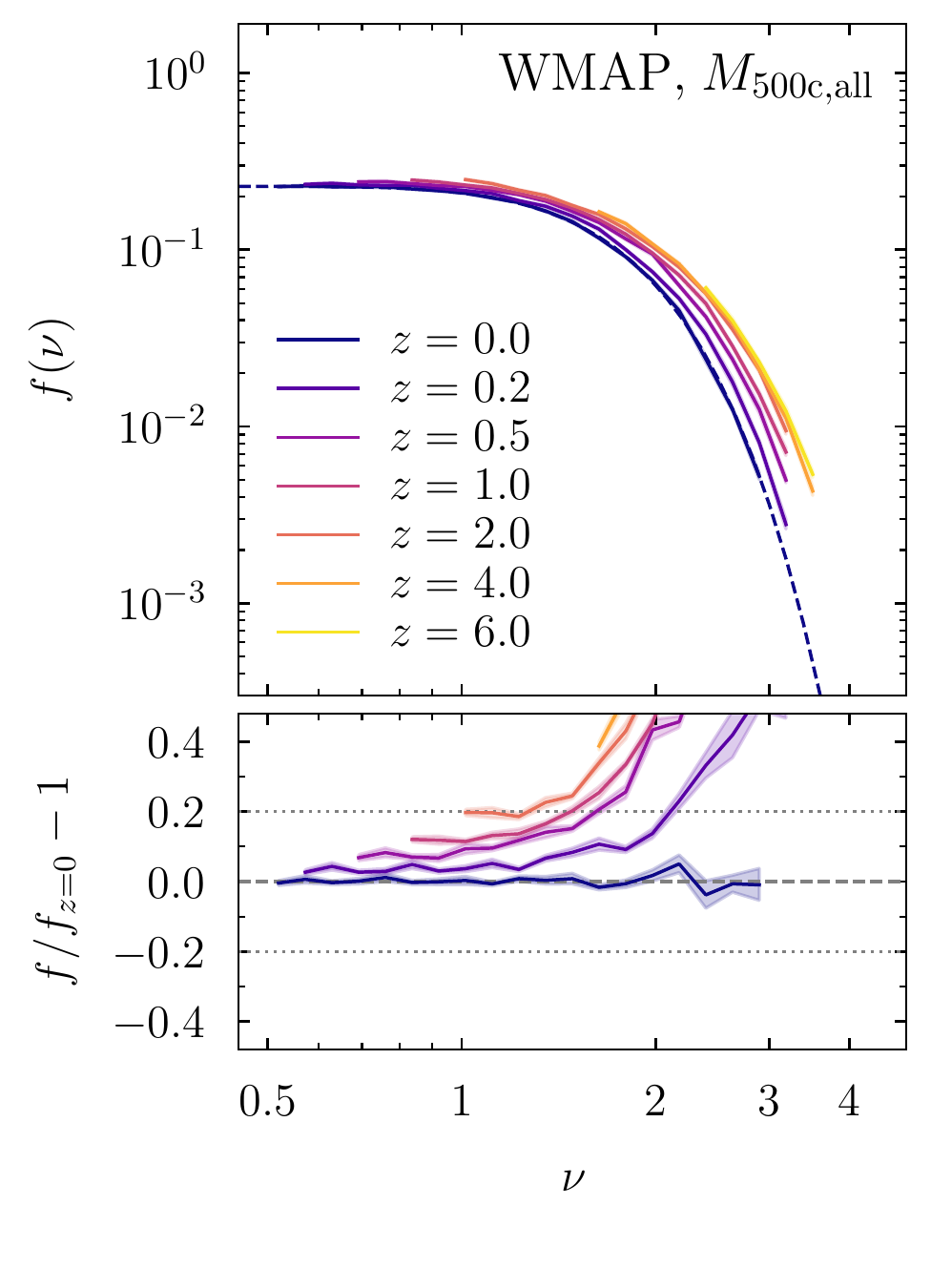}
\includegraphics[trim =  25mm 24mm 2mm 2mm, clip, scale=\figsize]{\figdir/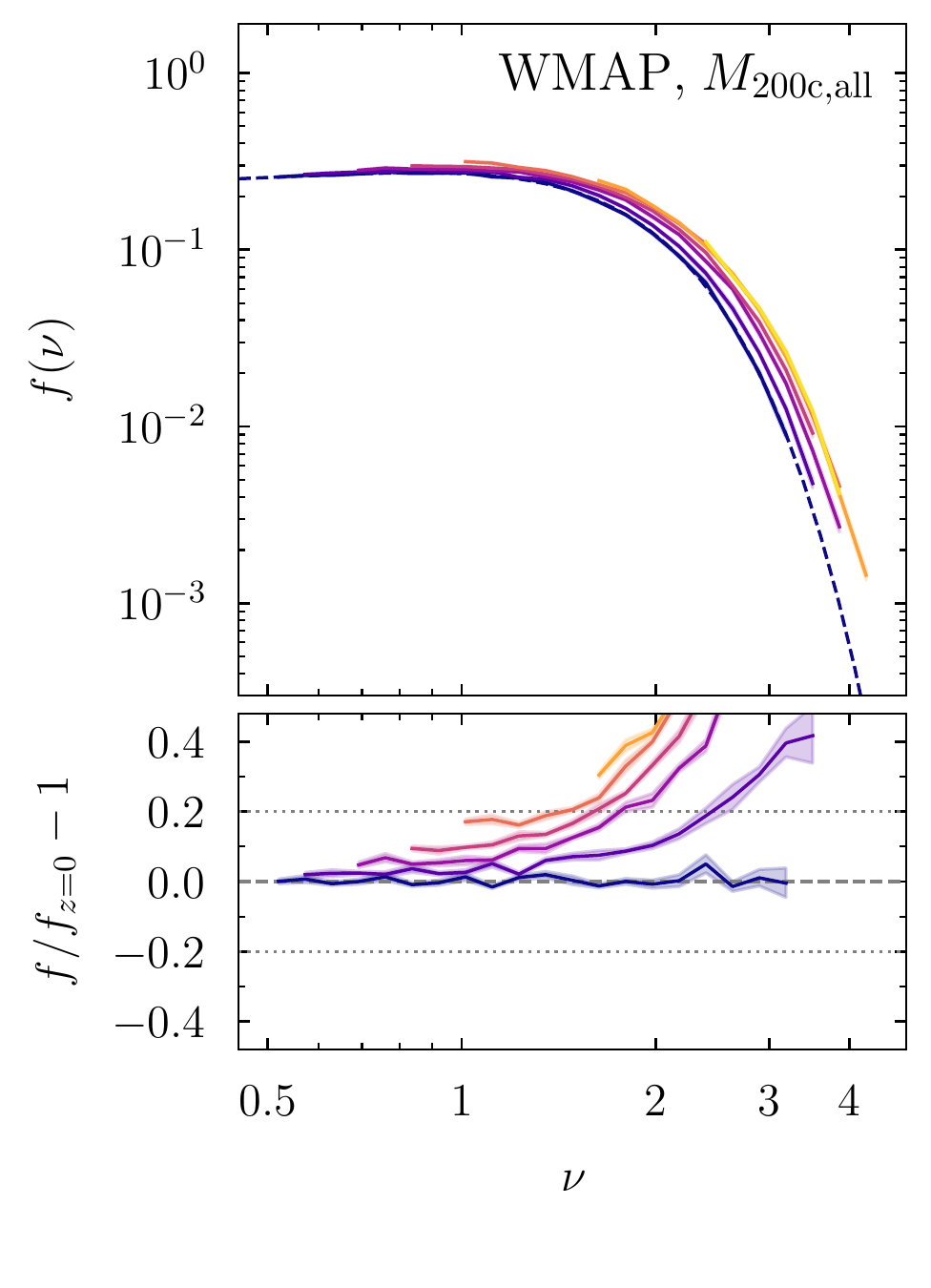}
\includegraphics[trim =  25mm 24mm 2mm 2mm, clip, scale=\figsize]{\figdir/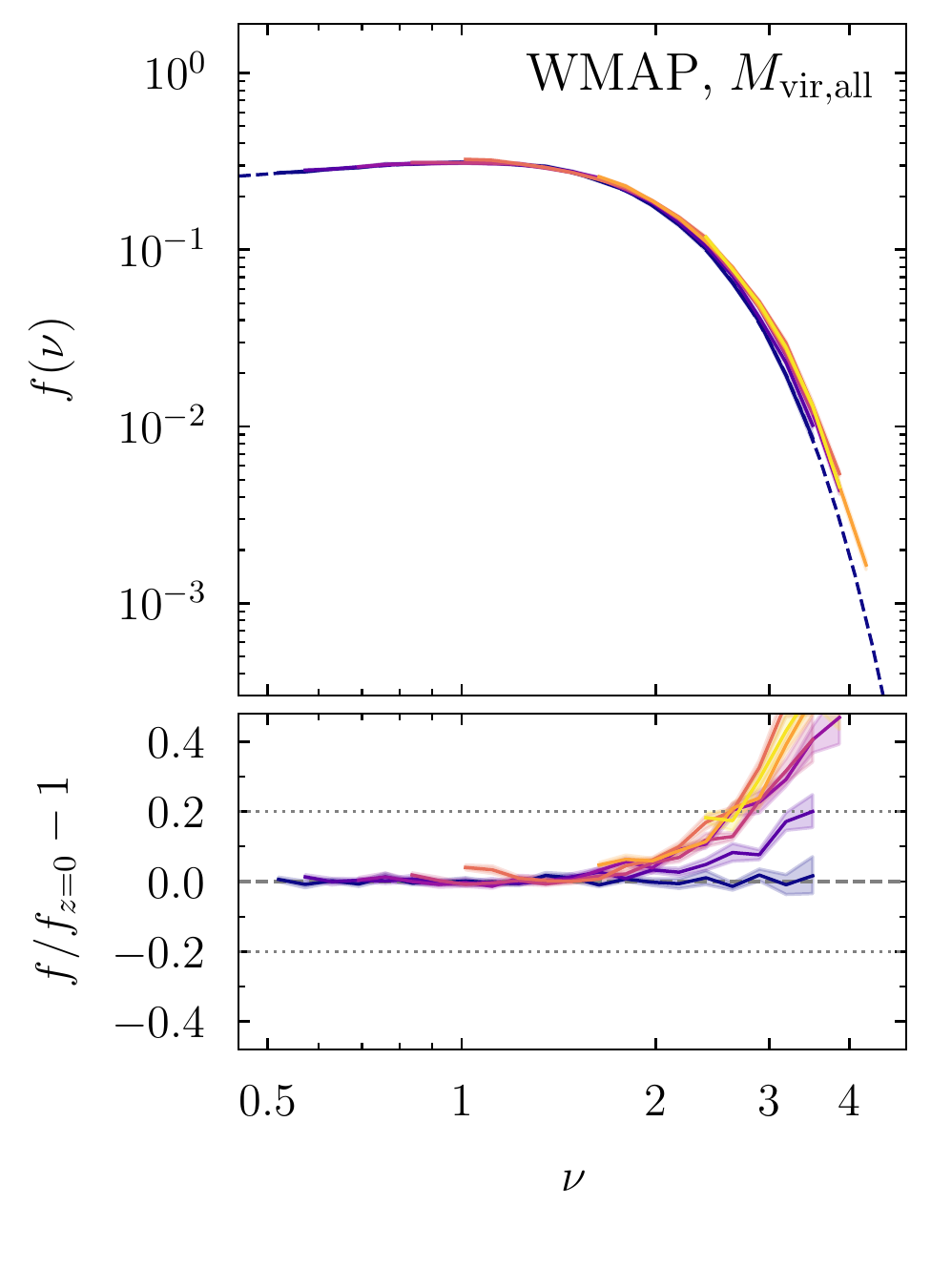}
\includegraphics[trim =  25mm 24mm 2mm 2mm, clip, scale=\figsize]{\figdir/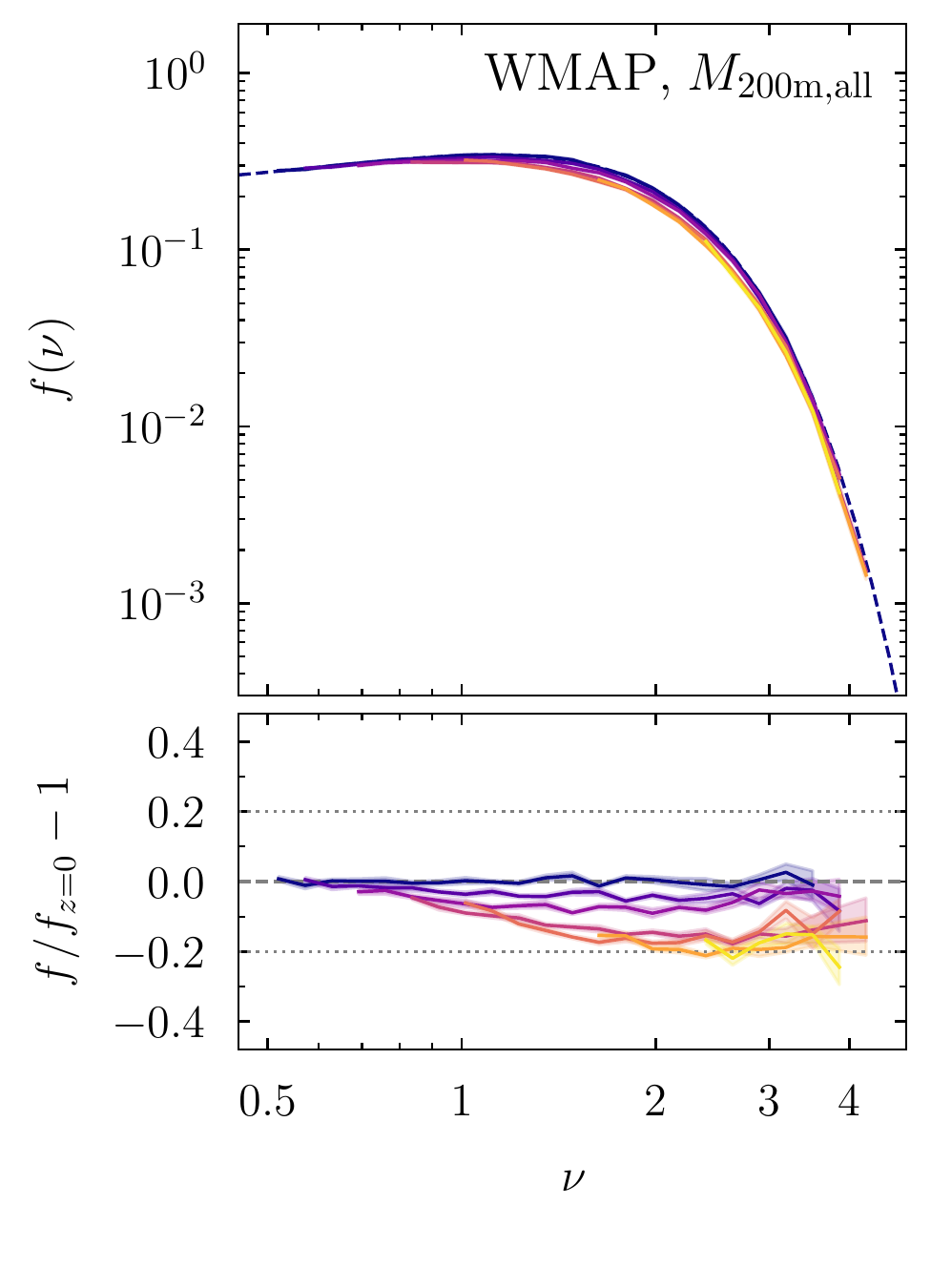}
\includegraphics[trim =  1mm 7mm 2mm 2mm, clip, scale=\figsize]{\figdir/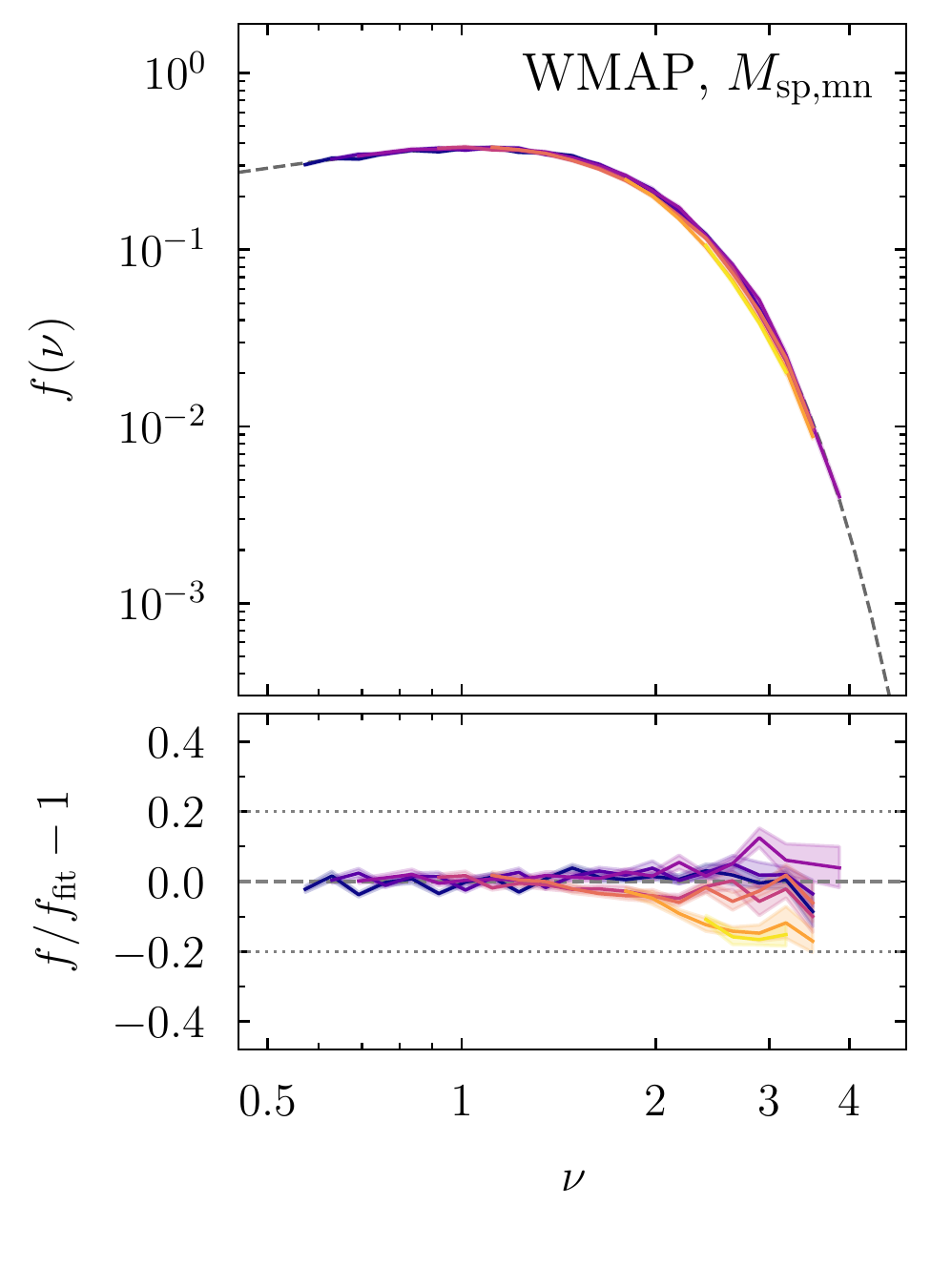}
\includegraphics[trim =  25mm 7mm 2mm 2mm, clip, scale=\figsize]{\figdir/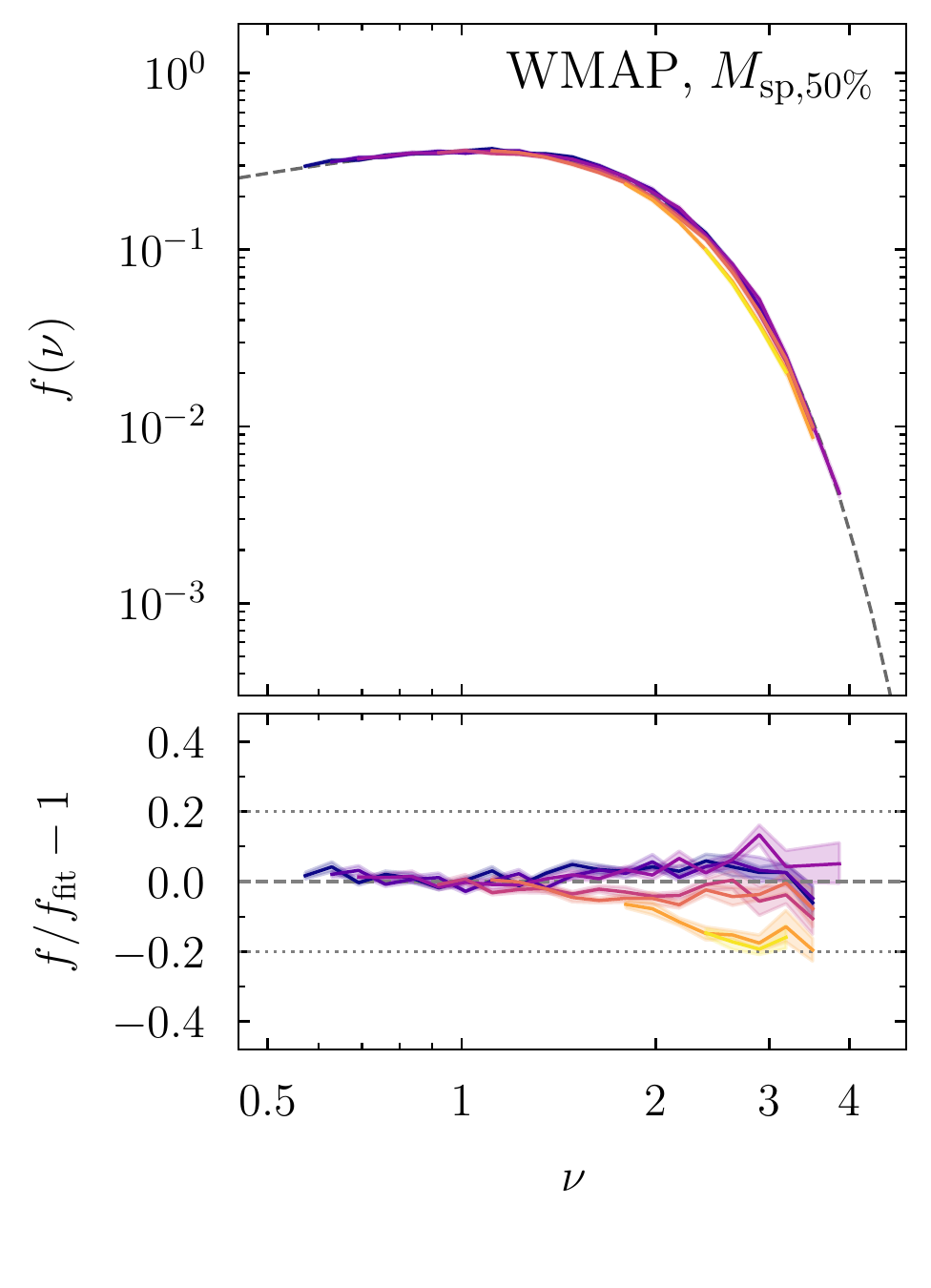}
\includegraphics[trim =  25mm 7mm 2mm 2mm, clip, scale=\figsize]{\figdir/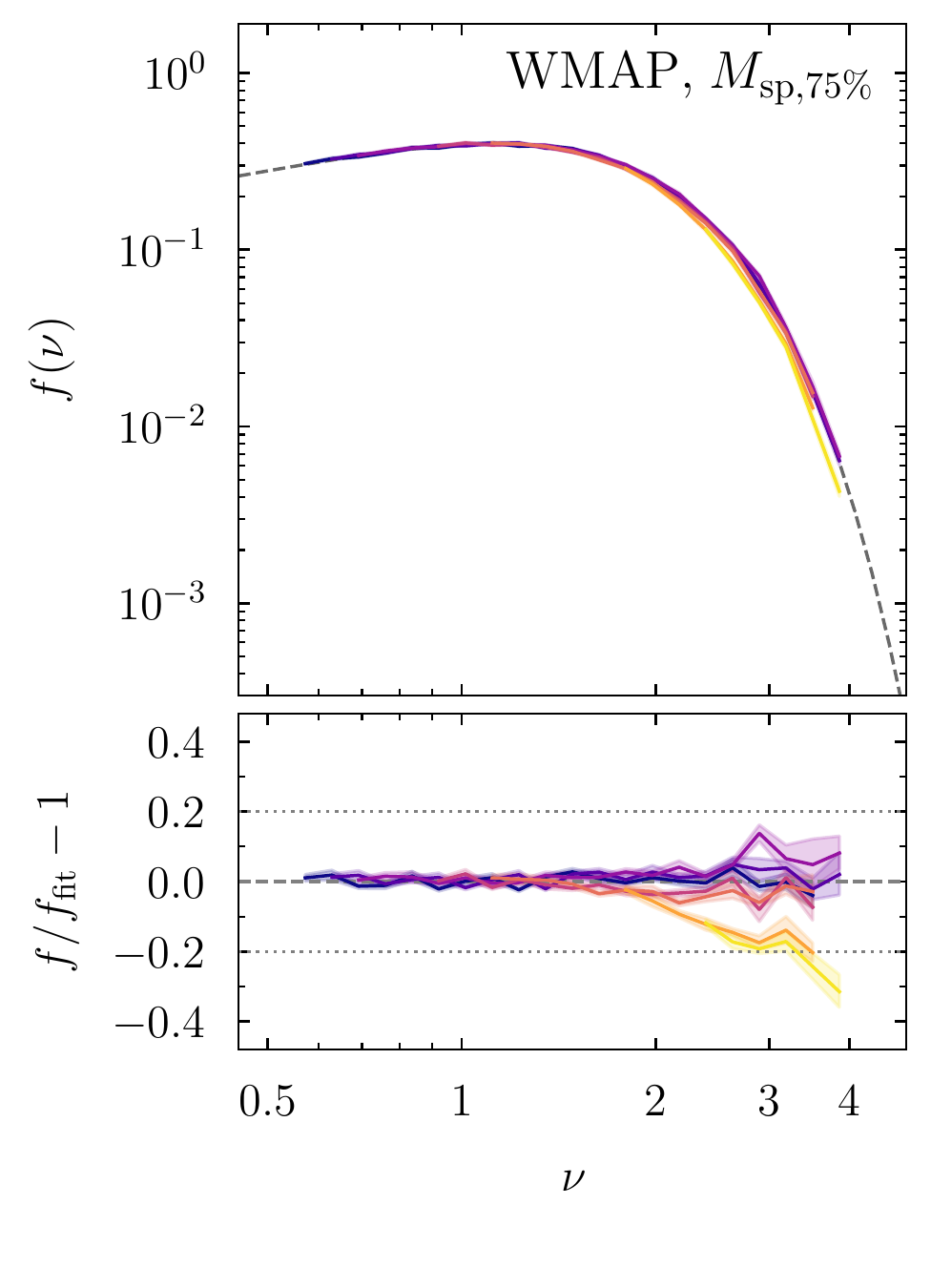}
\includegraphics[trim =  25mm 7mm 2mm 2mm, clip, scale=\figsize]{\figdir/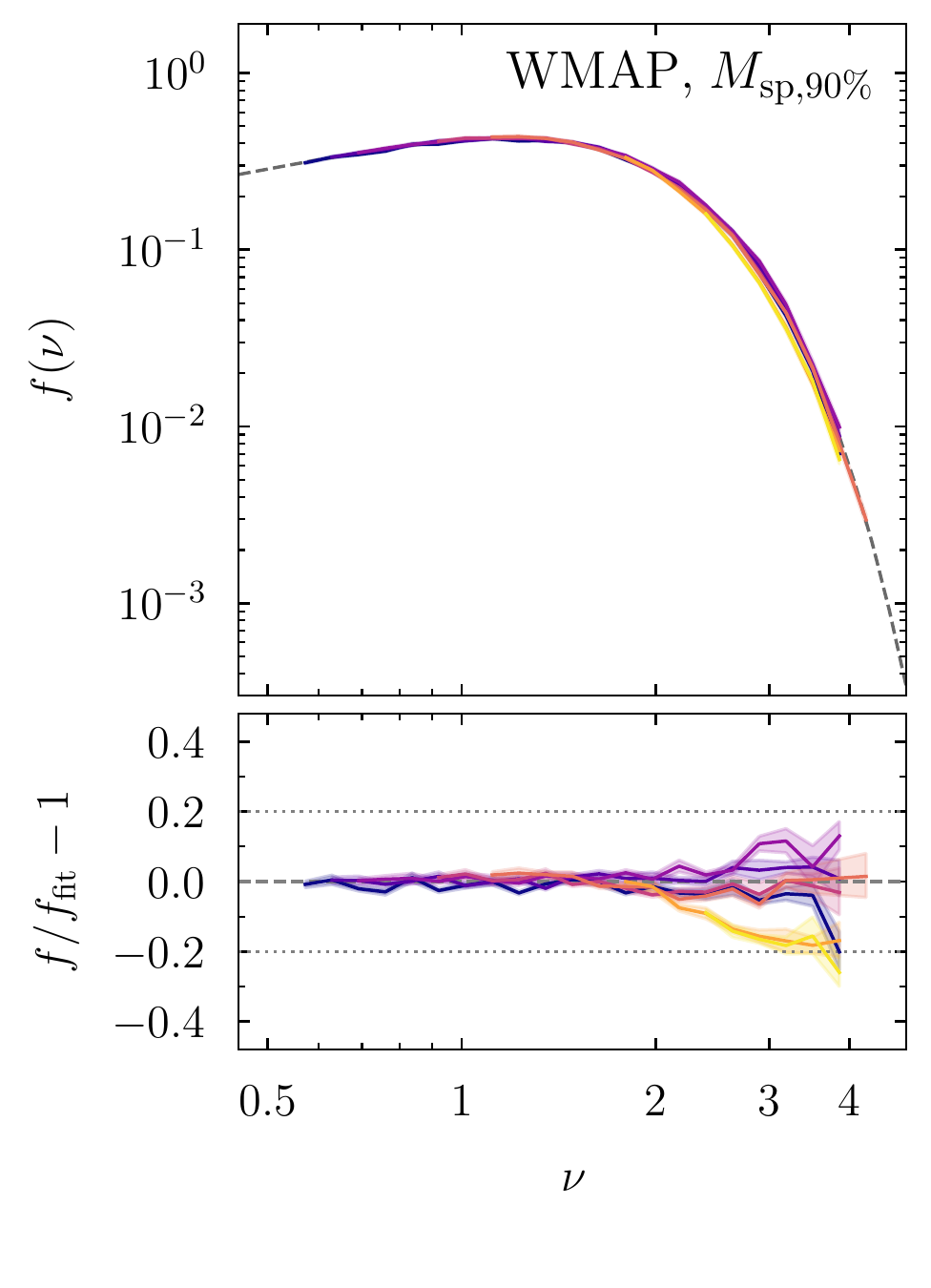}
\caption{Universality of the mass function with redshift in the \wmap cosmology. Each panel shows mass functions (lines) and statistical uncertainties (shaded areas) for one mass definition, the smaller bottom panels show a comparison to an empirical fit (dashed lines). The top row shows all-particle SO definitions. They are compared to a fit to the $z = 0$ data, which has no particular significance but serves as a smooth reference for comparison. All SO definitions exhibit significant non-universalities with redshift. The second row shows four splashback definitions, which we compare to our universal fitting function (dashed gray lines, Section~\ref{sec:results:fit}). The splashback mass functions exhibit non-universalities at high peak height and redshift but are universal to about 10\% at $z \leq 2$.}
\label{fig:univ_z}
\end{figure*}

For the self-similar simulations, we wish to combine different redshifts of the same simulation rather than different simulations at fixed redshift \citep[bottom-right panels of Figure~\ref{fig:convergence}; see, e.g.,][]{lacey_94, lee_99_mfunc}. Thus, the volume weighting in Equation~\ref{eq:dn} will not work, but we need to weight the different redshifts because they contribute very differently at different peak heights (with larger peak heights collapsing earlier). Thus, we replace $V_j$ with the number of halos, $n_{i,j}$ \citep{lacey_94}. This procedure is biased because bins that randomly fluctuate upwards in $n_{i,j}$ will receive systematically higher weight than bins that fluctuate downwards. In practice, this bias is negligible because most bins are dominated by one or two redshifts with large number counts and small statistical uncertainty. The uncertainty of the weighted average is then
\begin{equation}
\left. \frac{\sigma_f}{f} \right|_i = \pm \frac{\sqrt{ \sum_{j}^{N_z} \left( \sqrt{n_{i,j} + \frac{1}{4}} \pm \frac{1}{2} \right)^2}}{\sum_{j}^{N_z} n_{i,j}} \,,
\end{equation}
where $N_z$ is the number of redshifts we have averaged over \citep[see][for a more sophisticated method to test the convergence of self-similar mass functions]{leroy_20}. In principle, we could consider all snapshots of a self-similar simulation, but that would lead to highly correlated measurements. Instead, we assume that halo masses change significantly over one dynamical time (a crossing time defined as in \citetalias{diemer_17_sparta}). We avoid the final $0.3$ dynamical times due to the late-time correction to the splashback masses (Section~\ref{sec:method:sp}) and move backwards in steps of one dynamical time. The resulting redshifts are $z =$ [2.2, 3.1, 4.1, 5.5, 7.3, 9.5, 12.3, 16, 20] for $n = -1$, $z =$ [1.1, 1.7, 2.4, 3.3, 4.5, 6, 7.9, 10, 13] for $n = -1.5$, $z = $ [0.6, 1, 1.6, 2.3, 3.1, 4.2, 5.6] for $n = -2$, and $z = $ [0.07, 0.35, 0.7, 1.2, 1.8] for $n = -2.5$ (Figure~\ref{fig:convergence}).

In the following figures and fits, we omit bins where the mass function is constrained by fewer than $200$ halos to avoid noisy measurements. We plot each $\nu$ bin at its logarithmic center (as opposed to the mean peak height of halos in the bin).


\def\figsize{0.72}
\begin{figure*}
\centering
\includegraphics[trim =  1mm 5mm 2mm 3mm, clip, scale=\figsize]{\figdir/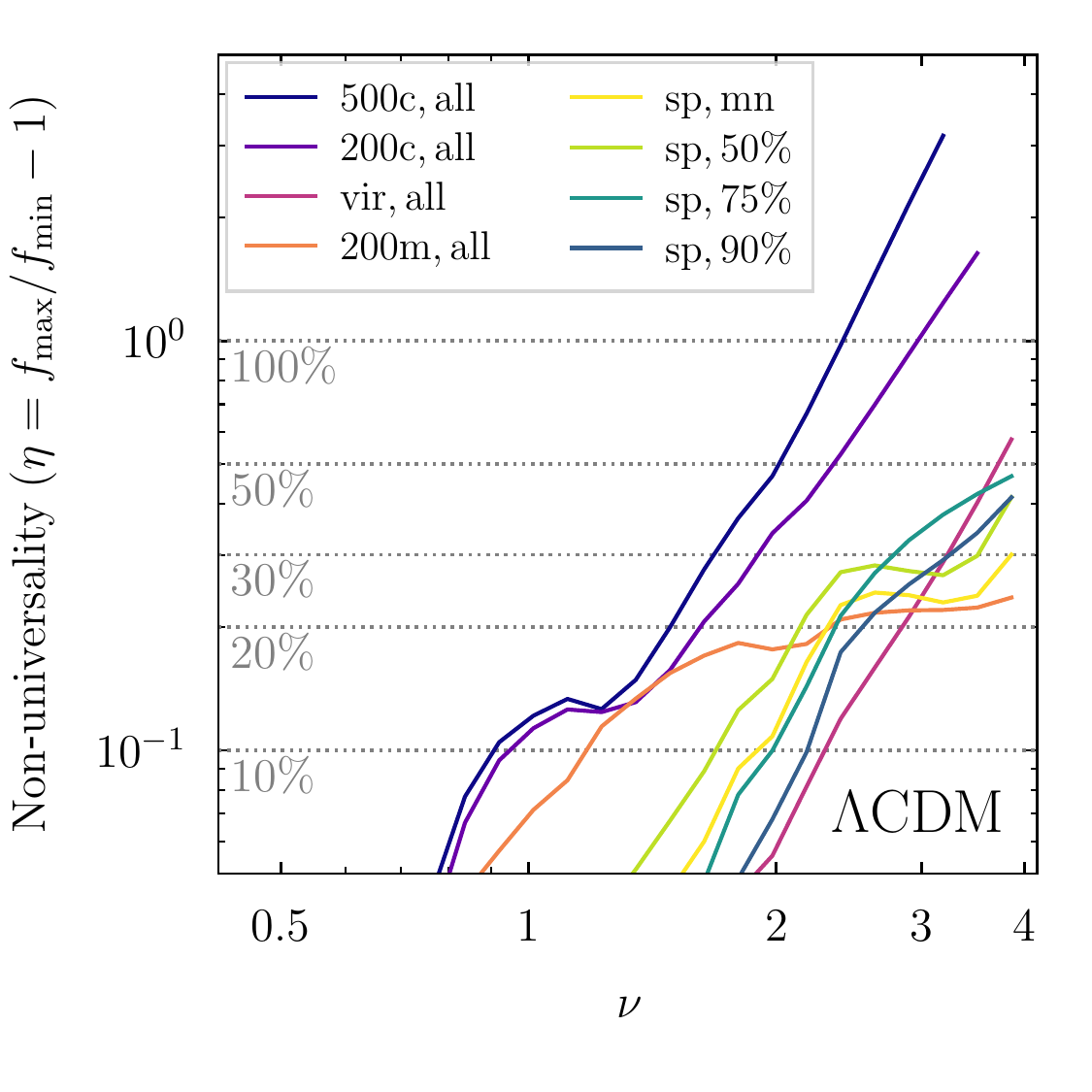}
\includegraphics[trim =  20mm 5mm 2mm 3mm, clip, scale=\figsize]{\figdir/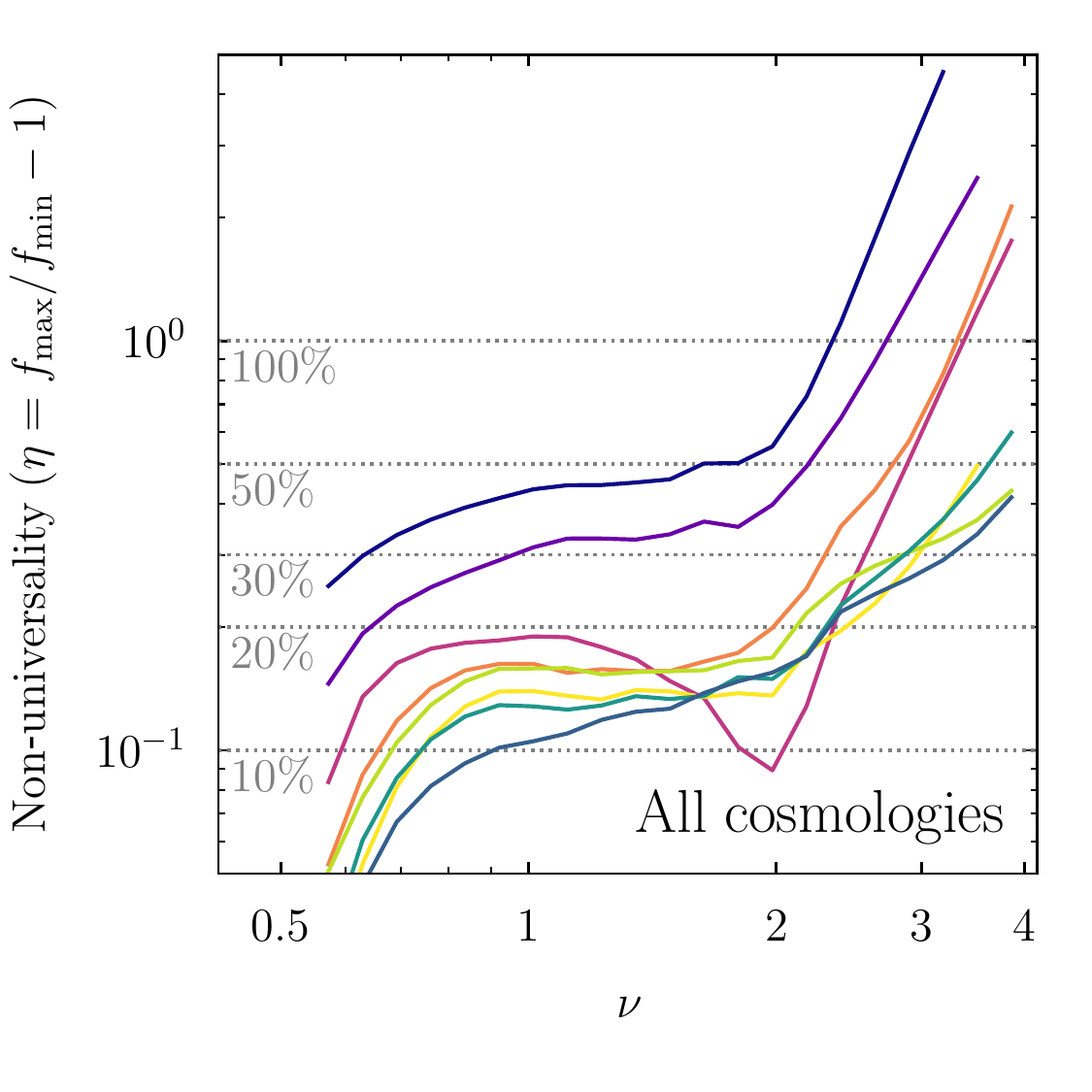}
\caption{Non-universality of the multiplicity function for different mass definitions, taking into account the \wmap and \planck cosmologies (left panel) and all cosmologies including self-similar universes (right panel). The relative differences appear slightly larger than in Figures~\ref{fig:univ_z} and \ref{fig:univ_all} because they are defined with respect to the minimum $f(\nu)$ rather than to some fitting function and because we include the \planck cosmology. In \LCDM, the non-universality grows with $\nu$ for all definitions, partly because high-redshift data are only available at higher peak height. $\mtom$ is non-universal to 20\% over a wide range of peak heights whereas $\mvir$ and the splashback definitions are more universal up to $\nu \approx 2$ and then rise to non-universalities between 30\% and 60\%. When we take into account self-similar cosmologies, the splashback definitions are clearly more universal than their SO counterparts.}
\label{fig:eta}
\end{figure*}

\section{Results}
\label{sec:results}

We are now in a position to investigate the universality of mass functions by comparing them between different redshifts and cosmologies (Sections~\ref{sec:results:univ_z} and \ref{sec:results:univ_cosmo}). We encapsulate the splashback mass functions in a simple, universal fitting function (Section~\ref{sec:results:fit}).

\subsection{Universality with Redshift}
\label{sec:results:univ_z}

We begin by comparing mass functions at different redshifts in the \wmap cosmology (Figure~\ref{fig:univ_z}). In each panel, we compare multiplicity functions for a given mass definition. The bottom panels show the fractional deviation from a reference function, for which we use a fit of Equation~\ref{eq:fit_global} (dashed lines) because the binned data do not span the entire $\nu$ range and because they suffer from shot noise. Any disagreements between the ratios shown in the bottom panels indicate non-universality.

The top row shows the all-particle versions of four SO definitions. The comparisons are very similar for the bound-only results or for the \planck cosmology (for which we have fewer simulations and thus less coverage of $\nu$-$z$ space). We compare the mass function at each redshift to a fit to the $z = 0$ mass function (blue dashed lines). This fit has no physical significance, it is merely a convenient point of comparison. Clearly, all SO mass functions suffer from significant non-universality with redshift. For all but $\mtom$, $f(\nu)$ increases with redshift, especially at the high-$\nu$ end. The trend is particularly strong between $z = 0.5$ and $z = 0$, indicating that it is caused by the accelerated decline of $\Omega_{\rm m}(z)$ due to dark energy. This trend holds, to a lesser extent, even for the evolving $\Delta_{\rm vir}$ threshold. $\mtom$ shows a distinctly different redshift evolution, where the normalization of the mass function slowly decreases with redshift, reaching an offset of 20\% at $z = 6$ (in agreement with \citealt{tinker_08}). The $\mtoc$ and $\mtom$ mass functions are identical at high redshift where $\rhoc \approx \rhom$, the difference is in their evolution toward low $z$. 

\def\figsize{0.57}
\begin{figure*}
\centering
\includegraphics[trim =  1mm 7mm 2mm 2mm, clip, scale=\figsize]{\figdir/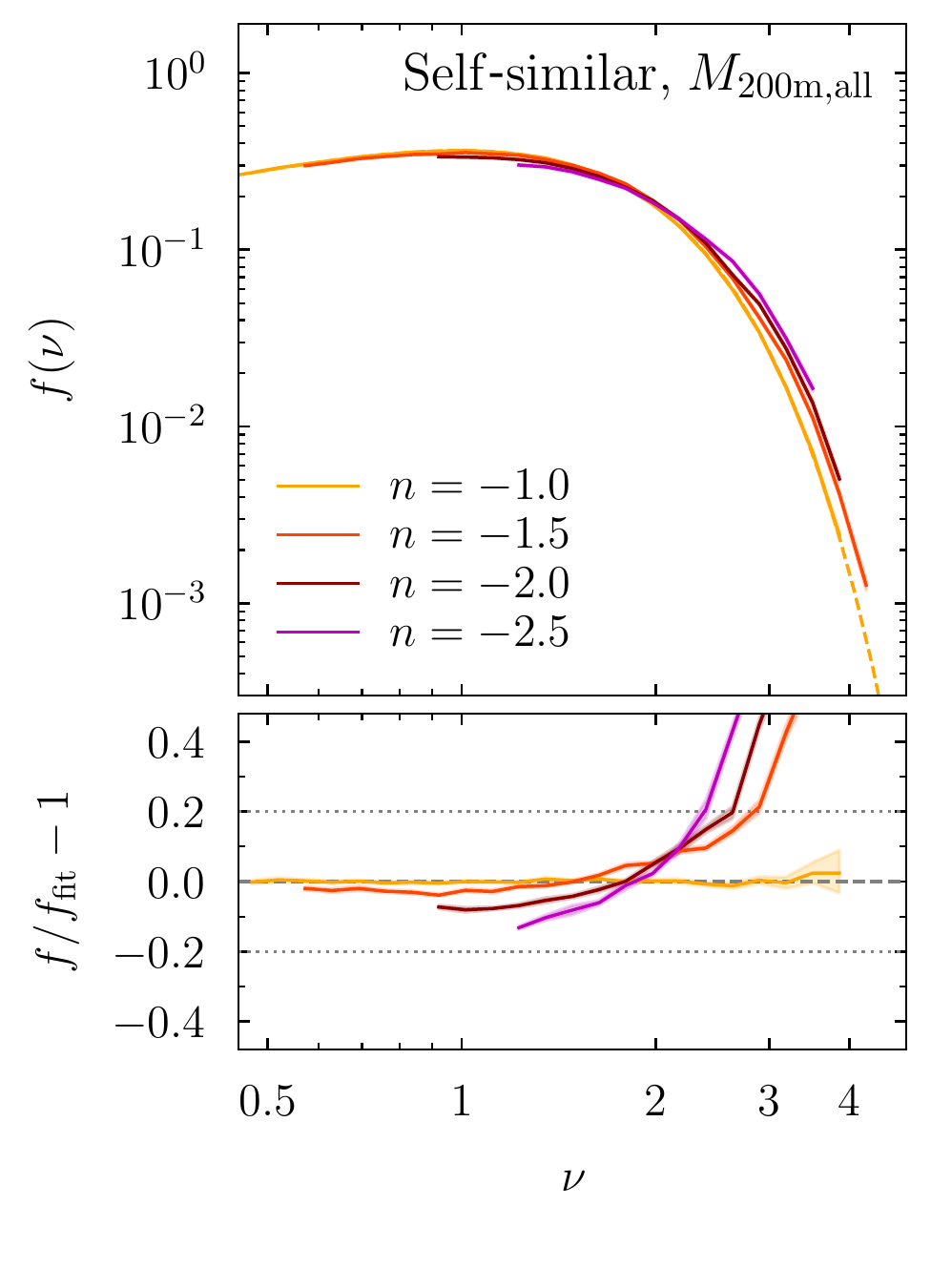}
\includegraphics[trim =  25mm 7mm 2mm 2mm, clip, scale=\figsize]{\figdir/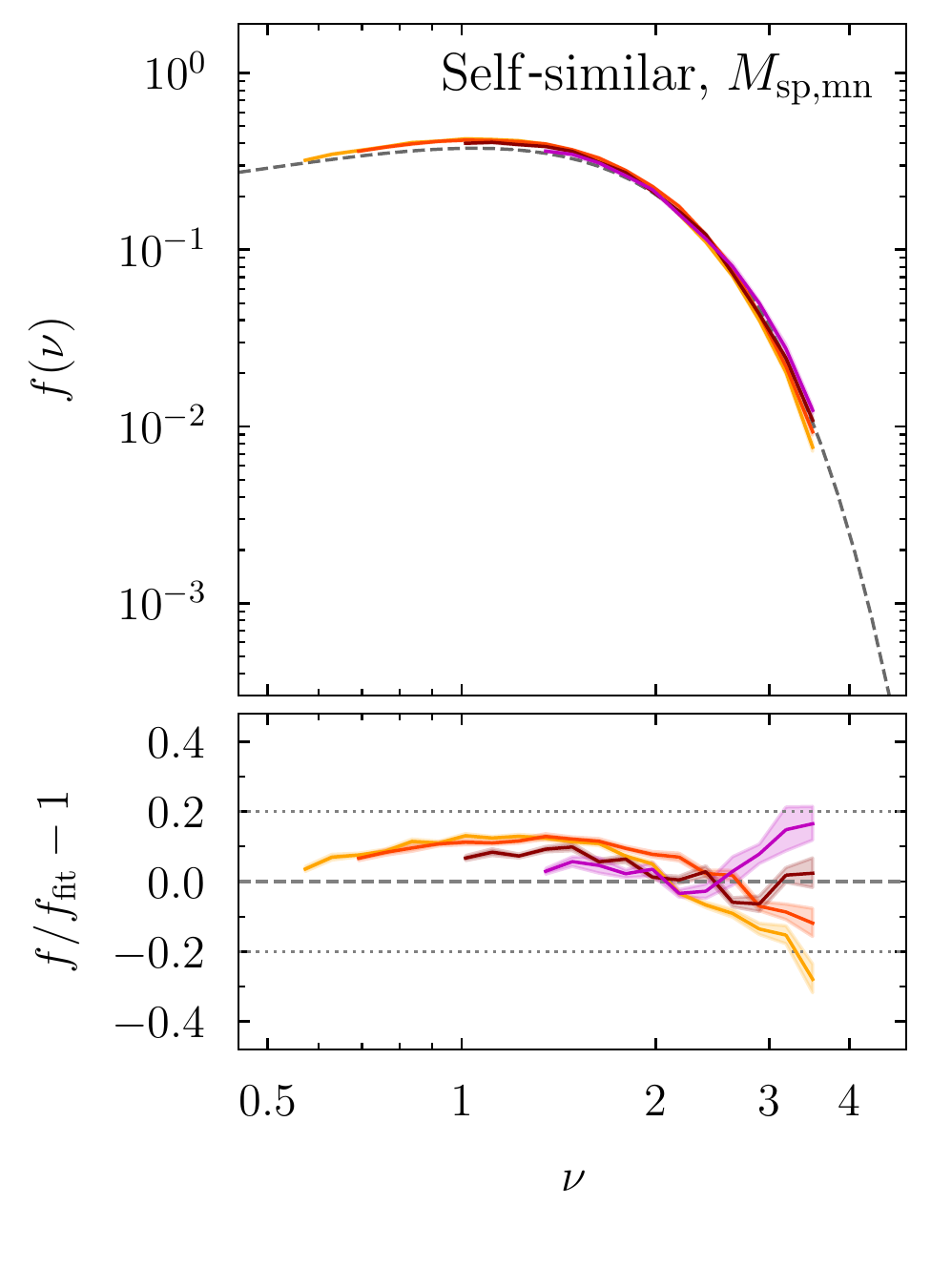}
\includegraphics[trim =  25mm 7mm 2mm 2mm, clip, scale=\figsize]{\figdir/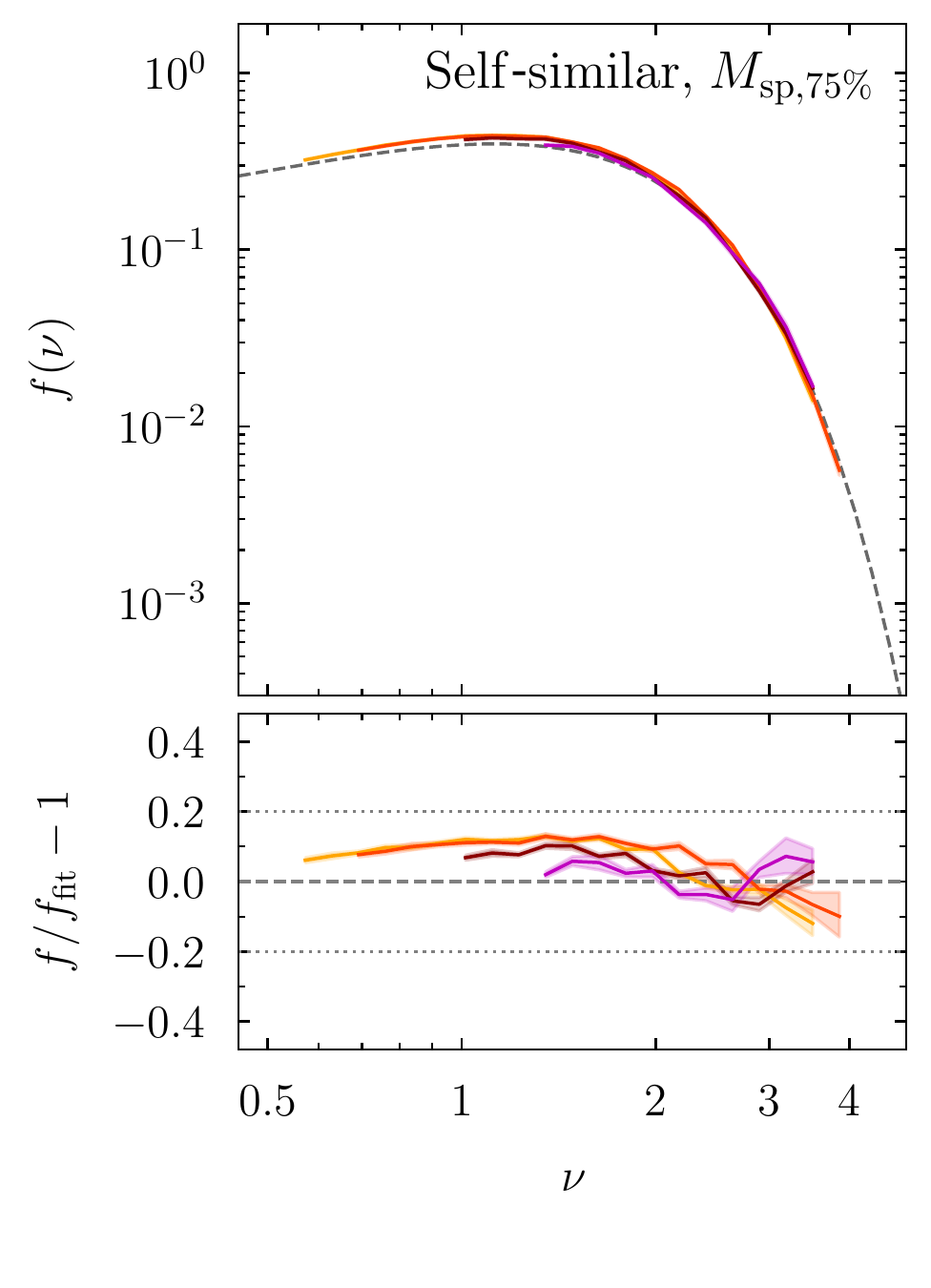}
\includegraphics[trim =  25mm 7mm 2mm 2mm, clip, scale=\figsize]{\figdir/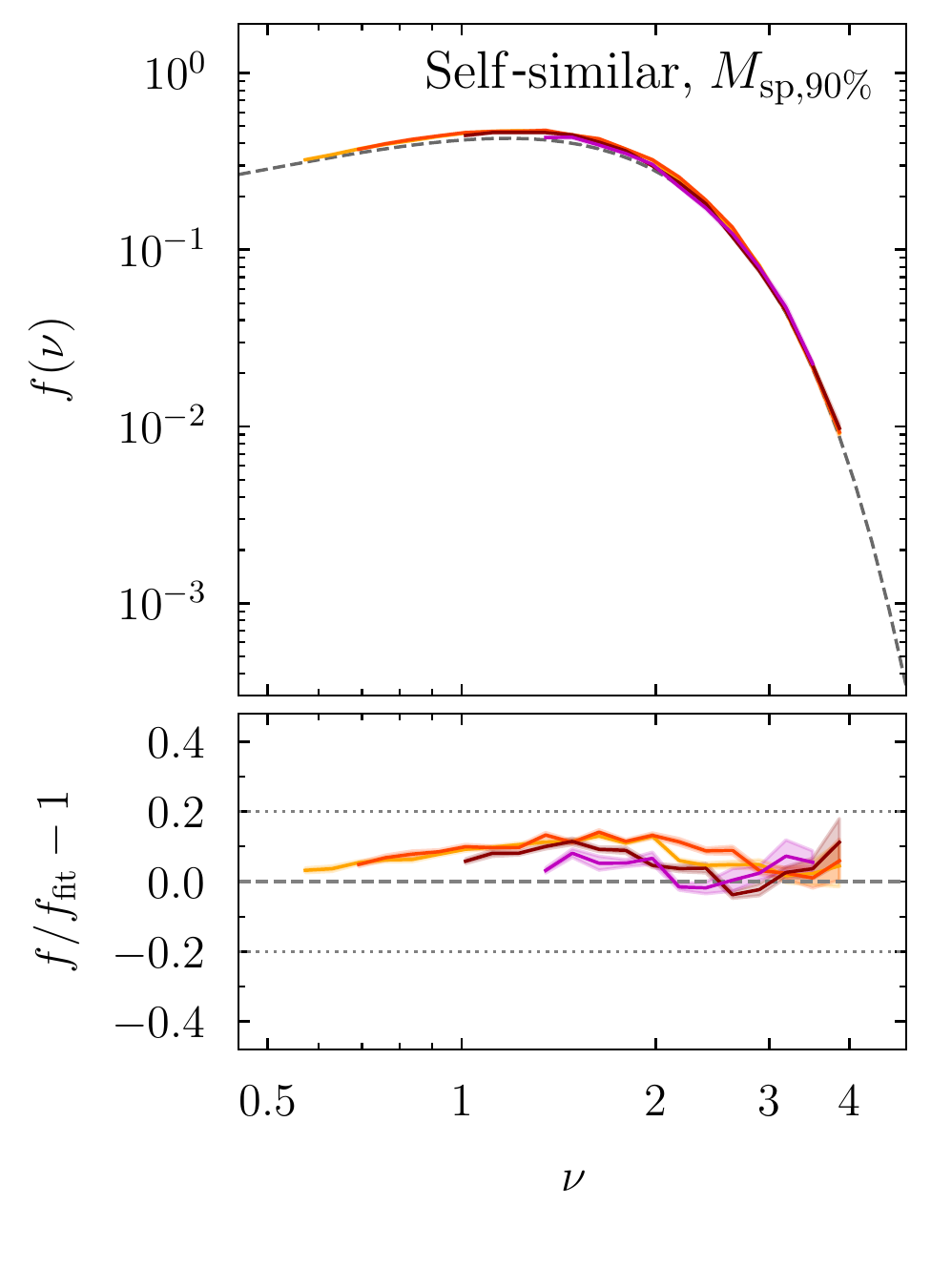}
\caption{Same as Figure~\ref{fig:univ_z} but for self-similar cosmologies. For $\mtom$ (left panel), we arbitrarily compare the mass functions to a fit to $n = -1$. For the splashback definitions (right three panels), we compare to the universal fitting function. While $\mtom$ is the most universal of the SO definitions (Figure~\ref{fig:univ_z}), its mass function varies significantly with power spectrum slope. The splashback mass functions are notably more universal than their SO counterparts. Their universality improves with percentile, meaning that larger radii lead to more universal mass functions.}
\label{fig:univ_ss}
\end{figure*}

However, it can be misleading to make quantitative inferences about universality based on Figure~\ref{fig:univ_z} because the eye is drawn to noisy outlier bins and deviations from the arbitrary fitting functions. To accurately quantify the differences, we define the non-universality $\eta$ as the fractional difference between the highest and lowest $f(\nu)$,
\begin{equation}
\eta \equiv {\rm max} \left( \frac{f_{\rm max} - f_{\rm min} - \sqrt{\sigma_{\rm max}^2 + \sigma_{\rm min}^2}}{f_{\rm min}}, \, 0 \right) \,,
\end{equation}
where $f_{\rm min}$ and $f_{\rm max}$ are the lowest and highest multiplicity functions in a given $\nu$ bin and $\sigma_{\rm min}$ and $\sigma_{\rm max}$ are their uncertainties. The error term ensures that only statistically significant differences are considered, i.e., that $\eta = 0$ if $f_{\rm min}$ and $f_{\rm max}$ are compatible within one standard deviation. Given that the uncertainties do not include systematic errors, $\eta$ should be seen as an upper limit. The left panel of Figure~\ref{fig:eta} shows $\eta$ as a function of peak height for our \LCDM cosmologies. We have included redshifts $0$--$6$ in the \wmap cosmology and $0$--$4$ in the \planck cosmology, the range where we have reliable data. We use the same peak height bins as in Figure~\ref{fig:univ_z}. In a given bin, we remove mass functions measured with fewer than $100$ halos as in Figure~\ref{fig:convergence}; compared to the stricter $200$ halo limit, this choice extends the curves slightly at high $\nu$. Moreover, we have removed a spurious drop of $\eta$ at the highest peak heights that can occur because the mass functions of certain redshifts are not constrained any longer. To avoid noisy curves, we smooth the results with a fourth-order \citet{savitzky_64} filter with a window size of $11$ bins. We have confirmed that this smoothing does not change any of the noteworthy features in Figure~\ref{fig:eta}. Most importantly, it does not systematically change the end points of the curves where they reach the largest non-universalities. Nevertheless, our definition of $\eta$ is far from unique. With higher-resolution simulations, for instance, we could push to higher redshift and might infer larger non-universalities. Figure~\ref{fig:eta} is thus intended as a relative comparison of the different mass definitions.

The $\eta$ curves in the left panel of Figure~\ref{fig:eta} reflect our previous conclusions regarding SO mass functions but further illuminate their behavior at high $\nu$: while $\mvir$ is universal to about 10\% up to $\nu \approx 2.5$, its non-universality quickly increases with $\nu$ and reaches about 50\% (in agreement with Figure~2 of \citealt{despali_16}). Figure~\ref{fig:eta} also allows us to quantify the non-universality of $\mtoc$ and $\mfoc$, which is apparent even at low $\nu$ and reaches factors of more than unity.

We now turn to the splashback definitions (second row of Figure~\ref{fig:univ_z}). Here, we compare to the universal fitting function introduced in Section~\ref{sec:results:fit} (gray dashed lines). In contrast to the SO definitions, this fit is performed at $z \leq 2$ rather than $z = 0$, but we are interested in the relative differences between mass functions, not in the fit quality (which we consider in Section~\ref{sec:results:fit}). For a fair comparison with the SO definitions, we use the $z = 0$ splashback masses rather than $z = 0.13$ despite the caveat mentioned in Section~\ref{sec:method:sp}; we have checked that the $z = 0.13$ results are very similar. All splashback mass functions exhibit excellent universality up to $z \approx 2$, with the redshifts agreeing to about 10\%. At $z \geq 3$ and $\nu > 2$, we notice a relative decline in the mass function. The nature of this high-$z$ decrease is fundamentally different from the redshift evolution of the SO mass functions in that it clearly does not depend on dark energy. Instead, it could indicate an issue in \sparta's algorithm at high $z$ or a genuine non-universality in the splashback mass functions, perhaps caused by the rapidly changing mass accretion rates at high $z$. Our conclusions are confirmed in Figure~\ref{fig:eta}, where the splashback definitions top out between 30\% and 45\% (though their differences at high $\nu$ are not particularly significant). Overall, $M_{\rm sp,90\%}$ exhibits the highest level of universality over most of the peak height range. The values of $\eta$ are a little larger than one might infer from Figure~\ref{fig:univ_z} because we are dividing the difference by the lowest $f(\nu)$ now instead of the the universal fitting function.

In summary, the mass functions of no definition are truly universal across all redshifts in \LCDM. While high-threshold SO definitions such as $\mtoc$ and $\mfoc$ exhibit a large redshift evolution, $\mtom$, $\mvir$, and $\msp$ lead to comparable levels of non-universality with somewhat different dependencies on peak height. However, the $\msp$ functions are noticeably more universal at $z \leq 2$.

\subsection{Universality with Cosmology}
\label{sec:results:univ_cosmo}

\def\figsize{0.57}
\begin{figure*}
\centering
\includegraphics[trim =  1mm 24mm 2mm 2mm, clip, scale=\figsize]{\figdir/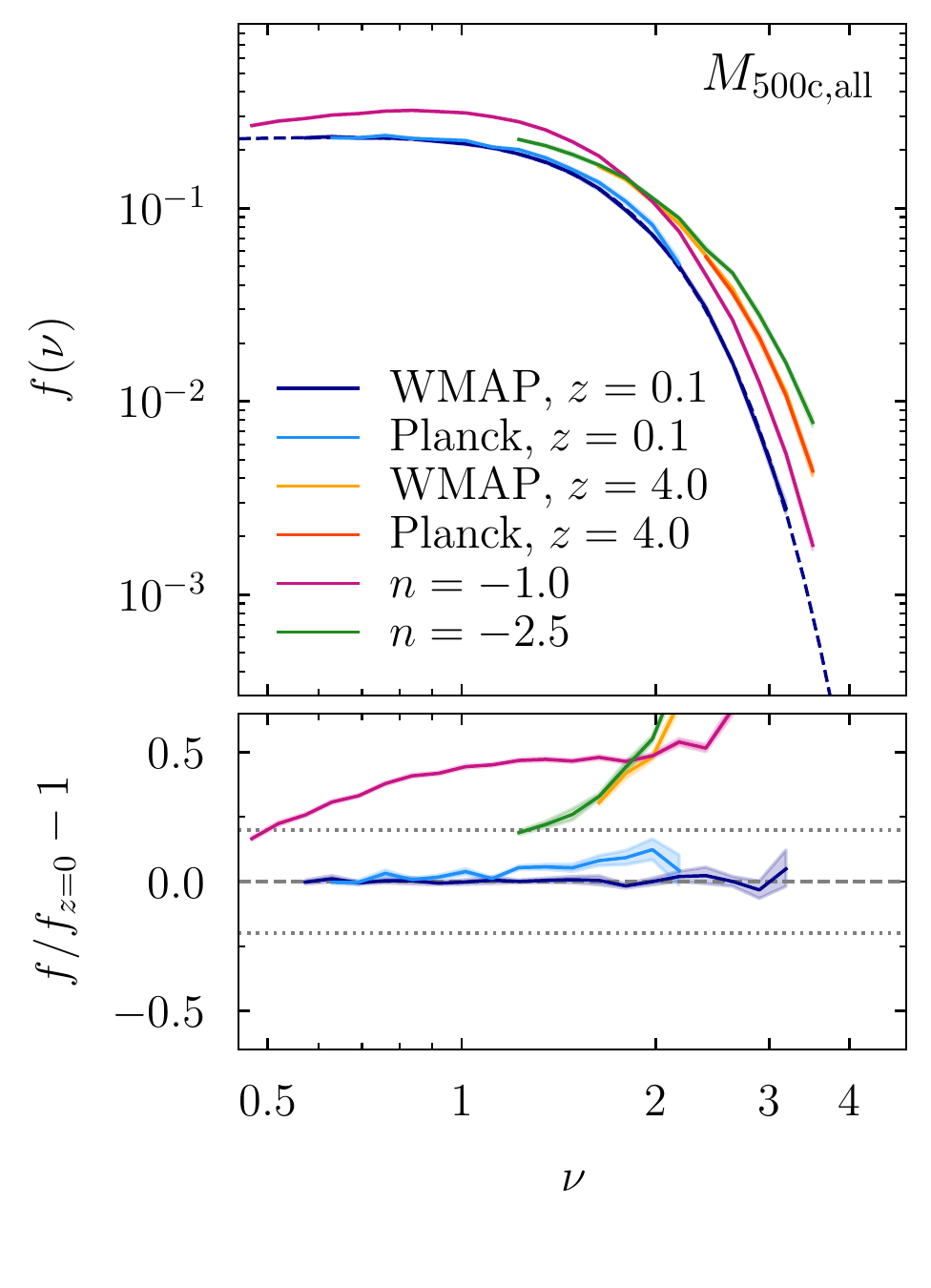}
\includegraphics[trim =  25mm 24mm 2mm 2mm, clip, scale=\figsize]{\figdir/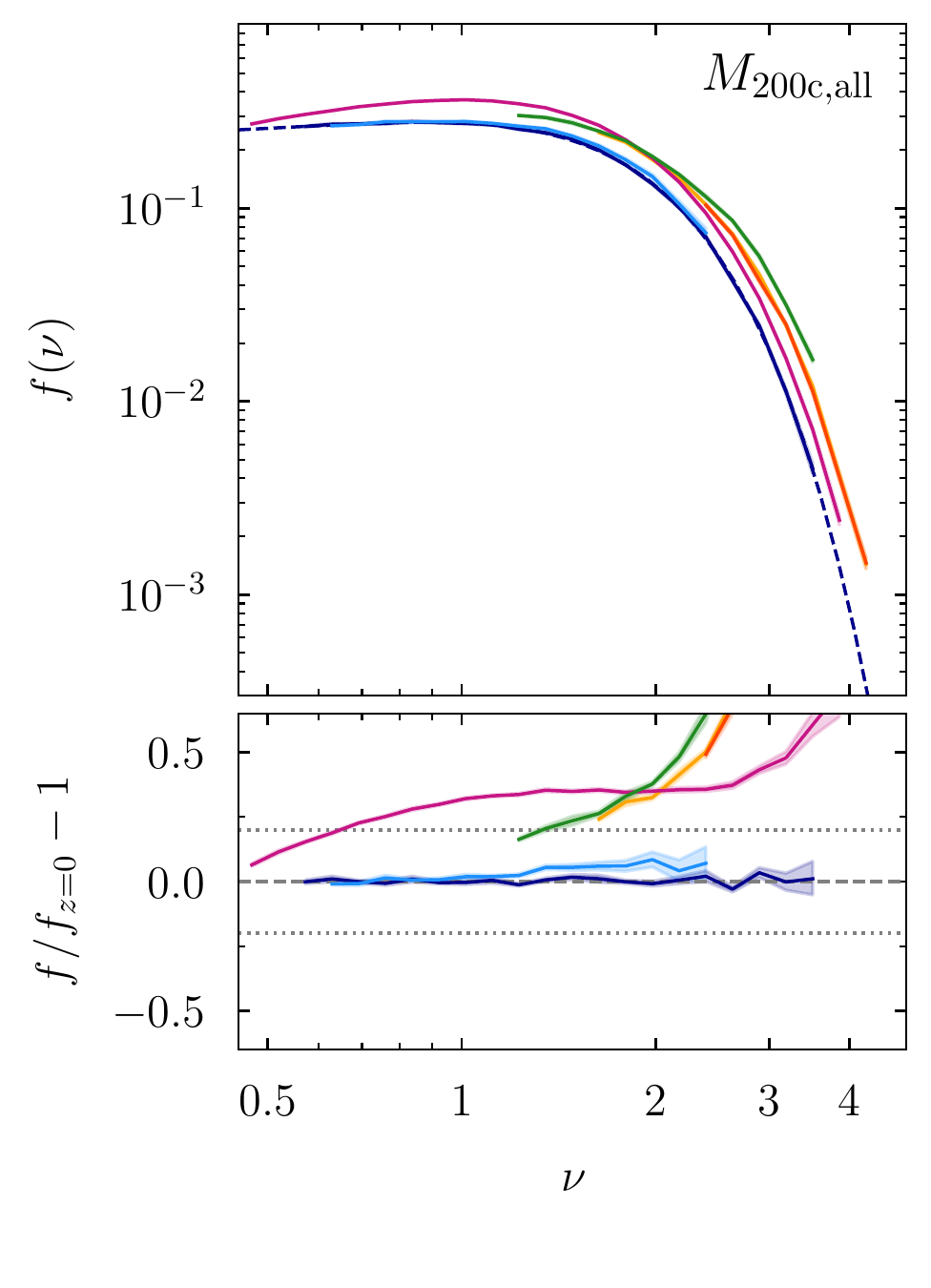}
\includegraphics[trim =  25mm 24mm 2mm 2mm, clip, scale=\figsize]{\figdir/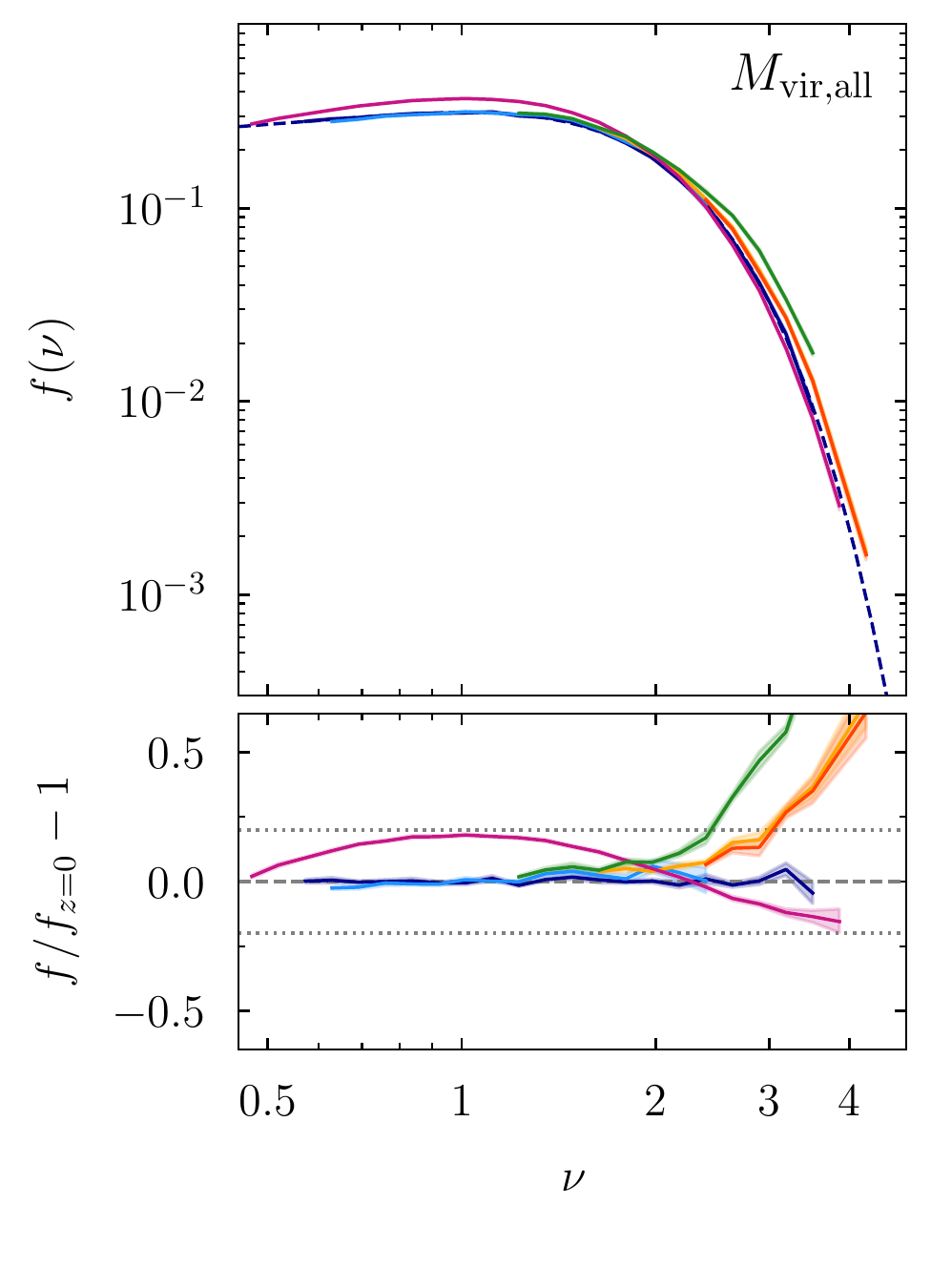}
\includegraphics[trim =  25mm 24mm 2mm 2mm, clip, scale=\figsize]{\figdir/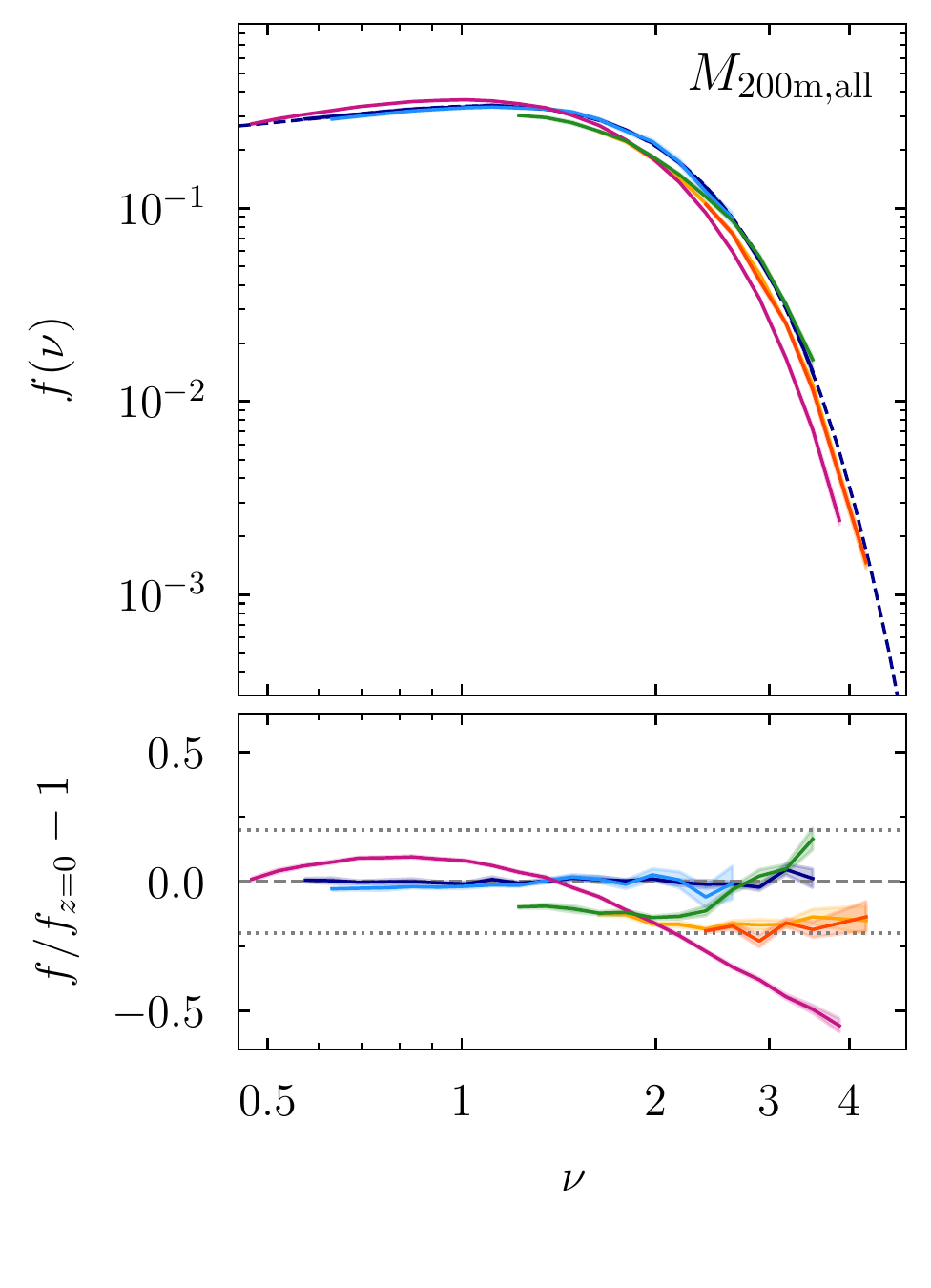}
\includegraphics[trim =  1mm 7mm 2mm 2mm, clip, scale=\figsize]{\figdir/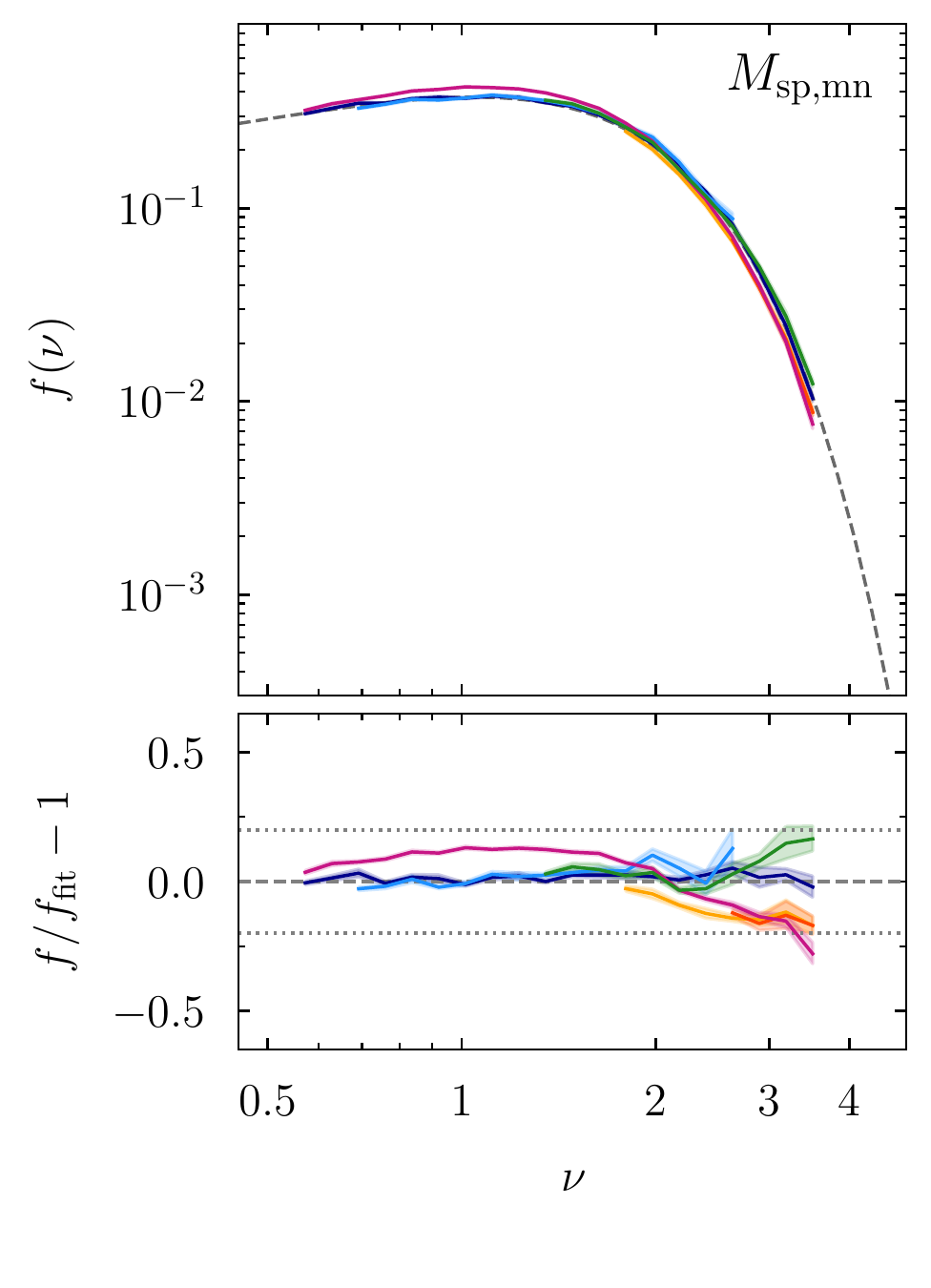}
\includegraphics[trim =  25mm 7mm 2mm 2mm, clip, scale=\figsize]{\figdir/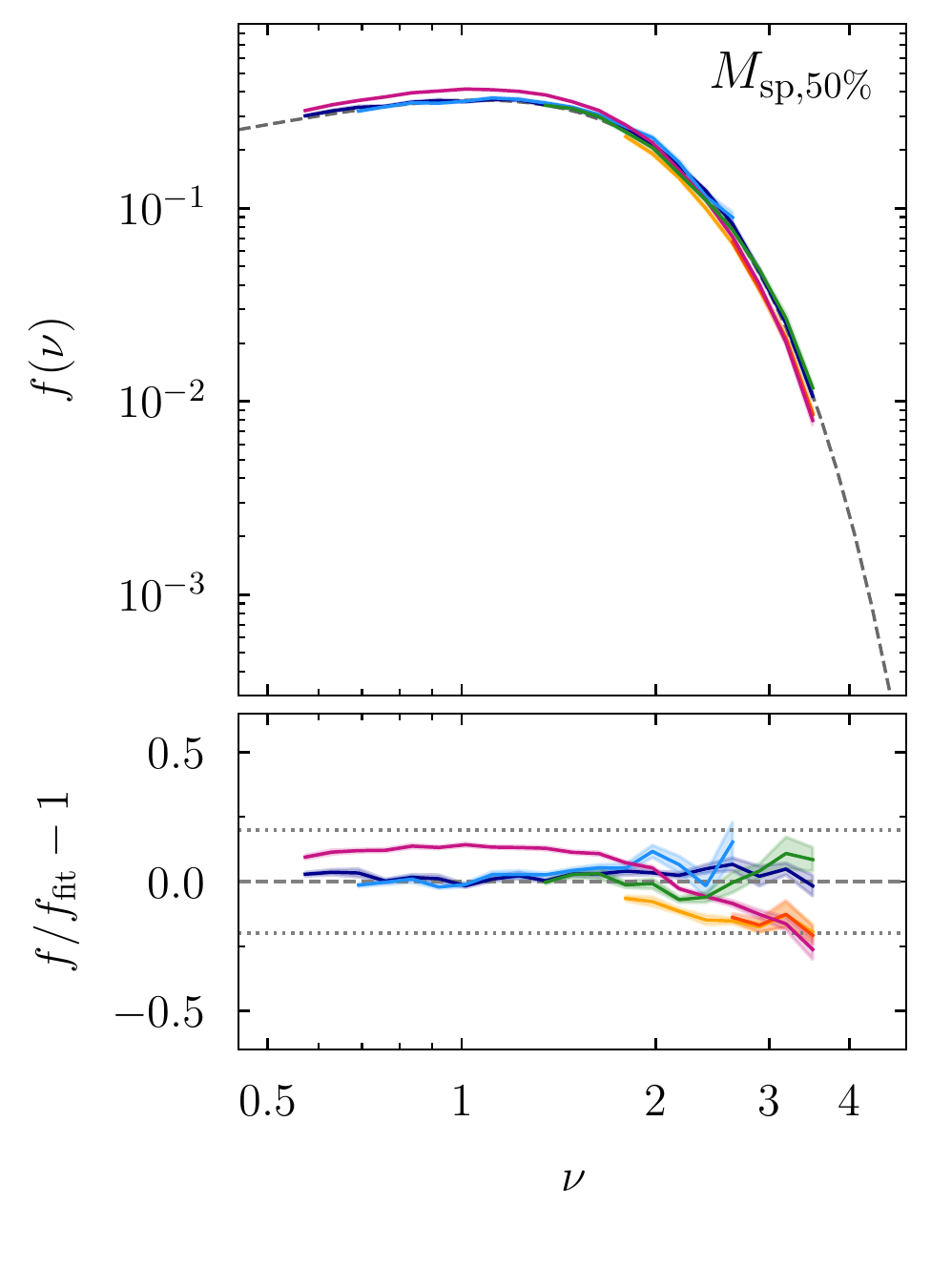}
\includegraphics[trim =  25mm 7mm 2mm 2mm, clip, scale=\figsize]{\figdir/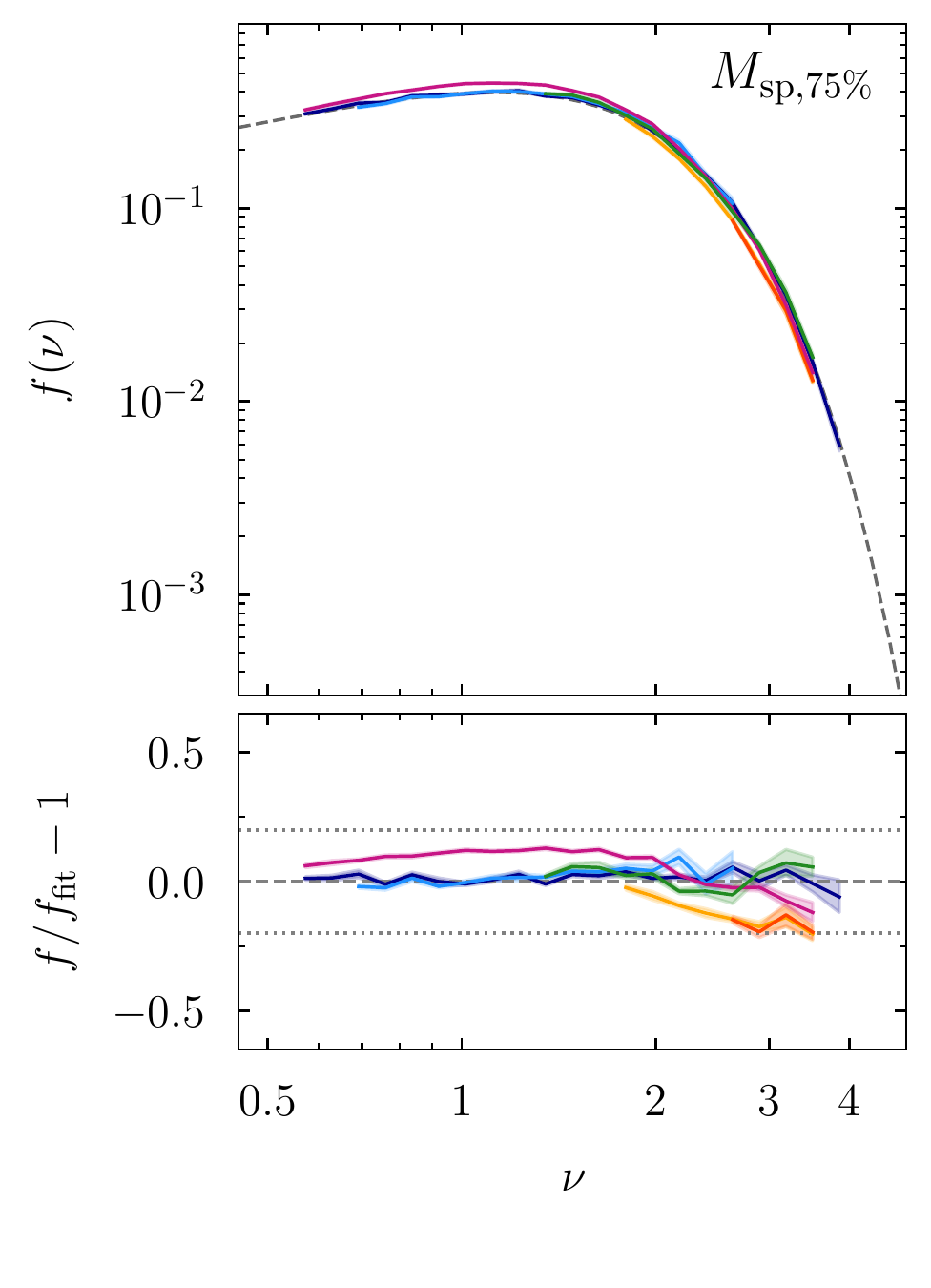}
\includegraphics[trim =  25mm 7mm 2mm 2mm, clip, scale=\figsize]{\figdir/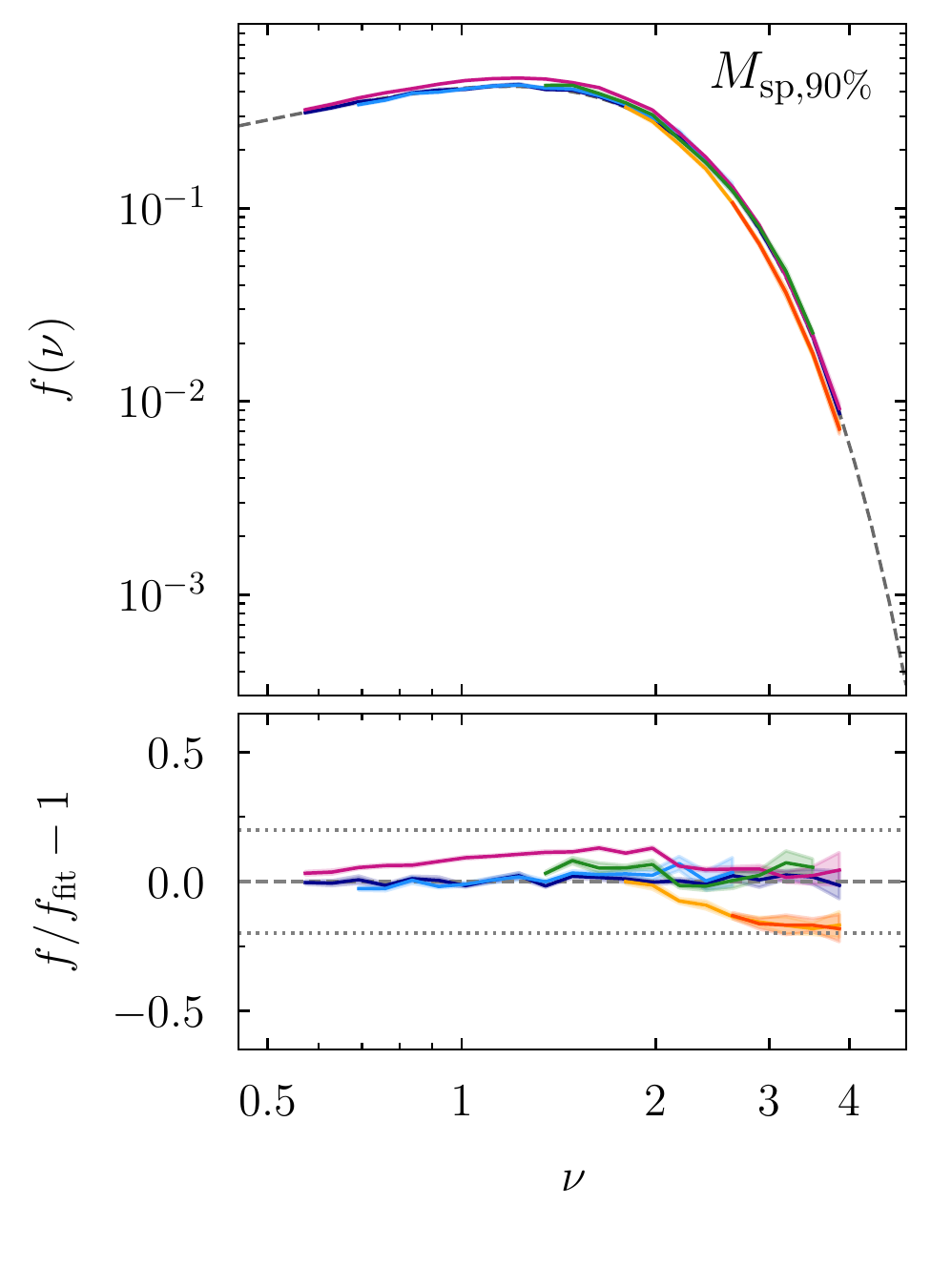}
\caption{Same as Figure~\ref{fig:univ_z} but combining different cosmologies and redshifts, namely the \wmap and \planck cosmologies at $z \approx 0$ and $z = 4$ as well as the self-similar cosmologies with the most extreme power spectra. This selection essentially brackets the parameter space we have investigated. The comparison conclusively demonstrates that none of the SO mass functions are universal; they either exhibit strong redshift evolution, are incompatible with the self-similar simulations, or both. The splashback mass functions are much closer to universality, especially for the highest percentiles.}
\label{fig:univ_all}
\end{figure*}

While universality with redshift is an important consideration, it is by no means sufficient: for $f(\nu)$ to be truly universal, it must hold across cosmologies with very different power spectra. \LCDM cosmologies such as \wmap and \planck do not differ nearly enough in their spectra to claim such universality \citep[e.g.,][]{brown_20}. Thus, we now consider the mass functions in our self-similar simulations (Figure~\ref{fig:univ_ss}). For $\mtom$ (left panel), we arbitrarily use a fit to the $n = -1$ case as a reference (dashed yellow line), for the splashback definitions we use the universal fitting function (dashed gray lines). All SO definitions exhibit significant non-universality with $n$; we show $\mtom$ because we found it to be most universal in the \LCDM case. In the self-similar simulations, $\mtom$ is also the most universal SO definition, although $\mtoc$ is the same by construction and $\mvir$ is very similar; the $\mfoc$ mass function is somewhat less universal. For all SO definitions, steeper power spectra result in shallower mass functions and thus a relative increase of the mass function at high masses. The splashback definitions fare much better, especially $M_{\rm sp,90\%}$ where the different cosmologies agree to about 15\% or better at fixed $\nu$. This agreement is remarkable given the very different structures formed in our self-similar universes (Figure~\ref{fig:viz_ss}).

Having established the basic trends, we now attempt a more ambitious test of universality: directly comparing the different cosmologies and redshifts to each other. To avoid crowding the plots with too many lines, Figure~\ref{fig:univ_all} shows a selection of the most extreme cases, namely low and high redshifts from \wmap and \planck and the self-similar simulations with the shallowest and steepest slopes. As before, we compare the SO mass functions to a fit to the $z \approx 0$ \wmap data and the splashback mass functions to the universal fitting function. This comparison highlights that the non-universality of the SO mass functions is more significant than suggested by Figures~\ref{fig:univ_z} and \ref{fig:univ_ss} individually: in all SO definitions, the mass functions in $n = -1$ and $n = -2.5$ behave qualitatively and quantitatively differently from the low-$z$ \LCDM case (top panels of Figure~\ref{fig:univ_all}). Even for $\mtom$, they strongly disagree at the high-$\nu$ end, and yet both disagree with \LCDM. Again, we note that the $\mtoc$ and $\mtom$ curves are the same for the self-similar simulations, they only appear different in relation to \LCDM. By comparison, the splashback definitions (bottom panels) are much more universal with cosmology: the differences between \LCDM and self-similar cosmologies are no larger than the differences between redshifts in \LCDM. The $n = -1$ mass functions tend to be slightly higher than \LCDM, especially at the low-$\nu$ end. This effect could be caused by their sparse snapshot sampling of only about $3.5$ snapshots per dynamical time, which was shown to cause a small bias \citepalias{diemer_17_sparta}, or by residual convergence issues in this particular simulation (Figure~\ref{fig:convergence}).

The conclusions based on Figure~\ref{fig:univ_all} are quantified by the $\eta$ curves shown in the right panel of Figure~\ref{fig:eta}. Below $\nu = 2$, both $\mvir$ and $\mtom$ are universal to about 20\% accuracy, unlike $\mfoc$ and $\mtoc$ which are clearly non-universal at all peak heights. At the highest masses, however, we observe non-universalities of 180\% or more in all SO definitions, with $\mvir$ and $\mtom$ behaving qualitatively similarly to $\mfoc$ and $\mtoc$. We conclude that the mass functions of any SO definition are not universal with cosmology. The splashback mass functions remain within $\eta \lsim 15\%$ below $\nu = 2$, while the non-universality increases toward high $\nu$ and reaches 40\% for $M_{\rm sp,90\%}$ and up to 60\% for the other percentiles. This peak non-universality is driven by the redshift differences in \LCDM, not by differences to the self-similar cosmologies. 

The universality of the splashback mass functions is remarkable given that they are computed with a fairly complicated algorithm. The slight improvement in universality toward the largest splashback radii suggests that the physically most meaningful halo boundary resides near the sharp drop in density in the outskirts \citep{mansfield_17, xhakaj_20}. We emphasize that \sparta was not tuned to result in a universal mass function; the algorithms described in \citetalias{diemer_20_catalogs} were implemented before measuring the mass functions for this paper.

\subsection{Fitting Function}
\label{sec:results:fit}

We now construct a universal fitting function for the splashback mass function that is valid regardless of redshift and cosmology (to the accuracy implied by Figures~\ref{fig:univ_z}, \ref{fig:univ_ss}, and \ref{fig:univ_all}, and for the range of cosmologies we consider). We use Equation~\ref{eq:fit_global} with parameters that vary depending on the definition of the splashback radius. The $M_{\rm sp, mn}$ mass function deviates from that of $M_{\rm sp, 50\%}$ by only 10\%, but its shape is systematically different. Thus, we provide one set of parameters for $M_{\rm sp,mn}$ and one for all higher percentiles. We parameterize the latter dependence as $p$, the percentile divided by $100$,
\begin{align}
\label{eq:fit_p}
a &= a_{0} + a_{\rm p} \times p^\alpha \nonumber \\
c &= c_{0} + c_{\rm p} \times p^\alpha \,.
\end{align}
The best-fit values of the seven free parameters are given in Table~\ref{table:fits}. They are derived from a fit to the binned mass functions at redshifts $0.13$, $0.5$, $1$, and $2$. We exclude $z = 0$ due to the late-time corrections (Section~\ref{sec:method:sp}) and higher redshifts due to their slight non-universality. We include both the \wmap and \planck samples with equal weight per bin, but we omit the self-similar simulations since we are most interested in a good fit to \LCDM. For the percentile-based definitions, we fit to all percentiles included in our halo catalogs (50, 70, 75, 80, 85, and 90). We have experimented with up-weighting the 50th percentile to account for the gap at low percentiles, but we find that it has little impact. With our dataset thus defined, we constrain all parameters simultaneously using a Levenberg--Marquart least-squares minimization. We add a systematic uncertainty of 3\% in quadrature to the statistical uncertainties shown in the figures. This extra error prevents the most statistically significant bins from dominating the fit and (artificially) results in a roughly unity $\chi^2$ per degree of freedom.

The accuracy of our fitting function can be read off the bottom panels in Figures~\ref{fig:univ_z}, \ref{fig:univ_ss}, and \ref{fig:univ_all}. In \LCDM and below $\nu \approx 2$, the formula is accurate to about 5\% or better for all splashback definitions. At the highest peak heights, the non-universality with redshift means that the accuracy is degraded to about 10\% for $z \lsim 2$ and about 20\% for $z > 2$. For the self-similar simulations, the accuracy depends more strongly on the definition, with $M_{\rm sp, mn}$ fit to about 20\% accuracy and $M_{\rm sp,90\%}$ to 15\% or better regardless of the power spectrum slope. We have implemented our function in the publicly available python code \colossus \citep{diemer_18_colossus}.

We note that the integral over our fit diverges because Equation~\ref{eq:fit_global} approaches $A$ as $\nu \to 0$. It is possible to reparameterize the function to avoid this divergence \citep{tinker_08}, but we do not attempt this conversion. Integrating the mass function from $\nu = 0.5$ to infinity, we find that between 50\% and 60\% of dark matter resides within the splashback radius, depending on the percentile. 

\begin{deluxetable}{lcc}
\tablecaption{Best-fit parameters for the splashback mass function model
\label{table:fits}}
\tablewidth{0.47\textwidth}
\tablehead{
\colhead{Parameter} &
\colhead{$M_{\rm sp,mn}$} &
\colhead{$M_{\rm sp,\%}$}
}
\startdata
\rule{3pt}{0pt} $A_0$           & $   0.1244$ & $   0.0919$ \\
\rule{3pt}{0pt} $a_0$           & $   1.1915$ & $   1.0883$ \\
\rule{3pt}{0pt} $b_0$           & $   0.3379$ & $   0.2421$ \\
\rule{3pt}{0pt} $c_0$           & $   0.4317$ & $   0.4453$ \\
\rule{3pt}{0pt} $a_{\rm p}$     &         $0$ & $   0.1675$ \\
\rule{3pt}{0pt} $c_{\rm p}$     &         $0$ & $  -0.0690$ \\
\rule{3pt}{0pt} $\alpha$        &         $0$ & $   1.7566$ 
\enddata
\end{deluxetable}


\section{Discussion}
\label{sec:discussion}

We have systematically investigated the universality of the multiplicity function and found it to strongly depend on the mass definition. As our results rely on an accurate conversion from mass to peak height, we investigate the subtleties of this calculation in Section~\ref{sec:discussion:nu}. We discuss the theoretical and practical implications of our findings in Section~\ref{sec:discussion:expectations} and consider possible directions for future research in Section~\ref{sec:discussion:future}.

\subsection{What Matters When Computing Peak Height?}
\label{sec:discussion:nu}

Throughout the paper, we have defined non-universality as differences in the multiplicity function at fixed peak height. Given the steep fall-off at high $\nu$, however, relative changes can equally be thought of as differences in $\nu$ at fixed number density. Thus, it is of paramount importance to calculate $\nu(M)$ in a physically meaningful and numerically accurate manner. Moreover, the details of the calculation vary across the literature, making it difficult to cross-compare results. In this section, we discuss the corrections we applied when computing $\sigma$ and $\deltac$, the filter function, and different strategies for computing the underlying power spectrum. Figure~\ref{fig:corrections} shows a graphical impression of their impact on the universality of $M_{\rm sp,90\%}$; the effect on other definitions is similar. 

The first panel shows the multiplicity functions at different redshifts in the \wmap cosmology, essentially the same as in the bottom-right panel of Figure~\ref{fig:univ_z}. Here, however, we have not corrected for finite box size and the collapse threshold is constant, $\deltac = \delta_{\rm c,EdS}$. In the second panel, we have applied the lower limit to the $\sigma$ integral (Section~\ref{sec:theory:corrections}), which accounts for the decreased variance on large scales due to missing low-$k$ modes. This correction increases $\nu$ for the most massive halos in a given box, and brings the redshifts into better agreement.

In the third panel, we additionally evolve $\deltac$ with redshift to account for dark energy (Equation~\ref{eq:deltac}). This correction makes the mass functions in all definitions moderately more universal at the high-$\nu$ end. The correction operates only at $z \lsim 2$ in \LCDM because $\Omega_{\rm m}(z)$ approaches unity at higher redshifts. Our prescription for $\deltac(z)$ is not unique; numerous works have experimented with different values of $\deltac$, its evolution, and dependencies on factors such as the density environment \citep[e.g.,][]{monaco_95, gross_98, tormen_98, lacey_94, desjacques_08, courtin_11, paranjape_13}. In this work, however, we do not treat $\deltac(z)$ as a free parameter. First, changing its normalization has no impact on the universality. Second, we are interested in how well different mass definitions capture the total mass that has collapsed into halos, not in modeling the physics that lead to particular functional forms of $\deltac(z)$. We thus impose the evolution predicted by the standard top-hat collapse collapse model in the presence of dark energy \citep{gunn_72, eke_96}. While this calculation can be complicated \citep[e.g.,][]{, courtin_11, corasaniti_11, benson_13_mergers}, the basic evolution of $\deltac(z)$ with $\Omega_{\rm m}(z)$ does not change even when taking into account minute algorithmic details \citep{pace_17}. Thus, alternative $\deltac(z)$ calculations could change our conclusions only if they include additional physics beyond the top-hat collapse model.

\def\figsize{0.72}
\begin{figure}
\centering
\includegraphics[trim =  3mm 24mm 2mm 75mm, clip, scale=\figsize]{\figdir/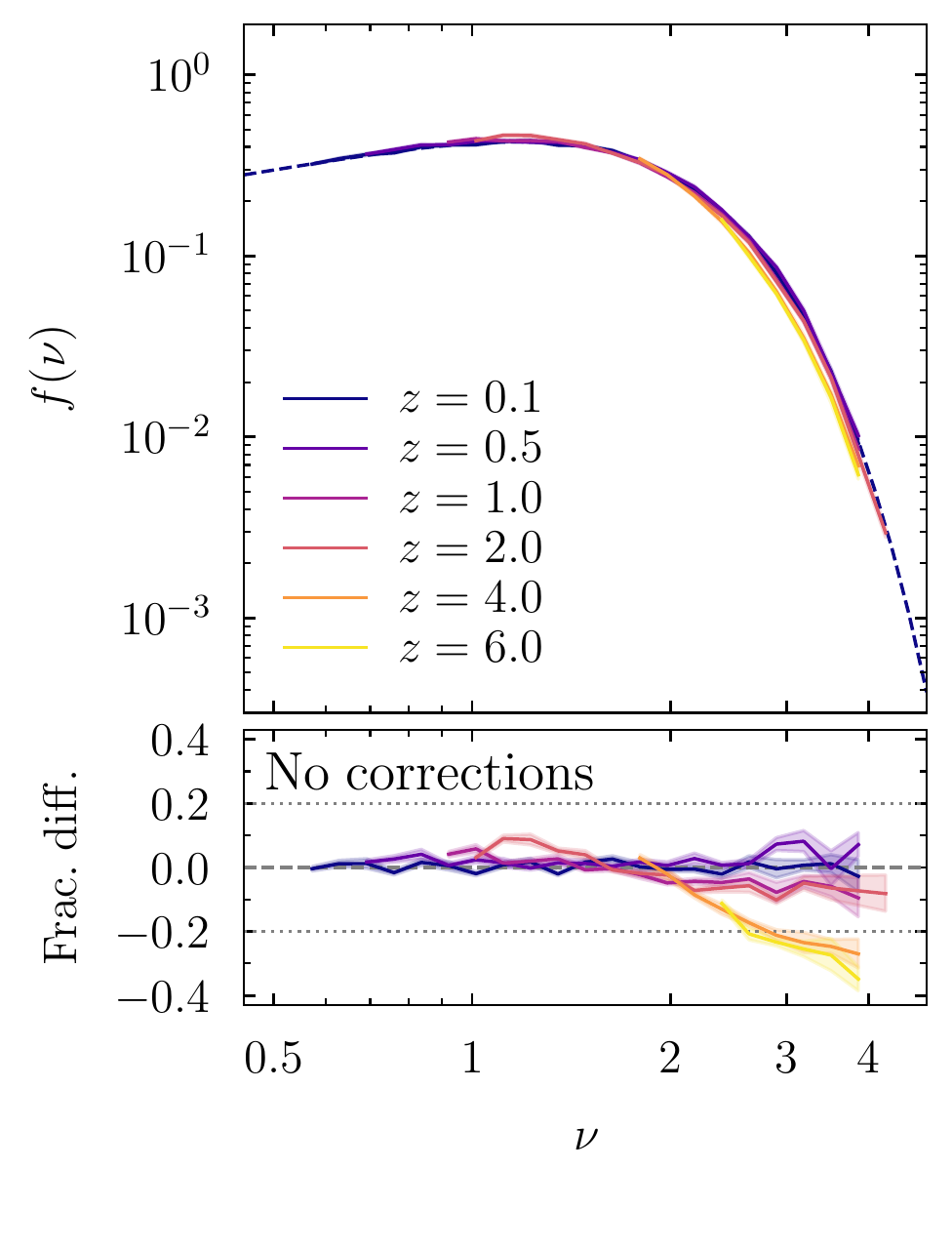}
\includegraphics[trim =  3mm 24mm 2mm 75mm, clip, scale=\figsize]{\figdir/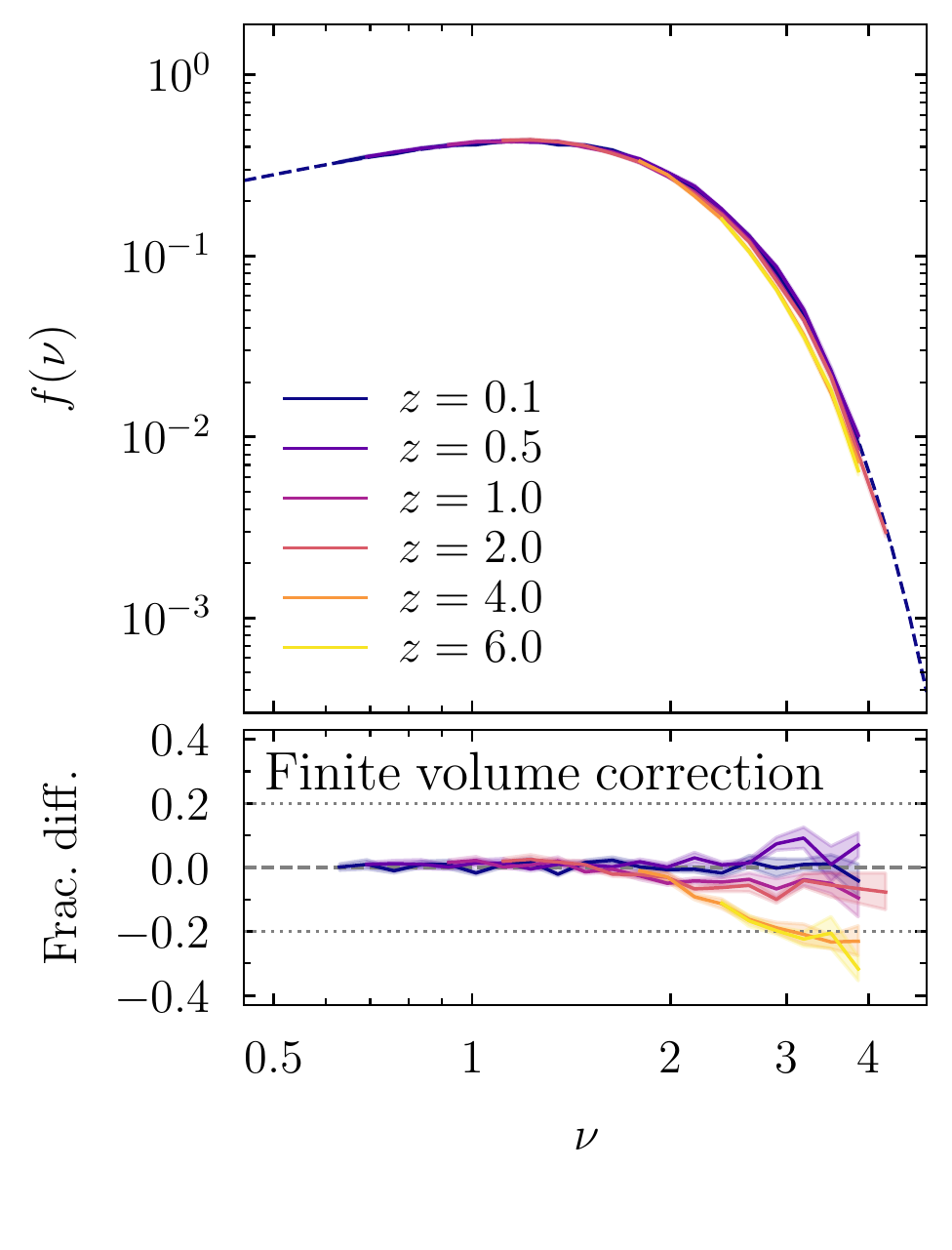}
\includegraphics[trim =  3mm 24mm 2mm 75mm, clip, scale=\figsize]{\figdir/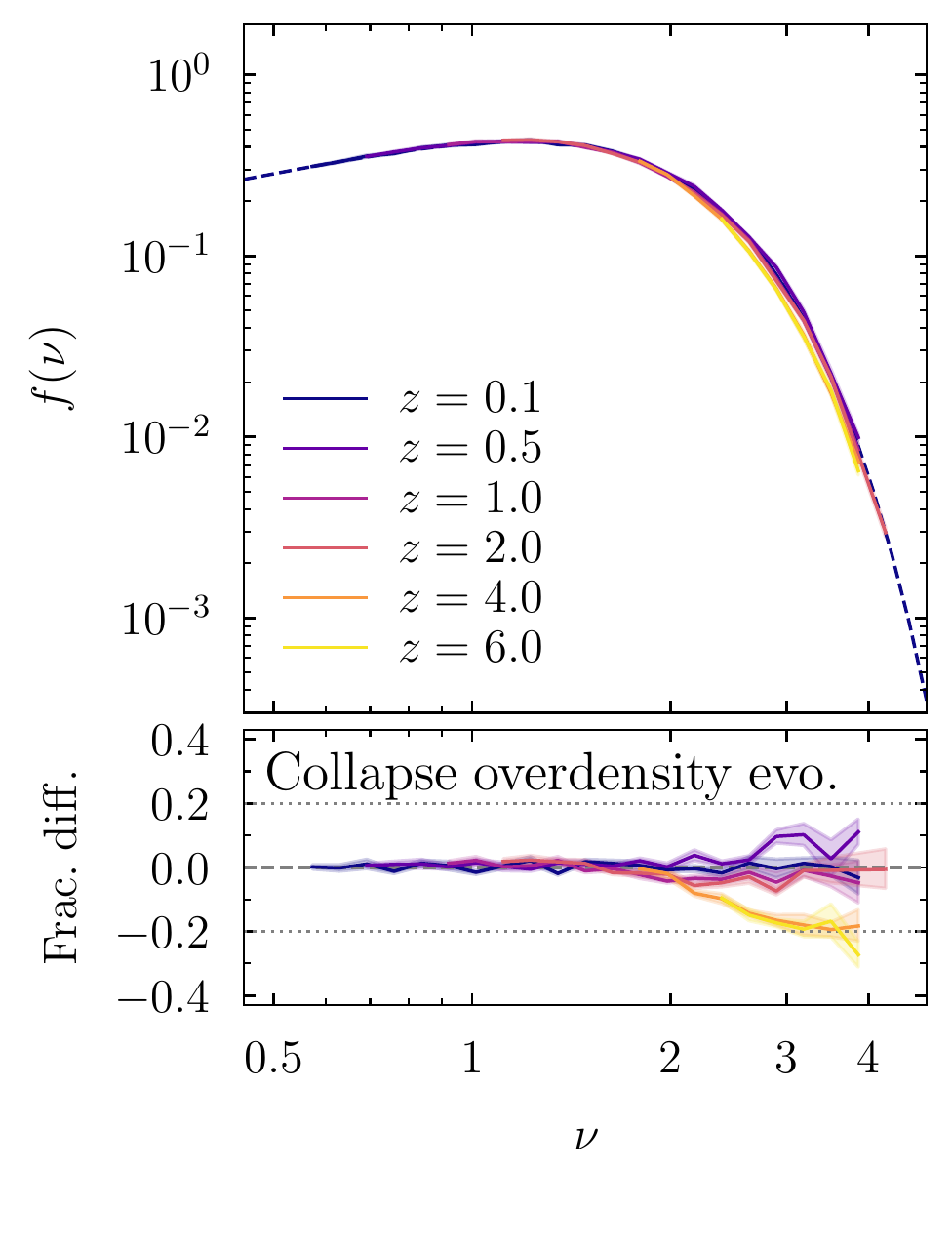}
\includegraphics[trim =  3mm 24mm 2mm 75mm, clip, scale=\figsize]{\figdir/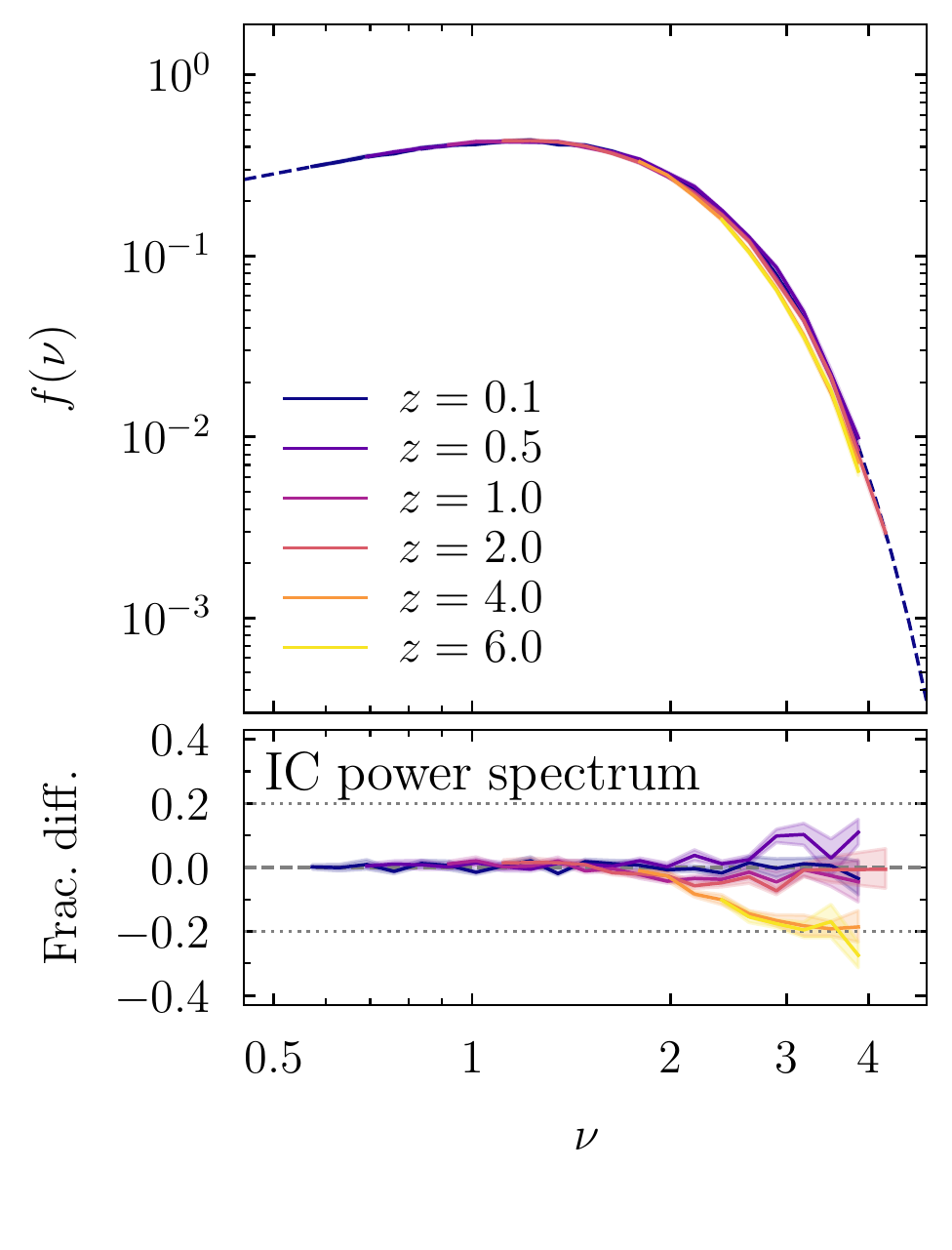}
\includegraphics[trim =  3mm 15mm 2mm 75mm, clip, scale=\figsize]{\figdir/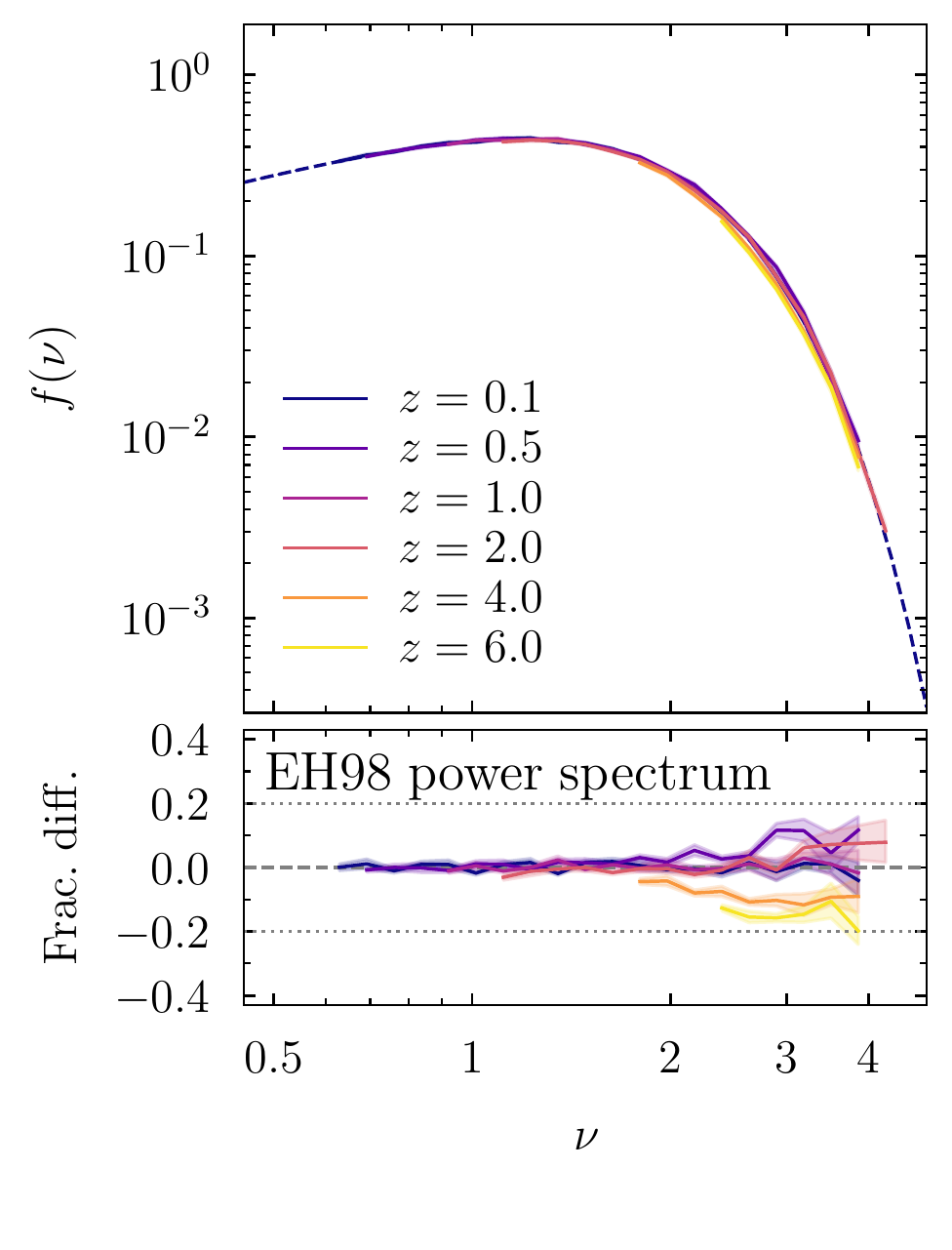}
\includegraphics[trim =  3mm 7mm 2mm 75mm, clip, scale=\figsize]{\figdir/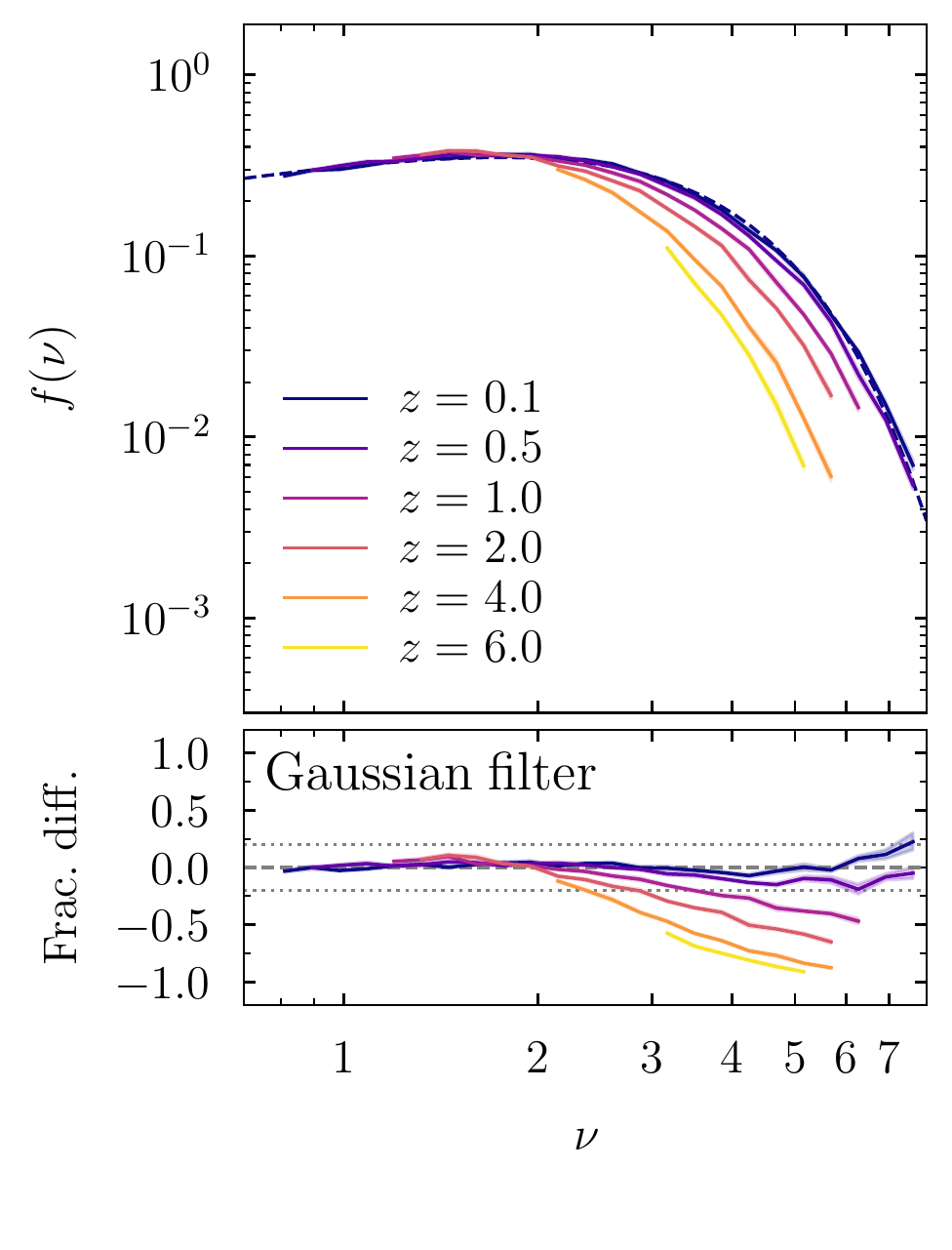}
\caption{Impact of details of the peak height calculation on the universality of $M_{\rm sp,90\%}$; the conclusions are similar for other mass definitions. Each panel shows the ratio of mass functions in the \wmap cosmology compared to a fit at $z = 0.13$, the colors have the same meaning as in Figure~\ref{fig:univ_z} (redshifts $0.13$, $0.5$, $1$, $2$, $4$, and $6$ from blue to yellow). In the top panel, we use a fixed $\deltac$ and no finite-volume correction. In the second panel, we impose a lower limit on the $\sigma$ integral (Section~\ref{sec:theory:corrections}), which has a visible impact. In the third panel, we also take into account the evolution of $\deltac$ in \LCDM cosmologies (Equation~\ref{eq:deltac}), which causes a slight relative shift between redshifts. In the fourth and fifth panels we explore different ways to calculate the power spectrum. First, we directly sum over the modes of the measured power spectrum of the initial conditions rather than the \textsc{Camb} input spectrum; the differences are negligible. In the fifth panel, we approximate the power spectrum with the transfer function of \citet{eisenstein_98}. Although this fit is accurate to about 5\%, it leads to significant relative changes in the mass functions. In the bottom panel, we replace the top-hat filter with a Gaussian, leading to strong non-universality (note the different scales on both axes). See Section~\ref{sec:discussion:nu} for a detailed discussion of these effects.}
\label{fig:corrections}
\end{figure}

In the fourth and fifth panels of Figure~\ref{fig:corrections}, we explore the effect of different power spectrum measurements. \citet{reed_07} advocated summing over the measured power spectrum of the initial conditions of each simulation to capture the random deviations of modes from the input power spectrum. Comparing the third and fourth panels, we find that this procedure makes virtually no difference; we thus use the \textsc{Camb} spectra throughout the paper \citep[see][for similar findings]{despali_16}. This conclusion, however, does not imply that the exact shape of the power spectrum is not important: in the fifth panel, we have approximated $P(k)$ using the \citet{eisenstein_98} transfer function, which is sufficiently accurate for most other applications but causes visible differences in the universality.

One final aspect of the calculation that has received renewed attention is the filter function in Equation~\ref{eq:sigma}. This filter ensures that $\sigma(M)$ refers to those spatial scales that actually contribute to the progenitor density fields of halos. By using the top-hat filter (Equation~\ref{eq:tophat}), we implicitly assume that the material in halos originated from spherical regions with sharp boundaries in real space. As a first check, we have computed $f(\nu)$ with a Gaussian filter, $W \propto \exp[-(k R)^2/2]$, which leads to drastically non-universal mass functions (bottom panel of Figure~\ref{fig:corrections}). The corresponding results for a sharp-$k$ filter (a top-hat in frequency space) are very similar. This result is somewhat expected because the initial peaks of simulated halos were found to have an intermediate shape somewhat closer to a top-hat than a Gaussian \citep{dalal_10, chan_17_filter}. Ideally, the filter should match this shape, leading \citet{chan_17_filter} to propose a filter that transitions between a top-hat and Gaussian. Their function has free parameters that depend on $\nu$, making the predictions somewhat difficult to compute. Nevertheless, we have experimented with the \citet{chan_17_filter} filter using the range of parameter values they suggest. We find that the mass functions move away from universality as the filter moves away from a top-hat. Similarly, \citet{leo_18} propose a smooth analog to the sharp-$k$ filter. We have tested their filter with a range of parameters recommended in their paper but find it to degrade the universality. In summary, the top-hat filter does not exactly capture the shape of peaks in the early universe, but it is difficult to improve upon, not least because the shape of peaks depends on their peak height \citep{bardeen_86}.

The preceding discussion highlights why self-similar universes are so useful in testing universality: many of the effects we have discussed do not apply to them. In particular, the input power spectrum is unambiguous, $\deltac$ is a constant, and finite-volume corrections barely affect universes with $n \gsim -2$ because the overall power is dominated by small scales. As a result, the comparisons of self-similar simulations in Figures~\ref{fig:univ_ss} and \ref{fig:univ_all} are highly robust to the details of the $\nu$ calculation. The filter function, however, matters even in self-similar universes. While their variance is always $\sigma^2 \propto R^{-n-3}$, its normalization depends on $n$. As in \LCDM, using a filter other than top-hat leads to much larger differences between different power spectrum slopes, confirming that the top-hat filter is a reasonable choice \citep[see][for similar experiments]{lacey_94}. Finally, the relative universality of the splashback mass functions in self-similar universes demonstrates that there is no need to include $n$ as a parameter in our fitting function \citep[cf.][]{reed_07}. 

\subsection{The Implications of Universality}
\label{sec:discussion:expectations}

Our key result is that the splashback mass function is remarkably universal, especially for definitions that enclose a large radius such as $R_{\rm sp,90\%}$. In this section, we consider the implications of this finding. In particular, we ask whether we have any reason to expect such universality, whether it could be replicated with other mass definitions, and what the practical uses of a universal mass function might be.

Throughout the paper, we have highlighted numerous reasons why we might not expect the mass function to be universal: the shape of the initial peaks influences their collapse and depends on peak properties beyond $\nu$ \citep{bardeen_86, bond_96_peakpatch}; the large extent of the filter function in $k$ space introduces a dependence on the power spectrum; the collapse of halos is more complicated than suggested by the spherical top-hat collapse model; and, not least, mass definitions may not capture the total mass collapsed into halos. Thus, even with a ``perfect'' definition of the halo boundary (in the sense of the spherical collapse model), we have no reason to expect a perfectly universal mass function --- which makes the agreement of $f(\nu)$ between different redshifts and cosmologies even more remarkable.

Could this success be replicated with other definitions of the halo boundary, such as a low-threshold SO mass?  Such an equivalency seems unlikely because the enclosed overdensity within $\rsp$ depends strongly on mass accretion rate, redshift, and halo mass (Section~\ref{sec:method:sp}). Low SO thresholds are also problematic in practice because they can merge neighboring halos, whereas overlapping splashback radii pose no issue. Another possibility is a threshold that varies with redshift like $\Delta_{\rm vir}$, but its evolution would not affect the universality in the self-similar cosmologies. Similarly, the agreement between FOF mass functions can be improved by changing the linking length, but the chosen value would need to depend on cosmology, mass, and redshift \citep{courtin_11, more_11_fof}.

What are the practical implications of a universal mass function? We might hope to compare counts of observed galaxy clusters to a universal mass function, which would obviate the need to simulate many different cosmologies because the different mass functions could simply be derived from one universal multiplicity function. In practice, however, this application seems unlikely because the accuracy of observational comparisons is limited by systematic uncertainties. While we can quantify simulated mass functions to high accuracy \citep{murray_13_mfunc}, we cannot actually measure the masses (SO or otherwise) of real clusters. Instead, they are inferred from proxies such as the X-ray signal or richness. The goal in cluster cosmology is to find proxies that correlate as tightly as possible with some definition of halo mass that can be measured in simulations and interpolated between cosmologies, for example using emulators \citep{mcclintock_19}. Thus, the uncertainties in the mass-observable relations will likely dominate the comparison for the foreseeable future \citep[e.g.,][]{kravtsov_12, des_20_clusters}.

Leaving aside observational inferences, a universal mass function would have significant theoretical ramifications. The key assumption underlying most models of the mass function is that the physics of halo collapse is insensitive to the power spectrum and that any further dependencies (e.g., on the shape of peaks) can be expressed as a function of only peak height. Our results indicate that real halo collapse is, to a striking degree, captured by this simplistic picture. However, our results do not necessarily support the specific Ansatz of \citet{press_74}, in which the mass in halos above a certain peak height is estimated as the fraction of the Lagrangian volume where the smoothed density field exceeds $\deltac$. While the Press \& Schechter prediction differs from our $M_{\rm sp,90\%}$ fit by only 20\% at low peak heights (as opposed to the infamous 50\% for FOF masses), it still drastically underpredicts the abundance of halos at high $\nu$. In related work, \citet{garcia_20} recently investigated a generic halo boundary optimized to fit the halo--matter correlation function, which appears to roughly match the splashback radius. They found that the corresponding mass function is closely approximated by \citet{press_74}, although for a non-standard value of $\deltac$. While this match is intriguing, our results demonstrate that the mass included within $\rsp$ of large halos is much larger than suggested by the \citet{press_74} model.

\subsection{Future Directions}
\label{sec:discussion:future}

The particularly accurate universality of the splashback mass function at low redshift motivates us to consider possible causes of the remaining non-universality at higher $z$, besides those discussed in Section~\ref{sec:discussion:expectations}. One such effect is non-sphericity: while real halos have long been known to be triaxial \citep{jing_02}, their masses are still routinely measured within spherical apertures. Numerous works have attempted to include non-sphericity via the collapse model calculation, typically expressed as a dependence of $\deltac$ on $\nu$ \citep[e.g.,][]{eisenstein_95, monaco_95, bond_96_peakpatch, audit_97, lee_98_mfunc, sheth_01, sheth_02}. Such corrections are justified by simulation results \citep{robertson_09}, but they affect the predicted shape of $f(\nu)$ rather than its universality. If, in reality, the average non-sphericity depends on redshift or cosmology as well, it could introduce a systematic shift in the mass measurements. Such a bias is likely to affect $\msp$ in particular because the splashback shells around halos can be strongly non-spherical and are poorly fitted by ellipsoids \citep{mansfield_17}. While the corresponding change in $\msp$ may be modest due to the low densities in the outskirts, non-spherical boundaries can drastically change the number of subhalos, which also influences the mass function \citep{mansfield_20_ab}. We note that splashback shells are not well approximated by isodensity contours because of filaments other structures. We plan to update our catalogs with non-spherical halo boundaries in the future.

Similarly, it remains to be seen whether the universality of the splashback mass function is affected by baryonic effects \citep{rudd_08, cui_12, velliscig_14}, although the large extent of the splashback radius should make it less sensitive to baryons than SO definitions. Baryonic simulations could also bridge the gap to observations of the galaxy cluster mass function (Section~\ref{sec:discussion:expectations}). In particular, it would be instructive to compare the scatter in scaling relations between observables and $\msp$ to those for SO masses. Since the scatter between splashback and SO masses is much larger than that in the scaling relations \citepalias{diemer_17_rsp}, it is not obvious that tight relations will emerge. For example, one might expect that small radii such as $\rfoc$ will correlate more tightly with X-ray properties measured near the cluster center. Richness, on the other hand, might correlate better with the splashback radius, given that it is designed to include all subhalos \citep{diemer_20_subs}.
 
Finally, we have not included simulations of more exotic universes with non-Gaussian random fields, modified gravity, or alternative dark matter models \citep[e.g.,][]{lokas_04, dalal_08_ng, pillepich_10, vonbraunbates_18, lovell_20}. We leave investigations of the splashback mass function in such cosmologies for future work.


\section{Conclusions}
\label{sec:conclusion}

We have systematically investigated the universality of halo mass functions in $N$-body simulations over a wide range of halo mass, redshift, and cosmology. For the first time, we have considered splashback masses as well as conventional SO definitions and find that their mass functions are remarkably universal. Our main conclusions are as follows.
\begin{enumerate}

\item We confirm that SO mass functions in \LCDM are generally not universal across redshifts. Out of the commonly used definitions, $\mtom$ is most universal but exhibits deviations of about 20\% across a large range of masses. $\mvir$ is most universal at low peak heights but reaches more than 50\% non-universality at the highest masses. In contrast, the $\mtoc$ and $\mfoc$ mass functions are non-universal at all masses. The splashback mass functions are universal to better than 10\% at $z \lsim 2$ but exhibit non-universalities around 40\% at the highest masses and redshifts. Some fraction of this non-universality might be due to non-sphericity and inaccuracies in the splashback measurements.

\item We compare mass functions across two \LCDM and four self-similar cosmologies with extremely different power spectra. Remarkably, we find that the modest non-universality of the splashback mass function (40\% for the largest radius definition) holds across cosmologies, whereas all SO definitions exhibit strong non-universalities between 180\% and 450\%.

\item Within the splashback definitions, mass functions based on the largest radii (defined by high percentiles of the particle splashback distribution) are most universal, indicating that physical definitions of the halo boundary should include the vast majority of particle and subhalo orbits.

\item We present a simple, universal fitting function for the splashback mass function in any CDM cosmology.

\end{enumerate}
Many details of the splashback mass function remain to be explored, most notably the impact of non-sphericity. The halo catalogs and merger trees that this work is based on are publicly available at \href{http://www.benediktdiemer.com/data}{benediktdiemer.com/data}.


\vspace{0.5cm}

I am deeply grateful to Susmita Adhikari, Shaun Brown, Neal Dalal, Giulia Despali, Michael Joyce, Andrey Kravtsov, Philip Mansfield, Joseph Mohr, and Darren Reed for illuminating discussions and detailed feedback on a draft. I thank the anonymous referee for their constructive suggestions. This work was partially completed during the Coronavirus lockdown and would not have been possible without the essential workers who did not enjoy the privilege of working from the safety of their homes. All computations were run on the \textsc{Midway} computing cluster provided by the University of Chicago Research Computing Center. This research made extensive use of the python packages \textsc{NumPy} \citep{code_numpy2}, \textsc{SciPy} \citep{code_scipy}, \textsc{Matplotlib} \citep{code_matplotlib}, and \colossus \citep{diemer_18_colossus}. Support for Program number HST-HF2-51406.001-A was provided by NASA through a grant from the Space Telescope Science Institute, which is operated by the Association of Universities for Research in Astronomy, Incorporated, under NASA contract NAS5-26555.


\bibliographystyle{aasjournal}
\bibliography{\includedir/bib_mine.bib,\includedir/bib_general.bib,\includedir/bib_structure.bib,\includedir/bib_galaxies.bib,\includedir/bib_clusters.bib,\includedir/bib_hydro_mhd_cr.bib}

\begin{thebibliography}{}
\expandafter\ifx\csname natexlab\endcsname\relax\def\natexlab#1{#1}\fi
\providecommand{\url}[1]{\href{#1}{#1}}

\bibitem[{{Abel} {et~al.}(2012){Abel}, {Hahn}, \& {Kaehler}}]{abel_12}
{Abel}, T., {Hahn}, O., \& {Kaehler}, R. 2012, \mnras, 427, 61

\bibitem[{{Adhikari} {et~al.}(2014){Adhikari}, {Dalal}, \&
  {Chamberlain}}]{adhikari_14}
{Adhikari}, S., {Dalal}, N., \& {Chamberlain}, R.~T. 2014, JCAP, 11, 19

\bibitem[{{Allen} {et~al.}(2011){Allen}, {Evrard}, \& {Mantz}}]{allen_11}
{Allen}, S.~W., {Evrard}, A.~E., \& {Mantz}, A.~B. 2011, \araa, 49, 409

\bibitem[{{Audit} {et~al.}(1997){Audit}, {Teyssier}, \& {Alimi}}]{audit_97}
{Audit}, E., {Teyssier}, R., \& {Alimi}, J.~M. 1997, \aap, 325, 439

\bibitem[{{Bagla} \& {Ray}(2005)}]{bagla_05}
{Bagla}, J.~S., \& {Ray}, S. 2005, \mnras, 358, 1076

\bibitem[{{Bardeen} {et~al.}(1986){Bardeen}, {Bond}, {Kaiser}, \&
  {Szalay}}]{bardeen_86}
{Bardeen}, J.~M., {Bond}, J.~R., {Kaiser}, N., \& {Szalay}, A.~S. 1986, \apj,
  304, 15

\bibitem[{{Behroozi} {et~al.}(2013{\natexlab{a}}){Behroozi}, {Wechsler}, \&
  {Wu}}]{behroozi_13_rockstar}
{Behroozi}, P.~S., {Wechsler}, R.~H., \& {Wu}, H.-Y. 2013{\natexlab{a}}, \apj,
  762, 109

\bibitem[{{Behroozi} {et~al.}(2013{\natexlab{b}}){Behroozi}, {Wechsler}, {Wu},
  {Busha}, {Klypin}, \& {Primack}}]{behroozi_13_trees}
{Behroozi}, P.~S., {Wechsler}, R.~H., {Wu}, H.-Y., {et~al.} 2013{\natexlab{b}},
  \apj, 763, 18

\bibitem[{{Benson}(2017)}]{benson_17}
{Benson}, A.~J. 2017, \mnras, 467, 3454

\bibitem[{{Benson} {et~al.}(2013){Benson}, {Farahi}, {Cole}, {Moustakas},
  {Jenkins}, {Lovell}, {Kennedy}, {Helly}, \& {Frenk}}]{benson_13_mergers}
{Benson}, A.~J., {Farahi}, A., {Cole}, S., {et~al.} 2013, \mnras, 428, 1774

\bibitem[{{Bertschinger}(1985)}]{bertschinger_85}
{Bertschinger}, E. 1985, \apjs, 58, 39

\bibitem[{{Bhattacharya} {et~al.}(2011){Bhattacharya}, {Heitmann}, {White},
  {Luki{\'c}}, {Wagner}, \& {Habib}}]{bhattacharya_11}
{Bhattacharya}, S., {Heitmann}, K., {White}, M., {et~al.} 2011, \apj, 732, 122

\bibitem[{{Bocquet} {et~al.}(2020){Bocquet}, {Heitmann}, {Habib}, {Lawrence},
  {Uram}, {Frontiere}, {Pope}, \& {Finkel}}]{bocquet_20}
{Bocquet}, S., {Heitmann}, K., {Habib}, S., {et~al.} 2020, \apj, 901, 5

\bibitem[{{Bocquet} {et~al.}(2016){Bocquet}, {Saro}, {Dolag}, \&
  {Mohr}}]{bocquet_16}
{Bocquet}, S., {Saro}, A., {Dolag}, K., \& {Mohr}, J.~J. 2016, \mnras, 456,
  2361

\bibitem[{{Bond} {et~al.}(1991){Bond}, {Cole}, {Efstathiou}, \&
  {Kaiser}}]{bond_91}
{Bond}, J.~R., {Cole}, S., {Efstathiou}, G., \& {Kaiser}, N. 1991, \apj, 379,
  440

\bibitem[{{Bond} \& {Myers}(1996)}]{bond_96_peakpatch}
{Bond}, J.~R., \& {Myers}, S.~T. 1996, \apjs, 103, 1

\bibitem[{{Brown} {et~al.}(2020){Brown}, {McCarthy}, {Diemer}, {Font},
  {Stafford}, \& {Pfiefer}}]{brown_20}
{Brown}, S.~T., {McCarthy}, I.~G., {Diemer}, B., {et~al.} 2020, \mnras, 495,
  4994

\bibitem[{{Bryan} \& {Norman}(1998)}]{bryan_98}
{Bryan}, G.~L., \& {Norman}, M.~L. 1998, \apj, 495, 80

\bibitem[{{Chan} {et~al.}(2017){Chan}, {Sheth}, \&
  {Scoccimarro}}]{chan_17_filter}
{Chan}, K.~C., {Sheth}, R.~K., \& {Scoccimarro}, R. 2017, \prd, 96, 103543

\bibitem[{{Corasaniti} \& {Achitouv}(2011)}]{corasaniti_11}
{Corasaniti}, P.~S., \& {Achitouv}, I. 2011, \prd, 84, 023009

\bibitem[{{Courtin} {et~al.}(2011){Courtin}, {Rasera}, {Alimi}, {Corasaniti},
  {Boucher}, \& {F{\"u}zfa}}]{courtin_11}
{Courtin}, J., {Rasera}, Y., {Alimi}, J.-M., {et~al.} 2011, \mnras, 410, 1911

\bibitem[{{Crocce} {et~al.}(2010){Crocce}, {Fosalba}, {Castander}, \&
  {Gazta{\~n}aga}}]{crocce_10}
{Crocce}, M., {Fosalba}, P., {Castander}, F.~J., \& {Gazta{\~n}aga}, E. 2010,
  \mnras, 403, 1353

\bibitem[{{Crocce} {et~al.}(2006){Crocce}, {Pueblas}, \&
  {Scoccimarro}}]{crocce_06}
{Crocce}, M., {Pueblas}, S., \& {Scoccimarro}, R. 2006, \mnras, 373, 369

\bibitem[{{Cui} {et~al.}(2012){Cui}, {Borgani}, {Dolag}, {Murante}, \&
  {Tornatore}}]{cui_12}
{Cui}, W., {Borgani}, S., {Dolag}, K., {Murante}, G., \& {Tornatore}, L. 2012,
  \mnras, 423, 2279

\bibitem[{{Dalal} {et~al.}(2008){Dalal}, {Dor{\'e}}, {Huterer}, \&
  {Shirokov}}]{dalal_08_ng}
{Dalal}, N., {Dor{\'e}}, O., {Huterer}, D., \& {Shirokov}, A. 2008, \prd, 77,
  123514

\bibitem[{{Dalal} {et~al.}(2010){Dalal}, {Lithwick}, \& {Kuhlen}}]{dalal_10}
{Dalal}, N., {Lithwick}, Y., \& {Kuhlen}, M. 2010, arXiv:1010.2539,
  arXiv:1010.2539

\bibitem[{{Davis} {et~al.}(1985){Davis}, {Efstathiou}, {Frenk}, \&
  {White}}]{davis_85}
{Davis}, M., {Efstathiou}, G., {Frenk}, C.~S., \& {White}, S.~D.~M. 1985, \apj,
  292, 371

\bibitem[{{DES Collaboration} {et~al.}(2020){DES Collaboration}, {Abbott},
  {Aguena}, {Alarcon}, {Allam}, {Allen}, {Annis}, {Avila}, {Bacon}, {Bermeo},
  {Bernstein}, {Bertin}, {Bhargava}, {Bocquet}, {Brooks}, {Brout},
  {Buckley-Geer}, {Burke}, {Carnero Rosell}, {Carrasco Kind}, {Carretero},
  {Castander}, {Cawthon}, {Chang}, {Chen}, {Choi}, {Costanzi}, {Crocce}, {da
  Costa}, {Davis}, {De Vicente}, {DeRose}, {Desai}, {Diehl}, {Dietrich},
  {Dodelson}, {Doel}, {Drlica-Wagner}, {Eckert}, {Eifler}, {Elvin-Poole},
  {Estrada}, {Everett}, {Evrard}, {Farahi}, {Ferrero}, {Flaugher}, {Fosalba},
  {Frieman}, {Garcia-Bellido}, {Gatti}, {Gaztanaga}, {Gerdes}, {Giannantonio},
  {Giles}, {Grandis}, {Gruen}, {Gruendl}, {Gschwend}, {Gutierrez}, {Hartley},
  {Hinton}, {Hollowood}, {Honscheid}, {Hoyle}, {Huterer}, {James}, {Jarvis},
  {Jeltema}, {Johnson}, {Kent}, {Krause}, {Kron}, {Kuehn}, {Kuropatkin},
  {Lahav}, {Li}, {Lidman}, {Lima}, {Lin}, {MacCrann}, {Maia}, {Mantz},
  {Marshall}, {Martini}, {Mayers}, {Melchior}, {Mena}, {Menanteau}, {Miquel},
  {Mohr}, {Nichol}, {Nord}, {Ogando}, {Palmese}, {Paz-Chinchon}, {Plazas
  Malag{\'o}n}, {Prat}, {Rau}, {Romer}, {Roodman}, {Rooney}, {Rozo}, {Rykoff},
  {Sako}, {Samuroff}, {Sanchez}, {Saro}, {Scarpine}, {Schubnell}, {Scolnic},
  {Serrano}, {Sevilla}, {Sheldon}, {Smith}, {Suchyta}, {Swanson}, {Tarle},
  {Thomas}, {To}, {Troxel}, {Tucker}, {Varga}, {von der Linden}, {Walker},
  {Wechsler}, {Weller}, {Wilkinson}, {Wu}, {Yanny}, {Zhang}, \&
  {Zuntz}}]{des_20_clusters}
{DES Collaboration}, {Abbott}, T., {Aguena}, M., {et~al.} 2020, arXiv e-prints,
  arXiv:2002.11124

\bibitem[{{Desjacques}(2008)}]{desjacques_08}
{Desjacques}, V. 2008, \mnras, 388, 638

\bibitem[{{Despali} {et~al.}(2016){Despali}, {Giocoli}, {Angulo}, {Tormen},
  {Sheth}, {Baso}, \& {Moscardini}}]{despali_16}
{Despali}, G., {Giocoli}, C., {Angulo}, R.~E., {et~al.} 2016, \mnras, 456, 2486

\bibitem[{{Diemer}(2017)}]{diemer_17_sparta}
{Diemer}, B. 2017, \apjs, 231, 5

\bibitem[{{Diemer}(2018)}]{diemer_18_colossus}
---. 2018, The Astrophysical Journal Supplement Series, 239, 35

\bibitem[{{Diemer}(2020{\natexlab{a}})}]{diemer_20_catalogs}
---. 2020{\natexlab{a}}, arXiv e-prints, arXiv:2007.09149

\bibitem[{{Diemer}(2020{\natexlab{b}})}]{diemer_20_subs}
---. 2020{\natexlab{b}}, arXiv e-prints, arXiv:2007.10992

\bibitem[{{Diemer} \& {Kravtsov}(2014)}]{diemer_14}
{Diemer}, B., \& {Kravtsov}, A.~V. 2014, \apj, 789, 1

\bibitem[{{Diemer} \& {Kravtsov}(2015)}]{diemer_15}
---. 2015, \apj, 799, 108

\bibitem[{{Diemer} {et~al.}(2013){Diemer}, {Kravtsov}, \&
  {More}}]{diemer_13_scalingrel}
{Diemer}, B., {Kravtsov}, A.~V., \& {More}, S. 2013, \apj, 779, 159

\bibitem[{{Diemer} {et~al.}(2017){Diemer}, {Mansfield}, {Kravtsov}, \&
  {More}}]{diemer_17_rsp}
{Diemer}, B., {Mansfield}, P., {Kravtsov}, A.~V., \& {More}, S. 2017, \apj,
  843, 140

\bibitem[{{Eddington}(1913)}]{eddington_13}
{Eddington}, A.~S. 1913, \mnras, 73, 359

\bibitem[{{Efstathiou} {et~al.}(1988){Efstathiou}, {Frenk}, {White}, \&
  {Davis}}]{efstathiou_88}
{Efstathiou}, G., {Frenk}, C.~S., {White}, S. D.~M., \& {Davis}, M. 1988,
  \mnras, 235, 715

\bibitem[{{Eisenstein} \& {Hu}(1998)}]{eisenstein_98}
{Eisenstein}, D.~J., \& {Hu}, W. 1998, \apj, 496, 605

\bibitem[{{Eisenstein} \& {Loeb}(1995)}]{eisenstein_95}
{Eisenstein}, D.~J., \& {Loeb}, A. 1995, \apj, 439, 520

\bibitem[{{Eke} {et~al.}(1996){Eke}, {Cole}, \& {Frenk}}]{eke_96}
{Eke}, V.~R., {Cole}, S., \& {Frenk}, C.~S. 1996, \mnras, 282, 263

\bibitem[{{Elahi}(2009)}]{elahi_09_thesis}
{Elahi}, P.~J. 2009, PhD thesis, Queen's University (Canada)

\bibitem[{{Elahi} {et~al.}(2009){Elahi}, {Thacker}, {Widrow}, \&
  {Scannapieco}}]{elahi_09_plsubs}
{Elahi}, P.~J., {Thacker}, R.~J., {Widrow}, L.~M., \& {Scannapieco}, E. 2009,
  \mnras, 395, 1950

\bibitem[{{Fillmore} \& {Goldreich}(1984)}]{fillmore_84}
{Fillmore}, J.~A., \& {Goldreich}, P. 1984, \apj, 281, 1

\bibitem[{{Garcia} {et~al.}(2020){Garcia}, {Rozo}, {Becker}, \&
  {More}}]{garcia_20}
{Garcia}, R., {Rozo}, E., {Becker}, M.~R., \& {More}, S. 2020, arXiv e-prints,
  arXiv:2006.12751

\bibitem[{{Gnedin} {et~al.}(2011){Gnedin}, {Kravtsov}, \&
  {Rudd}}]{gnedin_11_dc}
{Gnedin}, N.~Y., {Kravtsov}, A.~V., \& {Rudd}, D.~H. 2011, \apjs, 194, 46

\bibitem[{{Gross} {et~al.}(1998){Gross}, {Somerville}, {Primack}, {Holtzman},
  \& {Klypin}}]{gross_98}
{Gross}, M. A.~K., {Somerville}, R.~S., {Primack}, J.~R., {Holtzman}, J., \&
  {Klypin}, A. 1998, \mnras, 301, 81

\bibitem[{{Gunn} \& {Gott}(1972)}]{gunn_72}
{Gunn}, J.~E., \& {Gott}, III, J.~R. 1972, \apj, 176, 1

\bibitem[{{Hahn} {et~al.}(2013){Hahn}, {Abel}, \& {Kaehler}}]{hahn_13}
{Hahn}, O., {Abel}, T., \& {Kaehler}, R. 2013, \mnras, 434, 1171

\bibitem[{{Harris} {et~al.}(2020){Harris}, {Jarrod Millman}, {van der Walt},
  {Gommers}, {Virtanen}, {Cournapeau}, {Wieser}, {Taylor}, {Berg}, {Smith},
  {Kern}, {Picus}, {Hoyer}, {van Kerkwijk}, {Brett}, {Haldane}, {Fern{\'a}ndez
  del R{\'\i}o}, {Wiebe}, {Peterson}, {G{\'e}rard-Marchant}, {Sheppard},
  {Reddy}, {Weckesser}, {Abbasi}, {Gohlke}, \& {Oliphant}}]{code_numpy2}
{Harris}, C.~R., {Jarrod Millman}, K., {van der Walt}, S.~J., {et~al.} 2020,
  arXiv e-prints, arXiv:2006.10256

\bibitem[{{Heitmann} {et~al.}(2019){Heitmann}, {Finkel}, {Pope}, {Morozov},
  {Frontiere}, {Habib}, {Rangel}, {Uram}, {Korytov}, {Child}, {Flender},
  {Insley}, \& {Rizzi}}]{heitmann_19}
{Heitmann}, K., {Finkel}, H., {Pope}, A., {et~al.} 2019, \apjs, 245, 16

\bibitem[{{Hu} \& {Kravtsov}(2003)}]{hu_03}
{Hu}, W., \& {Kravtsov}, A.~V. 2003, \apj, 584, 702

\bibitem[{{Huchra} \& {Geller}(1982)}]{huchra_82}
{Huchra}, J.~P., \& {Geller}, M.~J. 1982, \apj, 257, 423

\bibitem[{Hunter(2007)}]{code_matplotlib}
Hunter, J.~D. 2007, Computing in Science Engineering, 9, 90

\bibitem[{{Jenkins} {et~al.}(2001){Jenkins}, {Frenk}, {White}, {Colberg},
  {Cole}, {Evrard}, {Couchman}, \& {Yoshida}}]{jenkins_01}
{Jenkins}, A., {Frenk}, C.~S., {White}, S.~D.~M., {et~al.} 2001, \mnras, 321,
  372

\bibitem[{{Jing} \& {Suto}(2002)}]{jing_02}
{Jing}, Y.~P., \& {Suto}, Y. 2002, \apj, 574, 538

\bibitem[{{Joyce} {et~al.}(2020){Joyce}, {Garrison}, \&
  {Eisenstein}}]{joyce_20}
{Joyce}, M., {Garrison}, L., \& {Eisenstein}, D. 2020, arXiv e-prints,
  arXiv:2004.07256

\bibitem[{{Juan} {et~al.}(2014){Juan}, {Salvador-Sol{\'e}}, {Dom{\`e}nech}, \&
  {Manrique}}]{juan_14}
{Juan}, E., {Salvador-Sol{\'e}}, E., {Dom{\`e}nech}, G., \& {Manrique}, A.
  2014, \mnras, 439, 3156

\bibitem[{{Kaehler} {et~al.}(2012){Kaehler}, {Hahn}, \& {Abel}}]{kaehler_12}
{Kaehler}, R., {Hahn}, O., \& {Abel}, T. 2012, ArXiv e-prints, arXiv:1208.3206

\bibitem[{{Kitayama} \& {Suto}(1996)}]{kitayama_96_2}
{Kitayama}, T., \& {Suto}, Y. 1996, \apj, 469, 480

\bibitem[{{Klypin} {et~al.}(2011){Klypin}, {Trujillo-Gomez}, \&
  {Primack}}]{klypin_11}
{Klypin}, A.~A., {Trujillo-Gomez}, S., \& {Primack}, J. 2011, \apj, 740, 102

\bibitem[{{Knollmann} {et~al.}(2008){Knollmann}, {Power}, \&
  {Knebe}}]{knollmann_08}
{Knollmann}, S.~R., {Power}, C., \& {Knebe}, A. 2008, \mnras, 385, 545

\bibitem[{{Komatsu} {et~al.}(2011){Komatsu}, {Smith}, {Dunkley}, {Bennett},
  {Gold}, {Hinshaw}, {Jarosik}, {Larson}, {Nolta}, {Page}, {Spergel},
  {Halpern}, {Hill}, {Kogut}, {Limon}, {Meyer}, {Odegard}, {Tucker}, {Weiland},
  {Wollack}, \& {Wright}}]{komatsu_11}
{Komatsu}, E., {Smith}, K.~M., {Dunkley}, J., {et~al.} 2011, \apjs, 192, 18

\bibitem[{{Kravtsov} \& {Borgani}(2012)}]{kravtsov_12}
{Kravtsov}, A.~V., \& {Borgani}, S. 2012, \araa, 50, 353

\bibitem[{{Lacey} \& {Cole}(1994)}]{lacey_94}
{Lacey}, C., \& {Cole}, S. 1994, \mnras, 271, 676

\bibitem[{{Lee} \& {Shandarin}(1998)}]{lee_98_mfunc}
{Lee}, J., \& {Shandarin}, S.~F. 1998, \apj, 500, 14

\bibitem[{{Lee} \& {Shandarin}(1999)}]{lee_99_mfunc}
---. 1999, \apjl, 517, L5

\bibitem[{{Leo} {et~al.}(2018){Leo}, {Baugh}, {Li}, \& {Pascoli}}]{leo_18}
{Leo}, M., {Baugh}, C.~M., {Li}, B., \& {Pascoli}, S. 2018, \jcap, 2018, 010

\bibitem[{{Leroy} {et~al.}(2020){Leroy}, {Garrison}, {Eisenstein}, {Joyce}, \&
  {Maleubre}}]{leroy_20}
{Leroy}, M., {Garrison}, L., {Eisenstein}, D., {Joyce}, M., \& {Maleubre}, S.
  2020, arXiv e-prints, arXiv:2004.08406

\bibitem[{{Lewis} {et~al.}(2000){Lewis}, {Challinor}, \& {Lasenby}}]{lewis_00}
{Lewis}, A., {Challinor}, A., \& {Lasenby}, A. 2000, \apj, 538, 473

\bibitem[{{{\L}okas} {et~al.}(2004){{\L}okas}, {Bode}, \& {Hoffman}}]{lokas_04}
{{\L}okas}, E.~L., {Bode}, P., \& {Hoffman}, Y. 2004, \mnras, 349, 595

\bibitem[{{Lovell}(2020)}]{lovell_20}
{Lovell}, M.~R. 2020, \mnras, 493, L11

\bibitem[{{Ludlow} \& {Angulo}(2017)}]{ludlow_17}
{Ludlow}, A.~D., \& {Angulo}, R.~E. 2017, \mnras, 465, L84

\bibitem[{{Luki{\'c}} {et~al.}(2007){Luki{\'c}}, {Heitmann}, {Habib},
  {Bashinsky}, \& {Ricker}}]{lukic_07}
{Luki{\'c}}, Z., {Heitmann}, K., {Habib}, S., {Bashinsky}, S., \& {Ricker},
  P.~M. 2007, \apj, 671, 1160

\bibitem[{{Luki{\'c}} {et~al.}(2009){Luki{\'c}}, {Reed}, {Habib}, \&
  {Heitmann}}]{lukic_09}
{Luki{\'c}}, Z., {Reed}, D., {Habib}, S., \& {Heitmann}, K. 2009, \apj, 692,
  217

\bibitem[{{Maggiore} \& {Riotto}(2010)}]{maggiore_10}
{Maggiore}, M., \& {Riotto}, A. 2010, \apj, 711, 907

\bibitem[{{Mansfield} \& {Avestruz}(2020)}]{mansfield_20_resolution}
{Mansfield}, P., \& {Avestruz}, C. 2020, arXiv e-prints, arXiv:2008.08591

\bibitem[{{Mansfield} \& {Kravtsov}(2020)}]{mansfield_20_ab}
{Mansfield}, P., \& {Kravtsov}, A.~V. 2020, \mnras, 493, 4763

\bibitem[{{Mansfield} {et~al.}(2017){Mansfield}, {Kravtsov}, \&
  {Diemer}}]{mansfield_17}
{Mansfield}, P., {Kravtsov}, A.~V., \& {Diemer}, B. 2017, \apj, 841, 34

\bibitem[{{McClintock} {et~al.}(2019){McClintock}, {Rozo}, {Becker}, {DeRose},
  {Mao}, {McLaughlin}, {Tinker}, {Wechsler}, \& {Zhai}}]{mcclintock_19}
{McClintock}, T., {Rozo}, E., {Becker}, M.~R., {et~al.} 2019, \apj, 872, 53

\bibitem[{{Mo} {et~al.}(2010){Mo}, {van den Bosch}, \& {White}}]{mo_10_book}
{Mo}, H., {van den Bosch}, F.~C., \& {White}, S. 2010, {Galaxy Formation and
  Evolution} (Cambridge University Press)

\bibitem[{{Monaco}(1995)}]{monaco_95}
{Monaco}, P. 1995, \apj, 447, 23

\bibitem[{{Monaco} {et~al.}(2002){Monaco}, {Theuns}, {Taffoni}, {Governato},
  {Quinn}, \& {Stadel}}]{monaco_02}
{Monaco}, P., {Theuns}, T., {Taffoni}, G., {et~al.} 2002, \apj, 564, 8

\bibitem[{{More} {et~al.}(2015){More}, {Diemer}, \& {Kravtsov}}]{more_15}
{More}, S., {Diemer}, B., \& {Kravtsov}, A.~V. 2015, \apj, 810, 36

\bibitem[{{More} {et~al.}(2011){More}, {Kravtsov}, {Dalal}, \&
  {Gottl{\"o}ber}}]{more_11_fof}
{More}, S., {Kravtsov}, A.~V., {Dalal}, N., \& {Gottl{\"o}ber}, S. 2011, \apjs,
  195, 4

\bibitem[{{Murray} {et~al.}(2013){Murray}, {Power}, \&
  {Robotham}}]{murray_13_mfunc}
{Murray}, S.~G., {Power}, C., \& {Robotham}, A.~S.~G. 2013, \mnras, 434, L61

\bibitem[{{Musso} \& {Sheth}(2012)}]{musso_12}
{Musso}, M., \& {Sheth}, R.~K. 2012, \mnras, 423, L102

\bibitem[{{Pace} {et~al.}(2017){Pace}, {Meyer}, \& {Bartelmann}}]{pace_17}
{Pace}, F., {Meyer}, S., \& {Bartelmann}, M. 2017, \jcap, 2017, 040

\bibitem[{{Paranjape} {et~al.}(2013){Paranjape}, {Sheth}, \&
  {Desjacques}}]{paranjape_13}
{Paranjape}, A., {Sheth}, R.~K., \& {Desjacques}, V. 2013, \mnras, 431, 1503

\bibitem[{{Pillepich} {et~al.}(2010){Pillepich}, {Porciani}, \&
  {Hahn}}]{pillepich_10}
{Pillepich}, A., {Porciani}, C., \& {Hahn}, O. 2010, \mnras, 402, 191

\bibitem[{{Planck Collaboration} {et~al.}(2014){Planck Collaboration}, {Ade},
  {Aghanim}, {Armitage-Caplan}, {Arnaud}, {Ashdown}, {Atrio-Barandela},
  {Aumont}, {Baccigalupi}, {Banday}, \& et~al.}]{planck_14}
{Planck Collaboration}, {Ade}, P.~A.~R., {Aghanim}, N., {et~al.} 2014, \aap,
  571, A16

\bibitem[{{Power} \& {Knebe}(2006)}]{power_06}
{Power}, C., \& {Knebe}, A. 2006, \mnras, 370, 691

\bibitem[{{Press} \& {Schechter}(1974)}]{press_74}
{Press}, W.~H., \& {Schechter}, P. 1974, \apj, 187, 425

\bibitem[{{Reed} {et~al.}(2003){Reed}, {Gardner}, {Quinn}, {Stadel}, {Fardal},
  {Lake}, \& {Governato}}]{reed_03}
{Reed}, D., {Gardner}, J., {Quinn}, T., {et~al.} 2003, \mnras, 346, 565

\bibitem[{{Reed} {et~al.}(2007){Reed}, {Bower}, {Frenk}, {Jenkins}, \&
  {Theuns}}]{reed_07}
{Reed}, D.~S., {Bower}, R., {Frenk}, C.~S., {Jenkins}, A., \& {Theuns}, T.
  2007, \mnras, 374, 2

\bibitem[{{Robertson} {et~al.}(2009){Robertson}, {Kravtsov}, {Tinker}, \&
  {Zentner}}]{robertson_09}
{Robertson}, B.~E., {Kravtsov}, A.~V., {Tinker}, J., \& {Zentner}, A.~R. 2009,
  \apj, 696, 636

\bibitem[{{Rudd} {et~al.}(2008){Rudd}, {Zentner}, \& {Kravtsov}}]{rudd_08}
{Rudd}, D.~H., {Zentner}, A.~R., \& {Kravtsov}, A.~V. 2008, \apj, 672, 19

\bibitem[{{Savitzky} \& {Golay}(1964)}]{savitzky_64}
{Savitzky}, A., \& {Golay}, M.~J.~E. 1964, Analytical Chemistry, 36, 1627

\bibitem[{{Schneider}(2015)}]{schneider_15}
{Schneider}, A. 2015, \mnras, 451, 3117

\bibitem[{{Sheth} {et~al.}(2001){Sheth}, {Mo}, \& {Tormen}}]{sheth_01}
{Sheth}, R.~K., {Mo}, H.~J., \& {Tormen}, G. 2001, \mnras, 323, 1

\bibitem[{{Sheth} \& {Tormen}(1999)}]{sheth_99}
{Sheth}, R.~K., \& {Tormen}, G. 1999, \mnras, 308, 119

\bibitem[{{Sheth} \& {Tormen}(2002)}]{sheth_02}
---. 2002, \mnras, 329, 61

\bibitem[{{Shi}(2016)}]{shi_16_rsp}
{Shi}, X. 2016, \mnras, 459, 3711

\bibitem[{{Springel}(2005)}]{springel_05_gadget2}
{Springel}, V. 2005, \mnras, 364, 1105

\bibitem[{{Tinker} {et~al.}(2008){Tinker}, {Kravtsov}, {Klypin}, {Abazajian},
  {Warren}, {Yepes}, {Gottl{\"o}ber}, \& {Holz}}]{tinker_08}
{Tinker}, J., {Kravtsov}, A.~V., {Klypin}, A., {et~al.} 2008, \apj, 688, 709

\bibitem[{{Tormen}(1998)}]{tormen_98}
{Tormen}, G. 1998, \mnras, 297, 648

\bibitem[{{Tormen} \& {Bertschinger}(1996)}]{tormen_96}
{Tormen}, G., \& {Bertschinger}, E. 1996, \apj, 472, 14

\bibitem[{{Trenti} {et~al.}(2010){Trenti}, {Smith}, {Hallman}, {Skillman}, \&
  {Shull}}]{trenti_10}
{Trenti}, M., {Smith}, B.~D., {Hallman}, E.~J., {Skillman}, S.~W., \& {Shull},
  J.~M. 2010, \apj, 711, 1198

\bibitem[{{Velliscig} {et~al.}(2014){Velliscig}, {van Daalen}, {Schaye},
  {McCarthy}, {Cacciato}, {Le Brun}, \& {Vecchia}}]{velliscig_14}
{Velliscig}, M., {van Daalen}, M.~P., {Schaye}, J., {et~al.} 2014, \mnras, 442,
  2641

\bibitem[{{Virtanen} {et~al.}(2019){Virtanen}, {Gommers}, {Oliphant},
  {Haberland}, {Reddy}, {Cournapeau}, {Burovski}, {Peterson}, {Weckesser},
  {Bright}, {van der Walt}, {Brett}, {Wilson}, {Jarrod Millman}, {Mayorov},
  {Nelson}, {Jones}, {Kern}, {Larson}, {Carey}, {Polat}, {Feng}, {Moore}, {Vand
  erPlas}, {Laxalde}, {Perktold}, {Cimrman}, {Henriksen}, {Quintero}, {Harris},
  {Archibald}, {Ribeiro}, {Pedregosa}, {van Mulbregt}, \&
  {Contributors}}]{code_scipy}
{Virtanen}, P., {Gommers}, R., {Oliphant}, T.~E., {et~al.} 2019, arXiv
  e-prints, arXiv:1907.10121

\bibitem[{{von Braun-Bates} \& {Devriendt}(2018)}]{vonbraunbates_18}
{von Braun-Bates}, F., \& {Devriendt}, J. 2018, \jcap, 2018, 028

\bibitem[{{Warren} {et~al.}(2006){Warren}, {Abazajian}, {Holz}, \&
  {Teodoro}}]{warren_06}
{Warren}, M.~S., {Abazajian}, K., {Holz}, D.~E., \& {Teodoro}, L. 2006, \apj,
  646, 881

\bibitem[{{Watson} {et~al.}(2013){Watson}, {Iliev}, {D'Aloisio}, {Knebe},
  {Shapiro}, \& {Yepes}}]{watson_13_mf}
{Watson}, W.~A., {Iliev}, I.~T., {D'Aloisio}, A., {et~al.} 2013, \mnras, 433,
  1230

\bibitem[{{White}(2001)}]{white_01_mass}
{White}, M. 2001, \aap, 367, 27

\bibitem[{{White}(2002)}]{white_02}
---. 2002, \apjs, 143, 241

\bibitem[{{Xhakaj} {et~al.}(2020){Xhakaj}, {Diemer}, {Leauthaud}, {Wasserman},
  {Huang}, {Luo}, {Adhikari}, \& {Singh}}]{xhakaj_20}
{Xhakaj}, E., {Diemer}, B., {Leauthaud}, A., {et~al.} 2020, \mnras, 499, 3534

\bibitem[{{Yoshida} {et~al.}(2003){Yoshida}, {Sokasian}, {Hernquist}, \&
  {Springel}}]{yoshida_03_running}
{Yoshida}, N., {Sokasian}, A., {Hernquist}, L., \& {Springel}, V. 2003, \apj,
  598, 73

\end{thebibliography}

\end{document}